# Does Google Scholar contain all highly cited documents (1950-2013)?

**Alberto Martín-Martín[1], Enrique Orduña-Malea[2], Juan Manuel Ayllón[1], Emilio Delgado López-Cózar[1]**

[1]EC3: Evaluación de la Ciencia y de la Comunicación Científica, Universidad de Granada (Spain)
[2]EC3: Evaluación de la Ciencia y de la Comunicación Científica, Universidad Politécnica de Valencia (Spain)

## ABSTRACT

The study of highly cited documents on Google Scholar (GS) has never been addressed to date in a comprehensive manner. The objective of this work is to identify the set of highly cited documents in Google Scholar and define their core characteristics: their languages, their file format, or how many of them can be accessed free of charge. We will also try to answer some additional questions that hopefully shed some light about the use of GS as a tool for assessing scientific impact through citations.

The decalogue of research questions is shown below:
1. Which are the most cited documents in GS?
2. Which are the most cited document types in GS?
3. What languages are the most cited documents written in GS?
4. How many highly cited documents are freely accessible?
   a. What file types are the most commonly used to store these highly cited documents?
   b. Which are the main providers of these documents?
5. How many of the highly cited documents indexed by GS are also indexed by WoS?
6. Is there a correlation between the number of citations that these highly cited documents have received in GS and the number of citations they have received in WoS?
7. How many versions of these highly cited documents has GS detected?
8. Is there a correlation between the number of versions GS has detected for these documents, and the number citations they have received?
9. Is there a correlation between the number of versions GS has detected for these documents, and their position in the search engine result pages?
10. Is there some relation between the positions these documents occupy in the search engine result pages, and the number of citations they have received?

To answer these questions, a set of 64,000 documents indexed in Google Scholar has been collected, after performing 64 queries by year (from 1950 to 2013) using Google Scholar's advanced search, and collecting the maximum number of records that GS displays for any given query, which as we know is always 1,000. These 64,000 documents receive 122,245,865 citations in Google Scholar and 35,182,077 in Web of Science Core Collection.

Full raw data available at: http://dx.doi.org/10.6084/m9.figshare.1224314

# 1. INTRODUCTION

## 1.1 About this title

The reason behind the title of this work and its structure as questions is not simply a rhetorical device intended to attract the reader's attention. It is a genuine statement of intentions, since there is no absolute empirical certainty that our sample contains all the highly cited documents present in Google Scholar (GS) at the moment we collected the data. If GS provided a feature that allowed us to sort documents according to number of citations, as traditional bibliometric databases do (Web of Science and Scopus), we wouldn't harbor any doubts about this matter. Since this is not the case, we can not be completely sure that when we make a query by year of publication in GS, it will show us the 1,000 most cited documents published during that range of years (as we know, 1,000 is the maximum number of results GS will display for any given query). In short, we are not entirely sure that the data we collected comprises only highly cited documents in GS, and therefore it is likely that some of these documents don't actually belong to the group of "upper crust" documents in GS for each of the years in the selected range (1950-2013).

Nevertheless, there is strong evidence suggesting that our sample contains a very large portion of the highly cited documents in GS:

**Firstly**, in its documentation, GS explicitly declares[1] that the number of citations received by a document is one of the factors involved in the calculation of the position this document will occupy on the results page, although they don't specify the overall weight of this factor in the calculation. A high correlation between the position documents occupy in the search engine results page (SERP) when they are sorted by Google Scholar's default relevance criteria, and the position they occupy when they are sorted simply by their number of citations (See question 10, Figure 24) would confirm that citation count is indeed the factor that is given the highest weight in Google Scholar's ranking algorithm, and therefore it would be safe to presume that the first positions of a query will always be occupied by the most cited documents that satisfy said query.

**Secondly**, we can see other evidences that support the validity of our sample: in order to verify that the documents in our sample were in fact highly cited documents, we retrieved the top 1,000 most cited documents on the Web of Science Core Collection for each year in the range 1950-2013 (as of October the 30th 2014), and compared the two sets of documents for each year. The results showed that, on average, 81% of the documents in our sample from GS with a link to a WoS record[2] were also present in the ranking of the top 1,000 most cited documents in WoS. With the WoS dataset, we could also learn how many highly cited documents in WoS were missing from our GS dataset. In this respect, the results show that the number of highly cited documents in WoS that are not present in our GS sample is insignificant. There are only 396 (1.3%) documents in our WoS sample that have received enough citations to be included among the 30,000 most cited documents in our GS sample, but that according to their document ID are not present in this sample. Likewise, if we consider the 40,000 most cited documents in our sample, this figure raises to 1,645 (4.1%). As we lower the citation threshold, this figure obviously increases (See Question 1). This result seems logical for two reasons:

---

[1] About Google Scholar: How are documents ranked?
http://scholar.google.com/intl/en/scholar/about.html [accessed on October 7th 2014]

[2] Collaboration between Google Scholar and Web of Science
http://wokinfo.com/googlescholar/ [accessed on October 24th 2014]

a) factor ranking: citations are the main ranking factor but not the only one. Therefore, for documents with the highest number of citations, the position achieved clearly correlated with citations. In contrast, in the lower positions, where the number of citations is also lower, the effect of other ranking factors is more evident.

b) statistical noise: in the first positions, the differences between the documents in terms of citations are high, so the statistical error must be very large to obtain documents in wrong positions. However, as we approach the border cut (1,000 documents), the differences between the documents are smaller, and therefore small errors can result in significant changes in positions over the lower ranks (especially for positions in the margin 800-1,200).

**Lastly**, our own experience, gained through the daily observation of hundreds of searches. Usually, the relevance ranking used by GS is reduced to simply placing the highest cited documents in the first results pages, with very rare exceptions. This is something anyone can check just by doing a search in Google Scholar. We encourage researchers to experience this for themselves.

To sum up, in this work we analyse the 1,000 documents that GS retrieves for each one of 64 queries by year, from 1950 until 2013. Presumably, among them we should be able to find the most cited documents published in each of those years.

## 1.2 Citation Classics: Highly Cited Documents

The idea of identifying the most influential documents in science using the number of citations they generate in the scientific literature was introduced, like many other bibliometric tools, by Eugène Garfield. On January $3^{rd}$ 1977, Garfield published an essay entitled "Introducing Citation Classics: the human side of scientific papers" (1977), which appeared in *Current Contents*. The candidates for Citation Classics were selected from a group of 500 most cited papers during the years 1961-1975. Many of these had been listed before in Current Contents. From 1977 to 1993, 400 Citation Classic Commentaries were published in Current Contents. The full texts of these mostly one-page articles are now available in an open access server at http://garfield.library.upenn.edu/classics.html.

From 2001, the Highly Cited Papers were integrated in a new product from Thomson Scientific: the Essential Science Indicators. Neither Scopus nor other databases have released alternatives to this product.

What we do have is an extensive scientific literature, published during the last few decades, on the matter of highly cited documents in different journals, subject areas, institutions or countries (Oppenheim & Renn 1978; Narin & Frame 1983; Plomp 1990; Glänzel & Czerwon 1992; Glänzel, & Schubert 1992a-b; Glänzel et al. 1995; Tijssen et al. 2002; Aksnes 2003; Aksnes & Sivertsen 2004; Kresge et al. 2005; Levitt & Thelwall 2009; Smith 2009; Persson 2010). Recently, the need of ranking any product of scientific activity according to its citation performance has caused the emergence of this kind of classifications (top 1%, 10%, 15%). The calculation of percentiles, previously proposed explicitly by Maltrás (2003), has recently been rediscovered by other authors (Bornmann 2010, Bornmann & Mutz 2011, Bornmann et al. 2011).

The appearance of Google Scholar opened up new possibilities in this field. Its birth at the end of 2004 signaled a revolution in the way scientific publications were searched, retrieved and accessed (Jacsó, 2005).

From the get-go, GS became not only a search engine for scientific and academic documents, but also for the citations these documents receive. Although it took five years to get over its "beta" stage, today we can say without a doubt that GS is not only the largest database of scientific, academic and technical information in the world (Orduña-Malea et al., 2014, Ortega 2014), but also the richest and most varied, since Google's crawlers systematically parse and process the whole academic web, not making distinctions based on subject areas, languages, or countries (Ortega 2014). Despite the limitations of its spiders and processing software, the lack of normalization processes and quality control filters, GS is an irreplaceable source of global scientific knowledge.

Studies about GS have been limited to: a) explain how it works, its features, limitations, errors, etc.; b) define its coverage and size; c) compare the number of citations received by documents of a given subject area in GS, to the citations they receive in other databases; and d) its growth and evolution over time. However, the study of highly cited documents regardless of their discipline or field has never been addressed in a comprehensive manner.

Therefore, the objective of this work is to identify the set of highly cited documents in GS and define their core characteristics: language, file format, and how many of them can be accessed to free of charge. We will also try to answer some additional questions that - hopefully - shed some light about the use of GS as a tool for assessing impact through citations.

In short, we intend to answer the following questions:

# 2. RESEARCH QUESTIONS

1. Which are the most cited documents in GS?
2. Which are the most cited document types in GS?
3. In what languages are the most cited documents in GS written?
4. How many highly cited documents are freely accessible?
   a. What file types are the most commonly used to store these highly cited documents?
   b. Which are the main providers of these documents?
5. How many of the highly cited documents indexed by GS are also indexed by WoS?
6. Is there a correlation between the number of citations that these highly cited documents have received in GS and the number of citations they have received in WoS?
7. How many versions of these highly cited documents has GS detected?
8. Is there a correlation between the number of versions GS has detected for these documents, and the number citations they have received?
9. Is there a correlation between the number of versions GS has detected for these documents, and their position in the search engine result pages?
10. Is there some relation between the positions these documents occupy in the search engine result pages, and the number of citations they have received?

# 3. MATERIALS AND METHODS

This longitudinal study describes a set of 64,000 documents indexed in Google Scholar, obtained after performing 64 queries by year (from 1950 to 2013) using Google Scholar's advanced search, and collecting the maximum number of records that GS displays for any given query, which as we know is always 1,000.

This process was carried out twice, with a few days between the first and the second download processes. In one case, it was done from a computer connected to our university's IP range (to obtain WoS data embedded in GS), and in the other case, from a computer with a normal Internet connection (to obtain data about open access links unadulterated by our university's subscriptions). Besides, this also worked as a reliability check, because we confirmed that the two datasets contained the same records. These processes took place on the 28th of May and 2nd of June, 2014.

We downloaded the source HTML code for each of the result pages in our queries, parsed them to extract all the relevant information, and saved it in spreadsheet, which is a format more appropriate for the analysis of data. The fields extracted were the following (Figure 1):

- **Publication year:** It is the year that was used in the query, and not that contained in the bibliographical description of the record retrieved.
- **Rank:** The position that each document occupies in the search engine results page of GS.
- **Full Text:** Only marked when GS found a freely accessible version of the document. Then, some additional fields were obtained:
    - **Domain:** The domain where GS has found a full text version of the document.
    - **Link:** Link to the full text of the document.
    - **Format:** File type of the full text version of the document.
- **Brackets:** Some records display text in square brackets before the title of the document. The most common occurrences are: "[BOOK]" (the record is a book) and "[CITATION]" (the record has only been found in the reference list of another document), "[PDF]" and "[HTML]" (to indicate that the document has been found in those formats).
- **Title:** Title of the document.
- **Title Link:** The URL pointing to where the record has been found (it is not a link to a freely accessible version of the full text, since the document may be behind a paywall).
- **Authors – Publication Source – Year – Domain/Publisher:** This field contains information about the authors, publication source, year of publication, and publisher of each document. However, not all this information is always displayed for all records, and it is usually cropped to fit one line.
    - **Authors:** List of authors. When the author has a public Google Scholar Citations profile, his/her name includes a link to his/her profile. When there are many authors, only the first two or three are displayed.
    - **Publication source:** Name of the source where the document has been published, and, sometimes, publication details (volume, issue, pages). This information is not always displayed, and when it is, it's not always complete.
    - **Year:** year when the document was published. This field has been proved to correspond with the field "Publication year", previously described.
    - **URL domain / Publisher:** Domain where this document has been found, or, sometimes, the name of its publisher (only for big publishers).
- **Abstract:** First lines of the abstract (it is also cropped to fit a fixed space).
- **GS Citations:** Number of citations the document has received according to GS.
- **Link to GS Citations:** URL pointing to the list of citing documents in Google Scholar.
- **Link to Related documents:** URL pointing to the list of related documents.
- **Versions:** Number of versions GS has found of the documents.

- **Link to Versions:** URL pointing to the list of versions GS has found of the same document.
- **Web of Science:** This data will only appear if the query is performed from a computer connected to an IP range with access to Thomson Reuters' Web of Science, and only for the documents that are indexed both in GS and WoS.
  - **WoS Citations:** Number of citations according to Web of Science.
  - **WoS accession number (UT)**: identification number of the document in Web of Science. This code allows us to accurately match a GS record with a WoS record.
  - **WoS Link:** URL pointing to the list of citing documents in Web of Science.

**Figure 1. Fields extracted from Google Scholar's SERP**

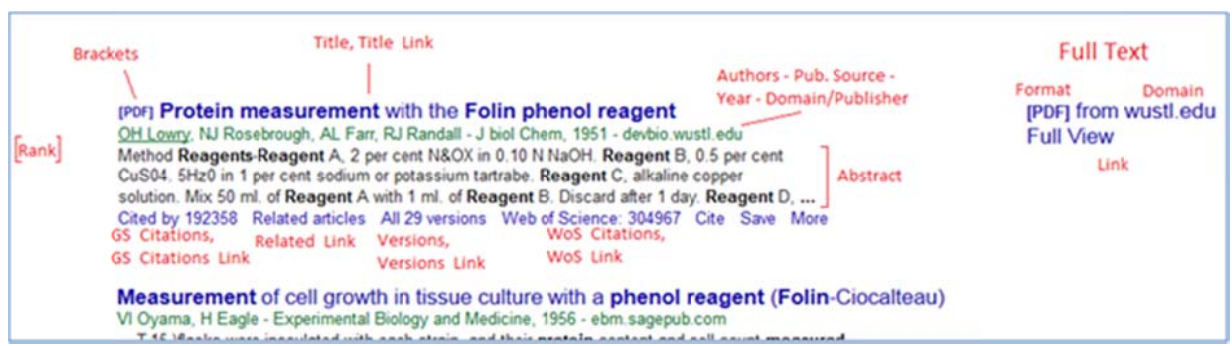

In addition to these fields, we added a few more in order to answer our questions related to: **type of the source publication**, and **language** of the document.

Given the difficulty of ascertaining the typologies of the documents indexed in Google Scholar (this information is not systematically provided by the search engine), we have devised three different strategies that, combined, have allowed us to know the type of a large portion of documents in our data set:

a) All documents where the field **Brackets = "**[BOOK]" have been considered as books (codified as "B").

b) For documents that were also indexed in WoS, GS data was merged with WoS data to obtain the document types. The correspondence is as follows:
  - Journal ("J"): "Article", "Letter", "Note", "Reviews".
  - Book ("B"): "Book", "Book Chapter".
  - Conference Proceedings ("C"): "Proceedings Papers".
  - Others ("O"): "Book Review", "Correction", "Correction, Addition", "Database Review", "Discussion", "Editorial Material", "Excerpt", "Meeting Abstract", "News Item", "Poetry", "Reprint", "Software Review".

  c) Lastly, we analysed the **publication source** (where possible), searching for keywords that could indicate the type of the source publication:
  - Journal ("J"): "Revista", "Anuario", "Cuadernos", "Journal", "Revue", "Bulletin", "Annuaire", "Anales", "Cahiers","Proceedings"[3].

---

○ Conference Proceedings ("C"): "Proceedings", "Congreso", "Jornada", "Seminar", "Simposio","Congrès", "Conference", "symposi", "meeting".

Combining these three strategies, we identified the document type for 71% of the 64,000 documents in our sample. We couldn't identify the document types for the remaining 29% because this would have required doing it manually for 18,590 documents, which would have taken an excessive amount of time. This information was saved in a new field called **Source Type**, and was codified as follows:

- B: Books or book chapters.
- J: Journal articles, reviews, letters and notes.
- C: Conference proceedings.
- O: Others (meeting abstracts, corrections, editorial material…).
- Unknown: we haven't been able to assign a source type (29% of the sample).

As regards the language of the documents (GS doesn't provide this information either), we used the language in which the title and abstract of the document were written, as well as WoS data (when available) as a basis for a new **Language** field.

In essence, we will show a sectional view (global results) as well as a longitudinal view (results by year, in order to detect potential changes) of this sample of documents.

The measures we have used to summarise the data are: absolute and relative frequencies of various aspects of the documents (questions 1-5), and the Pearson correlation (questions 6-10), with $p \leq 0.01$.

## 4. RESULTS

The structure we have followed to present the results of each research question is as follows: first we describe the results we have obtained, and after that, under a separate heading called "Discussion & limitations", we lay out and discuss possible inquiries and uncertainties raised by these findings.

# Question 1.
# Which are the most cited documents in Google Scholar?

In Table 1 we present the top 25 most cited documents in Google Scholar. Additionally, Appendix A shows the top 1% most cited documents in our sample (a total of 640 documents).

These lists are a faithful reflection of the all-encompassing indexing policies of Google Scholar: the academic/scientific/technical world against the scientific world displayed in traditional citation-based databases. In this respect, we can state that GS offers an original and different vision as regards what the most influential documents in the academic/scientific world are, from the perspective of their citation count. This is caused by several reasons:

First, its coverage is not limited to seminal research works in the entire spectrum of scientific fields, but it also covers greatly influential works directed not only to researchers but also to people who are training to become researchers or practitioners in their respective fields. This is testified by the presence of statistical manuals (*Handbook of Mathematical Functions with Formulas, Graphs, and Mathematical Tables; Biostatistical Analysis; Statistical Power Analysis for the Behavioral Science*), laboratory manuals (*Molecular cloning: a laboratory manual)*, manuals of research methodology (*Case study research: Design and methods*), and works that have become a *de facto* standard in professional practice (*Diagnostic and statistical manual of mental disorders, Numerical recipes: the art of scientific computing; Genetic algorithms in search, optimization, and machine learning*).

Second, a high proportion of the highly cited documents are books (a document type that is essential in the humanities and the social sciences as a vehicle for the communication of new results, and in the experimental sciences as a way to consolidate and disseminate knowledge). In fact, 62% of the top 1% most cited documents in our sample are books (see Appendix A). Moreover, books are the document type with a highest citation average: 2,700, against an average of 1,700 in journal articles, and 2,200 for conference proceedings. The importance of books and conference proceedings is therefore thoroughly proven.

Although the ranking is dominated by studies from the natural sciences, and within those, especially the life sciences, it also contains many works from the social sciences, especially from economics, psychology, sociology, education… and also from the Humanities (philosophy and history). For instance: *The structure of scientific revolutions; Diffusion of innovations;* and *Imagined communities. Reflections on the origin and spread of nationalism*).

Many of the works leading this ranking are clearly methodological in nature: they describe the steps of a certain procedure or how to handle basic tools to process and analyse all kinds of data. Precisely because they are essential to researchers, they reach such a high number of citations. This phenomenon is widely known in bibliometrics, where it has already been observed that works that deal with new data collecting and processing techniques or methodologies are more likely to receive a great number of citations.

Even though, as we comment before, GS presents a very different ranking of highly cited academic documents compared to the rankings offered by the traditional citation-based databases, in other aspects it presents a very similar portrait of the world of research to the one offered in traditional databases. This is so because the most cited scientific documents in GS match very closely with those that have been already identified as highly cited in the Web of Science (Garfield, 2005). This explains the high correlation found in the rankings of documents according to their number of citations in GS and WoS (See Question 6).

**8**

Therefore, it is not surprising that the most cited document according to GS is the already famous article written by Lowry, "*Protein measurement with the Folin phenol reagent"* published in 1951 in the *Journal of Biological Chemistry,* where he developed a new method to measure the concentration of a protein in a solution. The reasons for the success of this article were revealed by the author himself (Lowry, 1977), and in a short note published in the same journal on the occasion of its hundredth anniversary in 2005 (Kresge et al., 2005).[4]

We'll use this article as an example in the next section to comment some uncertainties and discuss the possible limitations of these results.

---

[4] See his profile on Google Scholar:
http://scholar.google.com/citations?user=YCS0XAcAAAAJ&hl=es

### Table 1. Top 25 most cited documents in Google Scholar (1950-2013)

| Document type | Bibliographic reference | 1st ed. Pub. Year | GS Citations |
|---|---|---|---|
| J | LOWRY, O.H. et al., (1951). Protein measurement with the Folin phenol reagent.The Journal of biological chemistry, 193(1), 265-275. | 1951 | 253671 |
| J | LAEMMLI, U.K. (1970). Cleavage of structural proteins during the assembly of the head of bacteriophage T4. Nature, 227(5259), 680-685. DOI: 10.1038/227680a0 | 1970 | 221680 |
| J | BRADFORD, M.M. (1976). A rapid and sensitive method for the quantitation of microgram quantities of protein using the principle of protein dye binding. Analytical Biochemistry, 72, 248-254. DOI: 10.1006/abio.1976.9999 | 1976 | 185749 |
| B | SAMBROOK, J., FRITSCH, E. F., & MANIATIS, T. (1982). Molecular cloning: a laboratory manual. New York, Cold Spring Harbor Laboratory Press. | 1982 | 171004 |
| B | AMERICAN PSYCHIATRIC ASSOCIATION. (1952). Diagnostic and statistical manual: mental disorders. Washington, American Psychiatric Assn., Mental Hospital Service. | 1952 | 129473 |
| B | PRESS, W. H. (1986). Numerical recipes: the art of scientific computing. Cambridge [Cambridgeshire], Cambridge University Press. | 1986 | 108956 |
| B | YIN, R. K. (1984). Case study research: design and methods. Beverly Hills, Calif, Sage Publications. | 1984 | 82538 |
| B | ABRAMOWITZ, M., & STEGUN, I. A. (1964). Handbook of mathematical functions: with formulas, graphs, and mathematical tables. Washington, Government printing office. | 1964 | 80482 |
| B | KUHN, T. S. (1962). The structure of scientific revolutions. Chicago, University of Chicago Press. | 1962 | 70662 |
| B | ZAR, J. H. (1974). Biostatistical analysis. Englewood Cliffs, Prentice Hall international. | 1974 | 68267 |
| J | SHANNON, C.E. (1948). A mathematical theory of communication. The Bell System Technical Journal, 27, 379-423. | 1948 | 66851 |
| J | CHOMCZYNSKI, , & SACCHI, N. (1987). Single-step method of RNA isolation by acid guanidinium thiocyanate-phenol-chloroform extraction. Analytical Biochemistry, 162, 156-159. DOI: 10.1006/abio.1987.9999 | 1987 | 63871 |
| J | SANGER F, NICKLEN S, & COULSON AR. (1977). DNA sequencing with chain-terminating inhibitors. Proceedings of the National Academy of Sciences of the United States of America. 74, 5463-7. DOI: 10.1073/pnas.74.12.5463 | 1977 | 63767 |
| B | COHEN, J. (1969). Statistical power analysis for the behavioral sciences. New York, Academic Press. | 1969 | 63766 |
| B | GLASER, B. G., & STRAUSS, A. L. (1967). The discovery of grounded theory: strategies for qualitative research. New York, Aldine de Gruyter. | 1967 | 61158 |
| B | NUNNALLY, J. C. (1967). Psychometric Theory. New York , McGraw-Hill. | 1967 | 60725 |
| B | GOLDBERG, D. E. (1989). Genetic algorithms in search, optimization, and machine learning. Reading, Mass, Addison-Wesley Pub. Co. | 1989 | 59764 |

## *Discussion & Limitations*

1.  *How confident are we that the 64,000 documents that make up our sample really contain the most cited documents in GS?*

    Although there are certain evidences that suggest that we have been able to collect the vast majority of the most cited documents in GS between 1950 and 2013 (as of the 28th of May 2014), as we already explained at the beginning of this study (see Introduction), there are still some questions that should be cleared up.

    To this end, first we have tried to find out if any of the documents in our sample aren't really highly cited documents, and second, if there are any highly cited documents that haven't been included in our sample. To do this, we have compared the 1,000 most cited papers in GS against the 1,000 most cited papers in WoS between 1950 and 2013 (Figure 2).

**Figure 2. Minimum number of citations received by top cited (1,000, 900, 890, 850) documents in Google Scholar and WoS (1950-2013)**

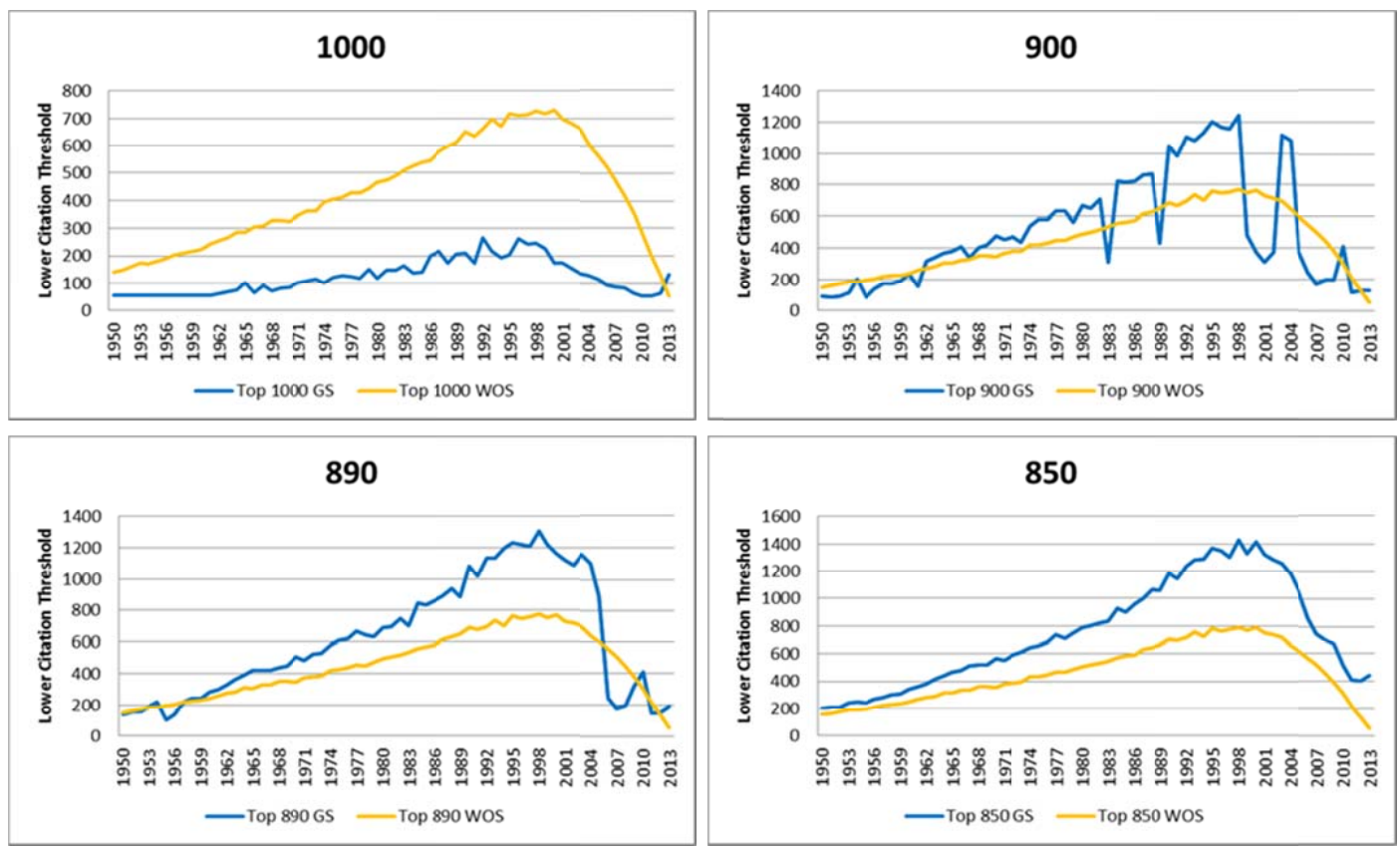

On the one hand, we have detected that the results displayed by GS to our queries become extremely erratic in terms of their citation count from about the 900th result onwards. This means that it is highly probable that approximately the last 100 documents for each year in our sample (a total of 6,400 documents) aren't actually highly cited documents, and therefore should be excluded from the sample.

In contrast, we also have checked that some documents in WoS with a number of citations that slightly exceed the threshold set by the 1,000 documents returned by GS, are not present in the first 1,000 results of the search engine.

Nonetheless, all these inconsistencies happen in the last 100 positions of each query for each year, whereas in the first 900 the consistency is high. To sum up, despite the various limitations described above, we can affirm that the majority of the documents in our sample are highly cited documents.

2.  *In order to be able to trust the results that our search strategy yielded, we must ask ourselves if the documents in our sample were really published in the year GS says they were published.*

To answer this question we carried out two different tests. In the first place, we tested the internal consistency of the search engine. We checked if the results displayed by GS met the requirements of our query. We found that the year of publication of the documents according to GS matched the year we entered in our query in practically 100% of the cases. Only two records out of 64,000 displayed a different year to the one we typed in the search box.

Secondly, we tested the external consistency. For those documents that had been linked to a WoS record (32,680 out of 64,000), we compared the publication year according to GS to the one displayed in the WoS record. Since WoS is a controlled database with a minimum error rate as regards its bibliographic information, we have used it as a benchmark. The results showed that the publication years in GS and WoS matched in 96.7% of the cases (31,600 documents). Curiously enough, the years where we detected more mismatches were 2012 and 2013. Consequently, we must conclude that the error rate in the publication years is very low for this subset of the sample.

**Figure 3. Publication year mismatches between journal articles in Google Scholar and Web of Science**

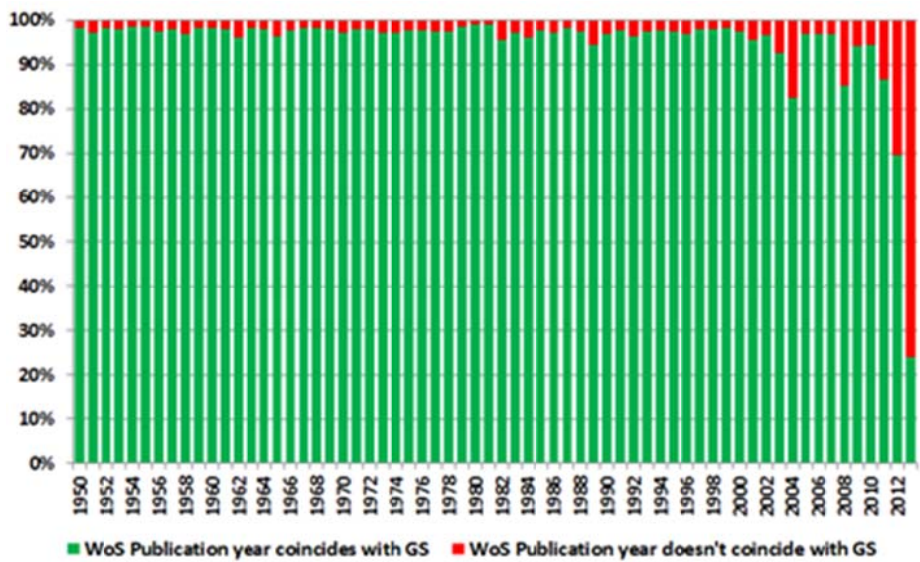

However, we have observed that, in the case of books, Google lumps together all the different editions of a same book, and systematically selects the latest edition of the book as the primary version. As a result, GS takes the publication date of the last edition (and not the publication date of the first edition) as the publication date of the book. This decision, as understandable as it is from a search point of view (users will probably want to access the latest edition of a book), obviously affects our sample. In Figure 4, the frequency distributions for both the publication year of the top 600 most cited books in our sample according to Google Scholar, and the publication year of the 1st edition of these books are displayed.

**Figure 4. Differences between the publication year of the top 600 most cited books according to Google Scholar, and the publication year of the 1st edition of these books**

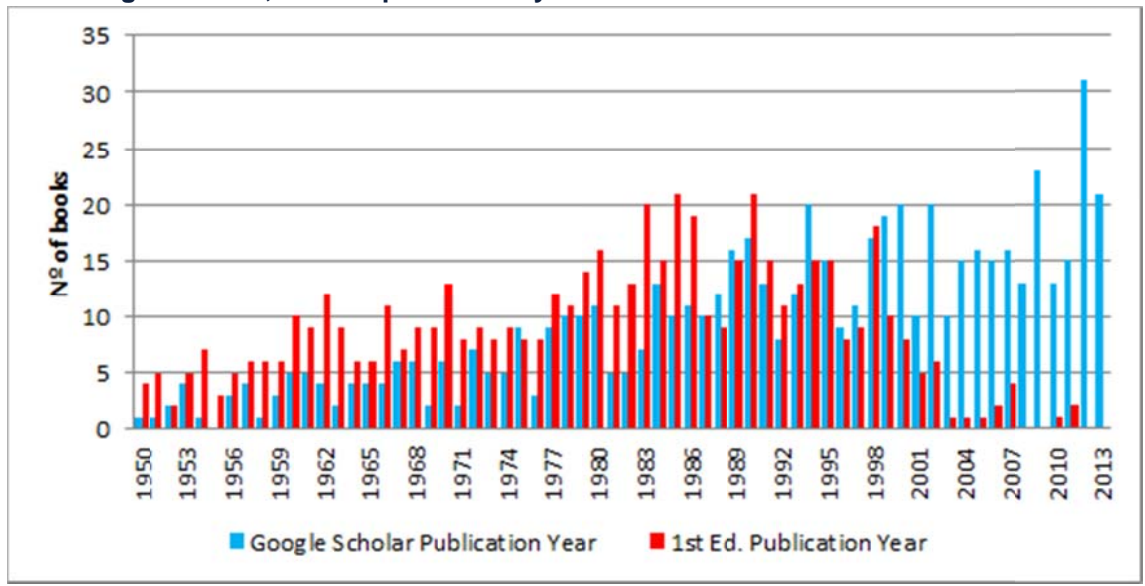

In any case, it should be noted that this limitation doesn't affect the status of these books as highly cited documents, only the year of publication assigned to them[5]. Moreover, this fact may be the reason behind the higher number of books in the last five year of the sample (see Question 2).

3.  When some time after collecting our sample, we checked again the number of citations to Lowry's article, we were taken by surprise by the result we found. As of the 21st of October, 2014, this study had 192,841 citations according to GS (Figure 5 top). However, on the 28th of May, 2014, when we collected our sample, this number was 253,671 (figure 5 middle). This means than within 5 months, Lowry's article has lost nothing less than 60,000 citations. Therefore, right now, it is not the highest cited article in GS, giving way to Laemli's work (Figure 5 bottom)

---

[5] With the exception, of the book *Mathematical theory of communication*, a special case study expanded and commented in Appendix B

**Figure 5. Citation loss of the most cited document in Google Scholar and Web of Science (Lowry, 1951)**

21st October 2014

28th May 2014

21st October 2014

The debate is served...

*How is it possible that the total number of citations of a document decreases over time? What are the reasons for these changes? Are the results offered by GS concerning citations stable and reliable, and consequently, the results concerning which the most cited documents are?*

There is an explanation for this phenomenon, although it's difficult to justify that a document presents a lower number of citations in the present than the number it presented in the past. The behavior of this document in WoS is more logical, since in these months it has accumulated a few more citations: as of the end of May 2014, it had 303,832 citations, and on October the 21st, 2014, it had 305,202 according to GS (Figure 5 top), and 305,248 according to WoS (Figure 6 bottom). WoS data in GS is updated regularly but not in real time.

**Figure 6. Citation of the most cited document in Google Scholar and Web of Science (Lowry, 1951)**

Why does this phenomenon occur in GS?

The answer is related to the dynamic nature of the Web: information is added and removed constantly, and therefore, GS always displays what is currently available on the Web. This is explained in Google Scholar's help pages[6], where they warn that "Google Scholar generally reflects the state of the web as it is currently visible to our search robots and to the majority of users". Presumably, this drastic change in citations took place when GS made a major "re-crawling" of the documents in its database earlier this year (around the third week of June 2014 according to our data).

---

[6] **My citation counts have gone down. Help!**
http://scholar.google.com/intl/en/scholar/help.html#corrections **[accessed on 24th October 2014]**

4. The consequences of this phenomenon in our study are self-evident: did we really collect the most cited documents?

To this end, we collected the entire sample again on the 4th of October, 2014, and compared the two samples to learn how many of the documents in our earlier sample are not present in the new sample (Table 2).

**Table 2. Comparison of two samples of 64,000 highly cited documents (May and October, 2014)**

| Position in rank | Nº of different documents | % |
|---|---|---|
| 1-100 | 402 | 0,6 |
| 101-200 | 340 | 0,5 |
| 201-300 | 319 | 0,5 |
| 301-400 | 373 | 0,6 |
| 401-500 | 450 | 0,7 |
| 501-600 | 588 | 0,9 |
| 601-700 | 778 | 1,2 |
| 701-800 | 1176 | 1,8 |
| 801-900 | 1802 | 2,8 |
| 901-1000 | 3174 | 5,0 |
| **TOTAL** | **9402** | **14,7** |

Only 14.7% of the 64,000 documents in the most recent sample were not also present in our earlier sample. Moreover, most of these new documents are placed in pretty low positions in Google Scholar's ranking of results.

5. Are we sure that all versions of a same document (not only different editions or reprints, but also translations to other languages) have been successfully merged, and that all their respective citations have been added, removing any possible duplicates?

GS has declared that they do exactly this (Verstak & Acharya, 2013), but we don't have empirical data to comment on the potential errors regarding this issue.

Nevertheless, it is not difficult to find obvious errors, like the case of the classic work in Molecular Biology "Molecular cloning: a laboratory manual" (Figure 7), where it is clear that there are still many different versions with a high number of citations that haven't been merged. This, of course, is an exceptional case. Normally, documents will not present as many versions as this example (See Question 7; Table 7), nor as many citations.

**Figure 7. A few versions of *Molecular cloning: a laboratory manual*, by J. Sambrook *et al*. that Google hasn't merged**

Lastly, a few well-known issues in bibliometrics (Garfield, 2005) should be kept in mind before proceeding to observe the ranking of the top 1% most cited documents in Google Scholar (see Appendix A). First, the citation windows: a document published in 1950 has had 64 years to receive citations, whereas a document published in 2013 has had only one year. Secondly, the different paces at which obsolescence takes place in the different scientific fields: generally, documents stop being cited at some point after their publication date. Thirdly, the exponential growth of production: as production volumes increase, the number of citations also increases.

# Question 2.
# Which are the most cited document types in Google Scholar?

### Document types and its evolution

The typologies of the documents in our sample are shown in Figure 8. As we stated in the methods section, we have been able to determine the typology of 45,410 documents in our sample (71%). The typologies of the remaining 29% are unknown.

**Figure 8. Document types of the highly cited documents in Google Scholar**

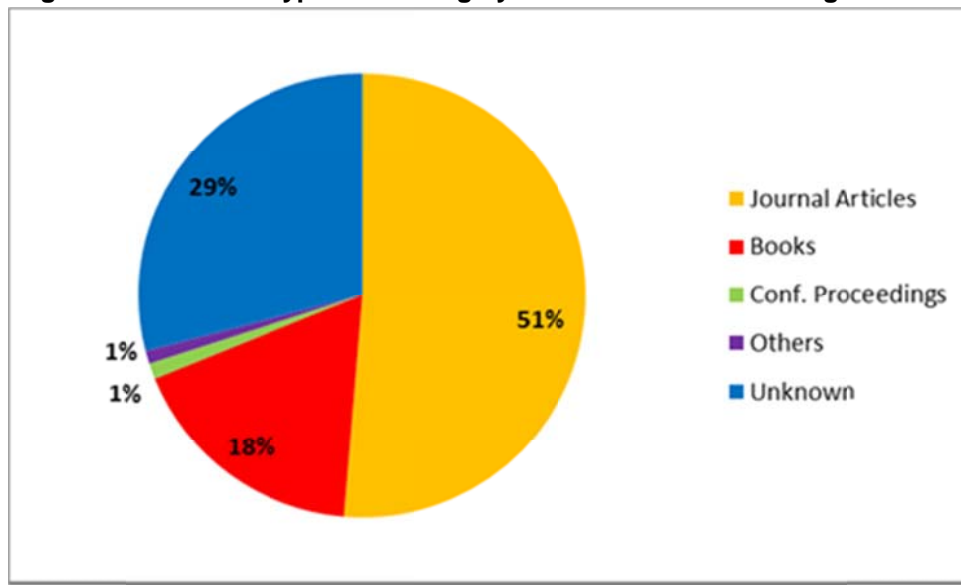

There is a clear predominance of journal articles, which make up a much higher fraction of the total than books and book chapters. The presence of conference proceedings is almost non-existent. Admittedly, this distribution might have been different if we could have defined the document type of the remaining 29% of our sample.

Figure 9 presents this distribution from a longitudinal perspective, where we find the following three phenomena:

- A steady decrease over time in the number of documents with an unknown document type.
- A constant increase in the number of books, which become the most frequent document type in the last five years (2009-2013). As an example, in the 1,000 results for the year 2013, we only find 27 journal articles. What's the reason for this obvious overrepresentation of the book format over the rest of the formats in the last years? We believe this phenomenon has very much to do with the decision of using the most recent edition of a book (and therefore, the most recent publication date), as the primary version of the document (See Question 1, Figures 3-4). This causes, for example, that a classic book originally published in 1965, and reprinted over the years with its latest edition published in 2012, will be considered as having been published in 2012. Since Google Scholar only presents 1,000 results for any given query, and we only collected information about the primary versions of the documents, these books are overshadowing other publications that have really been published in these years.
- Conference proceedings play an insignificant role in this sample, although they achieve greater presence during the last decade of the twentieth century.

**Figure 9. Document types of the highly cited documents in Google Scholar, broken down by years**

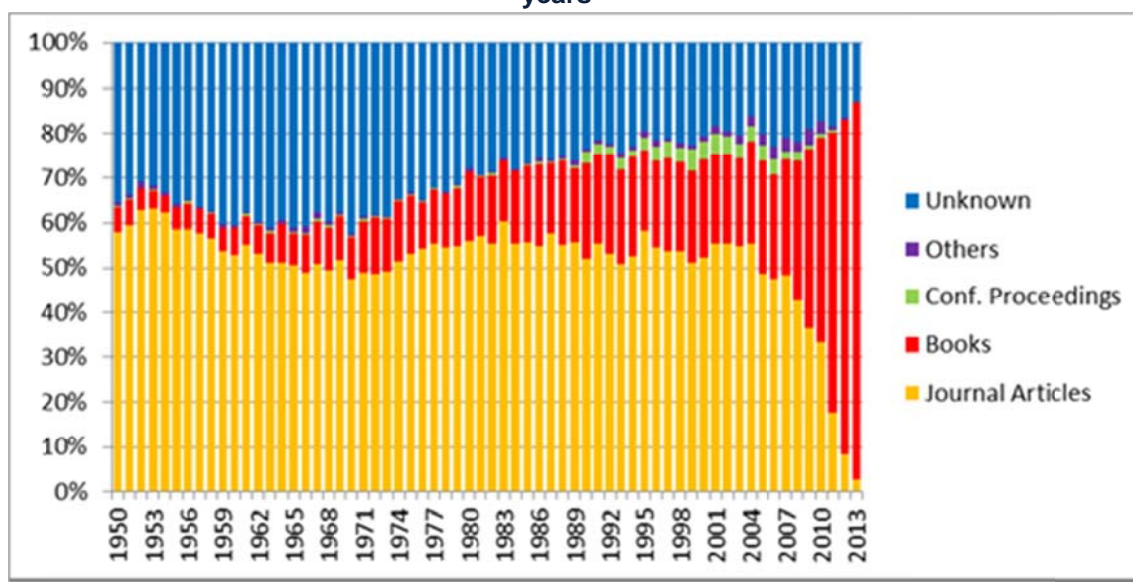

## *Citations and document types*

Books is the document type with a higher average citations per document (Table 3), followed by conference proceedings. Journal articles rank third in this list.

**Table 3. Citations according to document types**

| Document types | Millions of citations | Average citations per document |
|---|---|---|
| Journal Articles | 57,2 | 1700 |
| Books | 30 | 2700 |
| Conference proceedings | 1,6 | 2200 |
| Others | 1,2 | 2050 |

## *Journals containing highly cited documents (1950 y 2013)*

The articles contained in our sample have been published in a total of 3,131 different journals. In Table 4 we show the list of journals where the majority of articles are concentrated. As it could not be otherwise, multidisciplinary journals (Science and Nature) are the ones with the higher number of highly cited journals, followed by the major journals in the natural sciences (Physics and Chemistry). As regards the social sciences, only economics and psychology journals (American Economic Review, and Econometrica) are capable of reaching prominent positions.

**Table 4. Top 25 Most frequent journals in the highly cited documents in Google Scholar**

| Journal | Nº of articles | Area |
|---|---|---|
| NATURE | 1518 | Multidisciplinary |
| SCIENCE | 1437 | Multidisciplinary |
| NEW ENGLAND JOURNAL OF MEDICINE | 848 | Medicine |
| PHYSICAL REVIEW | 671 | Physics |
| PROCEEDINGS OF THE NATIONAL ACADEMY OF SCIENCES OF THE UNITED STATES OF AMERICA | 574 | Multidisciplinary |
| CELL | 483 | Biology |
| JOURNAL OF BIOLOGICAL CHEMISTRY | 452 | Biochemistry |
| PHYSICAL REVIEW LETTERS | 432 | Physics |
| LANCET | 363 | Medicine |
| JOURNAL OF THE AMERICAN CHEMICAL SOCIETY | 328 | Chemistry |
| JAMA-JOURNAL OF THE AMERICAN MEDICAL ASSOCIATION | 251 | Medicine |
| AMERICAN ECONOMIC REVIEW | 244 | Economics |
| ECONOMETRICA | 217 | Economics |
| PSYCHOLOGICAL REVIEW | 210 | Psychology |
| REVIEWS OF MODERN PHYSICS | 206 | Physics |
| CHEMICAL REVIEWS | 203 | Chemistry |
| JOURNAL OF POLITICAL ECONOMY | 200 | Economics |
| JOURNAL OF PHYSIOLOGY-LONDON | 200 | Medicine |
| PSYCHOLOGICAL BULLETIN | 194 | Psychology |
| JOURNAL OF CHEMICAL PHYSICS | 187 | Physics |
| ASTROPHYSICAL JOURNAL | 183 | Physics |
| BIOCHEMICAL JOURNAL | 180 | Biochemistry |
| PROCEEDINGS OF THE ROYAL SOCIETY OF LONDON SERIES A-MATHEMATICAL AND PHYSICAL SCIENCES | 180 | Mathematics; Physics |
| CIRCULATION | 174 | Medicine |
| JOURNAL OF CLINICAL INVESTIGATION | 164 | Medicine |

## *Discussions & Limitations*

1. *Google Scholar does not provide document type information systematically for all its documents (only for books).*

   Because of this, we could not determine the document types of the entire data set, since this would have required a manual inspection of the remaining 18,590 documents. If we did this, our guess is that the fraction of books and book chapters would increase, since this is the typology that GS has more trouble identifying.

2. *Would the weight of the book format be different over the years, had Google Scholar decided to take the first edition of books as their primary version?*

   Without a doubt, yes (see Question 1; Figure 4).

**20**

## Question 3.
## In what languages are the most cited documents in Google Scholar written?

In Figure 10 we show the document distribution according to language. As we can see, English dominates over the rest of languages as the most widely used language for scientific communication, accounting for 92.5% of all the documents in our sample. The second and third places are occupied by Spanish and Portuguese respectively, but neither of them reach even 2% of the total.

**Figure 10. Distribution of languages used in the highly cited documents in GS**

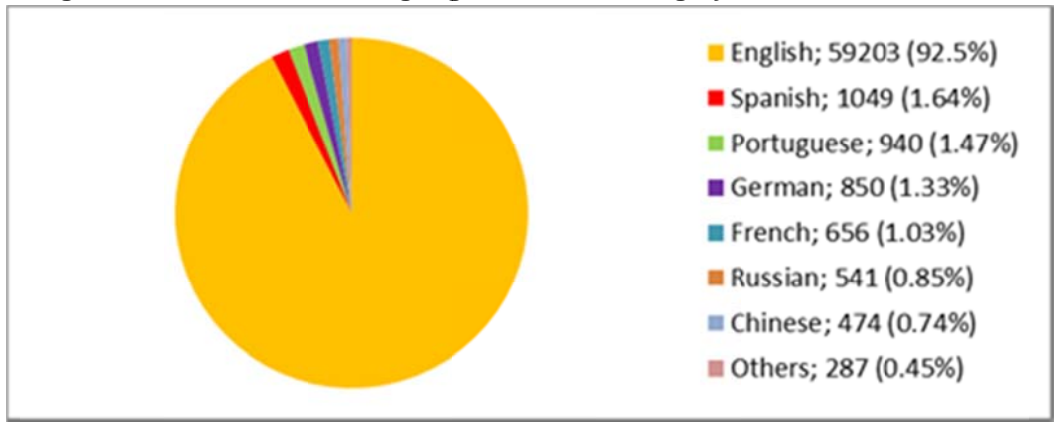

- English; 59203 (92.5%)
- Spanish; 1049 (1.64%)
- Portuguese; 940 (1.47%)
- German; 850 (1.33%)
- French; 656 (1.03%)
- Russian; 541 (0.85%)
- Chinese; 474 (0.74%)
- Others; 287 (0.45%)

In Figure 11 we can observe the same data broken down by years. The results for the language variable are much more stable through the years than the ones found for the document types. In this case, the English language predominates in every year, with an oscillation between its maximum and minimum value of less than 10% (87% in 2013, and 95% in 1991).

**Figure 11. Distribution of languages in the highly cited documents in GS by years of publication**

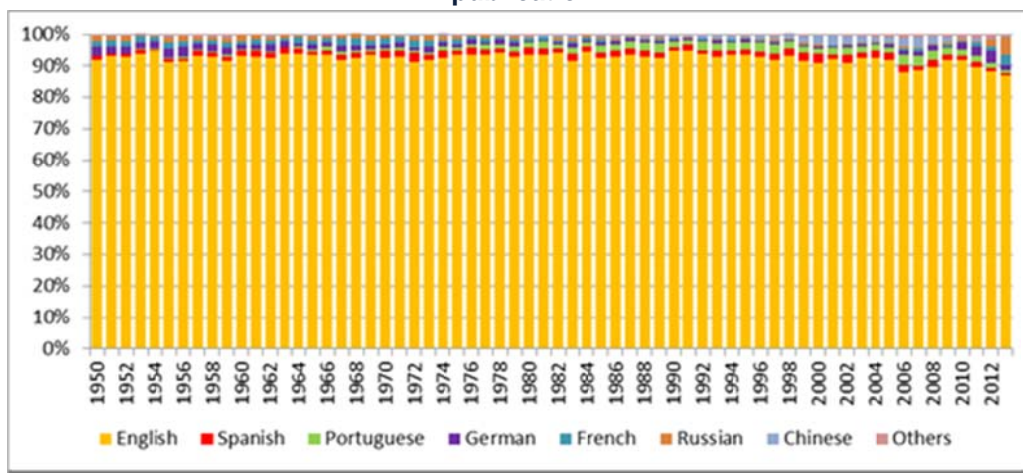

The "Others" category includes the following languages: Italian, Swedish, Indonesian, Finnish, Danish, Bulgarian, Polish, Norwegian, Turkish, Latin, Slovenian, Serbian, Dutch, Macedonian, Malayan, Japanese, Czech, Estonian, Slovak, Mongolian, Catalan, Croatian, Lithuanian, and Ukrainian.

## *Discussions & Limitations*

1. *As with document types, Google Scholar does not provide information about the languages in which the documents it indexes are written.*

   Because of this, we developed a strategy to determine this information, using WoS data where possible (around 50% of the cases), and the title and abstract of the document in all the other cases. This approach, however, may have introduced an overrepresentation of the English language, since it is usual for a document written in a language other than English to provide its title and abstract in English as well, for the purpose of being indexed in international databases.

2. *Additionally, our sample may contain records that are in fact translations of other documents (which may also be present in our sample).*

   As we pointed out in previous studies (Martín et al. 2014), Google Scholar usually fails to group together different translations of a same document. This is the case of journals that are published both in English and in other language, or books that are translated into various languages (see Figure 12). This issue has an immediate effect for the works affected by this problem: their citations are scattered across different records, and this could affect their status as highly cited documents.

**Figure 12. Example of language versions (Chinese, English, German, Spanish, French)**
***The structure of scientific revolutions*, by Kuhn**

allintitle: "Die Struktur wissenschaftlicher Revolutionen" author:Kuhn ▾

About 27 results (0.03 sec)

Tip: Search for **English** results only. You can specify your search language in Scholar Settings.

[BOOK] **Die struktur wissenschaftlicher revolutionen**
TS **Kuhn**, K Simon - 1967 - theodor-rieh.de
Thomas S. **Kuhn**, Professor für Wissenschaftstheorie und Wissenschaftsgeschichte in
Princeton und gelernter Physiker, unternimmt in diesem Buch den Versuch, den
Mechanismus wissenschaftlichen Fortschritts darzustellen. Im Klappentext der deutschen ...
Cited by 4501   Related articles   All 2 versions   Import into BibTeX   Save   More

[CITATION] **Die Struktur wissenschaftlicher Revolutionen**. Zweite revidierte und um das
Postskriptum von 1969 ergänzte Auflage
TS **Kuhn** - Frankfurt am Main: Suhrkamp, 1976
Cited by 17   Related articles   Import into BibTeX   Save   More

[CITATION] **Die Struktur wissenschaftlicher Revolutionen**
K ThS - Suhrkamp, Taschenbuch der Wissenschaft, 1969
Cited by 14   Related articles   Import into BibTeX   Save   More

[CITATION] **Die Struktur wissenschaftlicher Revolutionen**
K Thomas - Suhrkamp Taschenbuch, 1976
Cited by 9   Related articles   Import into BibTeX   Save   More

the structure of scientific revolutions author:Kuhn ▾

About 833 results (0.07 sec)

[BOOK] The **structure** of **scientific revolutions**
TS **Kuhn** - 2012 - books.google.com
A good book may have the power to change the way we see the world, but a great book
actually becomes part of our daily consciousness, pervading our thinking to the point that we
take it for granted, and we forget how provocative and challenging its ideas once were— ...
Cited by 74018   Related articles   All 79 versions   Import into BibTeX   Save   More

[CITATION] 77ie **structure** of **scientific revolutions**
TS **Kuhn** - Aufl. Chicago, 1970
Cited by 245   Related articles   Import into BibTeX   Save   More

The road since **structure**
T **Kuhn** - 2000 - philpapers.org
... Mind 121 (484):1031-1046. Thomas AC Reydon & Paul Hoyningen-Huene (2010). Discussion:
**Kuhn's** Evolutionary Analogy in the **Structure** of **Scientific Revolutions** and "the Road Since
**Structure**". Philosophy of **Science** 77 (3):468-476. Rupert Read (2004). ...
Cited by 372   Related articles   Import into BibTeX   Save   More

[PDF] The **structure** of **scientific revolutions**
TS **Kuhn** - 1962 - math-info.univ-paris5.fr
The essay that follows is the first full published report on a project originally conceived
almost fifteen years ago. At that time I was a graduate student in theoretical physics already
within sight of the end of my dissertation. A fortunate involvement with an experimental ...
Cited by 109   Import into BibTeX   Save   More

allintitle: "la estructura de las revoluciones científicas" author:Kuhn ▾

About 32 results (0.04 sec)

Tip: Search for **English** results only. You can specify your search language in Scholar Settings.

[CITATION] **La estructura de las revoluciones científicas**
T **Kuhn** - Méxic: Fondo de Cultura Económica, 1975
Cited by 39   Related articles   Import into BibTeX   Save   More

[CITATION] **La estructura de las revoluciones científicas**, fce
T **Kuhn** - 1981 - Madrid
Cited by 19   Related articles   Import into BibTeX   Save   More

[CITATION] **La estructura de las revoluciones científicas**
K Th-S - 1971 - FCE México
Cited by 19   Related articles   Import into BibTeX   Save   More

[CITATION] **La Estructura de las Revoluciones Científicas**
KT Samuel - Editorial FCE, México, 1971
Cited by 6   Related articles   Import into BibTeX   Save   More allintitle: "la structure des révolutions scientifiques" author:Kuhn

12 results (0.03 sec)

Tip: Search for **English** results only. You can specify your search language in Scholar Settings.

[CITATION] **La structure des révolutions scientifiques**
TS Kuhn - 1972 - cds.cern.ch
**...** Information; Discussion (0); Files; Holdings. Book. Title, **La structure des révolutions scientifiques**. Author(s). **Kuhn**, Thomas S. Publication. Paris : Flammarion, 1972. - 284 p. Subject code, 93:5. Subject category, Biography, Geography, History. CERN library copies - Purchase it for **...**
Cited by 2078   Related articles   Import into BibTeX   Save   More

[CITATION] **La structure des révolutions scientifiques**
K Thomas - Paris: Flammarion (Chicago: 1962).* LE GRAND Jean- ..., 1983
Cited by 133   Related articles   Import into BibTeX   Save   More

[CITATION] **La structure des révolutions scientifiques**
TS Kuhn - Paris: Flammarion, 1983
Cited by 36   Related articles   Import into BibTeX   Save   More

[CITATION] **La structure des révolutions scientifiques**
KT Samuel - Traduction. Française, Paris: Flammarion, 1972
Cited by 17   Related articles   Import into BibTeX   Save   More

# Question 4.
# How many highly cited documents are freely accessible?

The percentage of documents for which Google Scholar provides a freely accessible full text link can be observed in Figure 13. Over 40% of the documents in our sample provided a full text link, and these links are mostly concentrated in the last two decades. The lower rate of records with an open access link in the last four years might be explained by journal's and publisher's embargo policies. Additionally, the high percentage of books in the last 5 years of the sample may influence as well.

**Figure 13. Percentage of freely accessible highly cited documents in Google Scholar. Global results for the 1950-2013 period (left), and broken down by decades (right)**

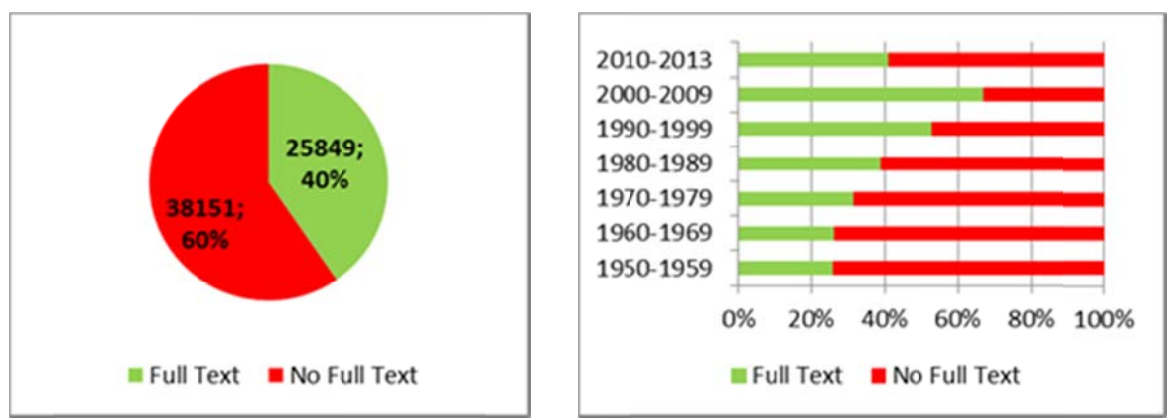

These results are consistent with those published by Archambault et al. in 2013, (since they also found that over 40% of the articles from their sample were freely accessible from Google Scholar), and much higher than the results obtained by Khabsa and Giles (2014), and Björk et al. (2010), who found only a 24% and 20.4% of open access documents respectively.

*What file types are the most commonly used to store these highly cited documents?*

Full text links point to documents in a variety of formats. The most common one is the PDF format, followed by the HTML format. Figure 14 presents the distribution of these formats for all the documents that provide a Full Text Link. These results confirm the data previously identified, among others, by Aguillo, Ortega, Fernández & Utrilla (2010) and Orduña-Malea, Serrano-Cobos & Lloret-Romero, N. (2009).

**Figure 14. File Formats of the highly cited documents in Google Scholar freely accessible (1950-2013)**

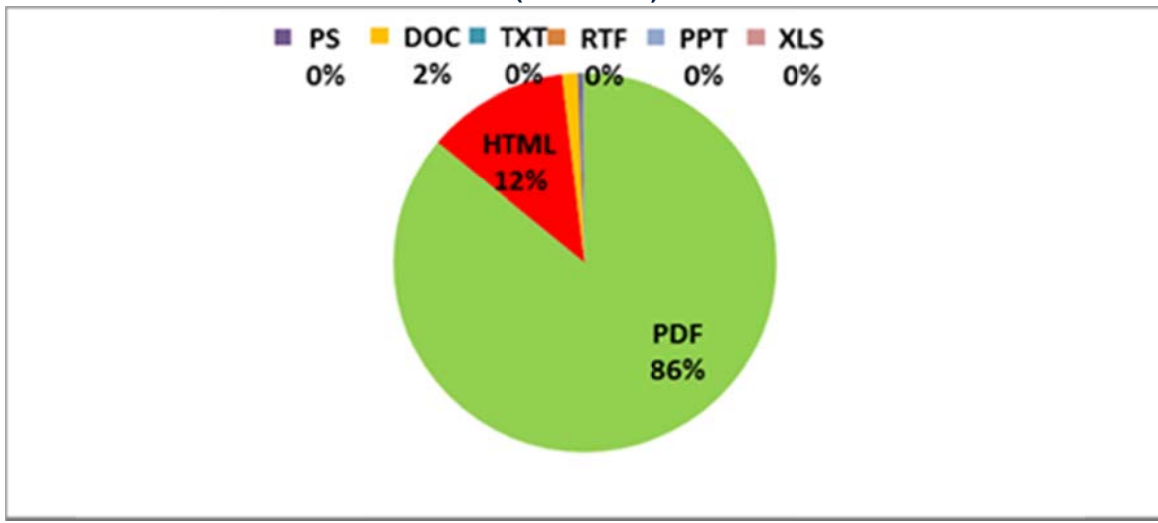

Figure 15 shows the same data broken down by years. We can see that the predominance of the PDF format is present throughout the entire range of years. However, it is also noteworthy that the HTML format has started gaining more presence for documents published in the last 25 years, with a peak of almost 20% of the share in 2010.

**Figure 15. File Formats of the highly cited documents in Google Scholar that are freely accessible, broken down by years (1950-2013)**

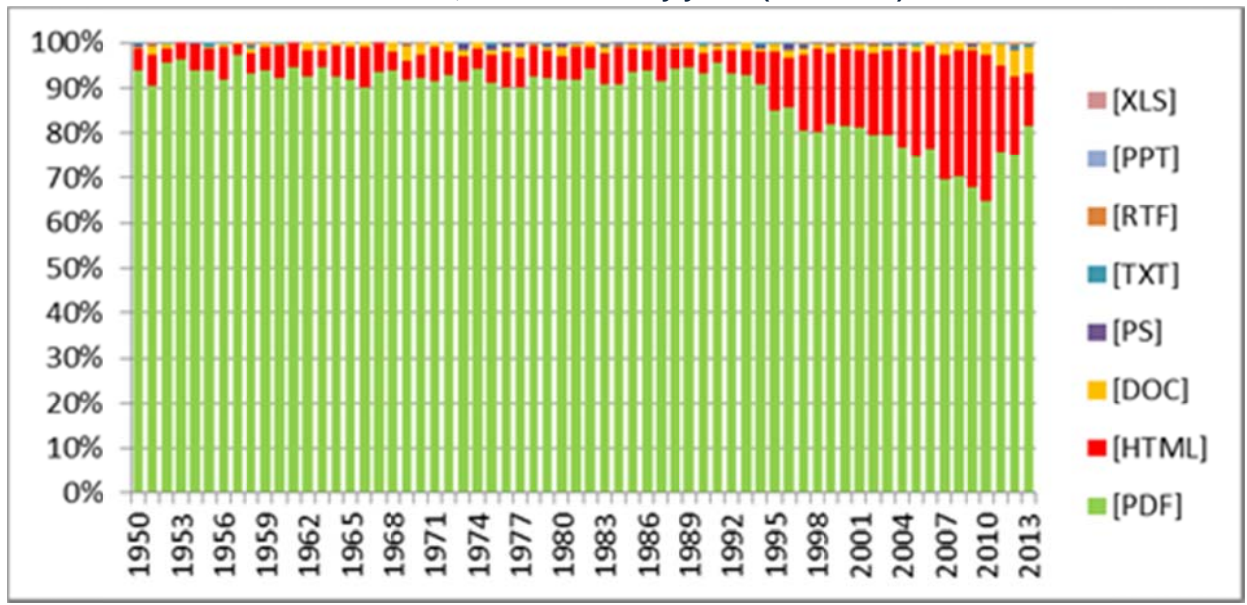

### *Which are the main providers of these documents?*

We have found a total of 5,715 different providers of Full Text Links in our sample. However, a group of 35 providers account for more than a third of all the links. Table 5 shows these main providers.

**Table 5. Full Text provider**

| Provider | Nº of Full Text Links | Type of entity |
|---|---|---|
| nih.gov | 1405 | Public administration |
| researchgate.net | 815 | Social network |
| harvard.edu | 495 | University |
| pnas.org | 478 | Scientific society |
| oxfordjournals.org | 466 | Publisher |
| psu.edu | 424 | University |
| arxiv.org | 423 | Repository |
| jbc.org | 414 | Journal |
| sciencedirect.com | 394 | Publisher |
| wiley.com | 324 | Publisher |
| jstor.org | 322 | Digital library |
| rupress.org | 304 | University |
| royalsocietypublishing.org | 266 | Scientific society |
| ahajournals.org | 218 | Scientific society |
| dtic.mil | 208 | Public administration |
| stanford.edu | 203 | University |
| google.com | 188 | Company |
| mit.edu | 180 | University |
| tu-darmstadt.de | 177 | University |
| nature.com | 161 | Publisher |
| yale.edu | 141 | University |
| caltech.edu | 140 | University |
| physoc.org | 140 | Scientific society |
| cmu.edu | 122 | University |
| umich.edu | 120 | University |
| duke.edu | 118 | University |
| princeton.edu | 116 | University |
| wisc.edu | 113 | University |
| ucsd.edu | 112 | University |
| asm.org | 112 | Scientific society |
| berkeley.edu | 107 | University |
| upenn.edu | 104 | University |
| washington.edu | 103 | University |
| columbia.edu | 102 | University |
| yimg.com | 101 | Company |
| **TOTAL** | 9616 | |

If we analyse the top-level domains of these links, the most frequent are academic institutions (.edu) and organizations (.org). Moreover, the number of links provided by academic institutions is probably higher than 6,136, because there are many universities that use national top-level domains instead of .edu. Table 6 shows the 20 most frequent top-level domains.

This means that GS feeds highly cited documents mainly, at least as far as our sample is concerned, from universities (institutional repositories) and public organizations (working papers, grey literature), and not from commercial publishers. Of special note is the role of the scientific social network ResearchGate, where researchers often upload their publications.

**Table 6. Main top-level domains contributing Full Text links in Google Scholar**

| Domain | Nº of Full Text Links |
|--------|--------|
| .edu | 6136 |
| .org | 5528 |
| .com | 3466 |
| .gov | 1712 |
| .net | 1345 |
| .de | 678 |
| .cn | 489 |
| .uk | 485 |
| .ca | 404 |
| .ru | 374 |
| .fr | 357 |
| .br | 343 |
| .it | 275 |
| .ch | 214 |
| .mil | 210 |
| .nl | 186 |
| .es | 145 |
| .tw | 136 |
| .au | 131 |
| .in | 118 |
| Others | 3117 |
| **TOTAL** | **25849** |

## *Discussions & Limitations*

1.  Do these links really point to full text versions of the documents?

    More rigorous analyses should be carried out in order to determine if there are false positives among these links. For example, a freely accessible PDF document containing a review of a book, or just the cover and the table of contents of a book could be mistaken for the book itself.

    Moreover, the dynamic nature of the web means that a link that was accessible some time ago may no longer be available. How often does Google Scholar checks that these links are still functioning properly?

2.  Our analysis deals only with the full text link provided for the version of the document GS considers as the primary version.

    However, when the primary version of a document is not freely accessible, GS points the user to any other free version if available. Figure 16 is an example of a case where the primary version is the publisher's edition of a journal article, but the Full Text link is a preprint from arXiv.

**Figure 16. Primary version, Publisher and Full Text provider**

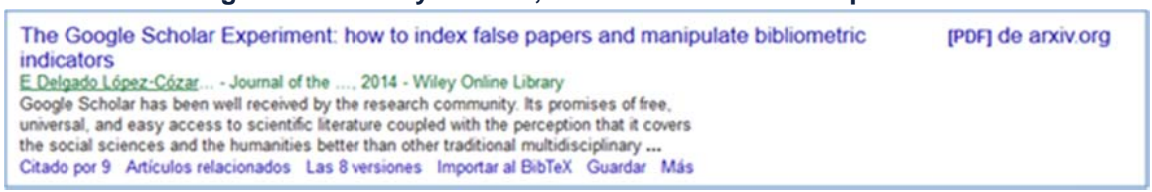

3.  For documents with more than one version, there may be more than one full text version of the document.

    These versions may be hosted in other domains. Again, we want to stress that we only study the Full Text Links displayed for the primary versions of the documents.

# Question 5.
# How many of the highly cited documents indexed by GS are also indexed by WoS?

Almost half of the highly cited documents according to Google Scholar are not indexed on the Web of Science (Figure 17).

**Figure 17. Percentage of highly cited documents in Google Scholar that are also indexed in the Web of Science (1950-2013)**

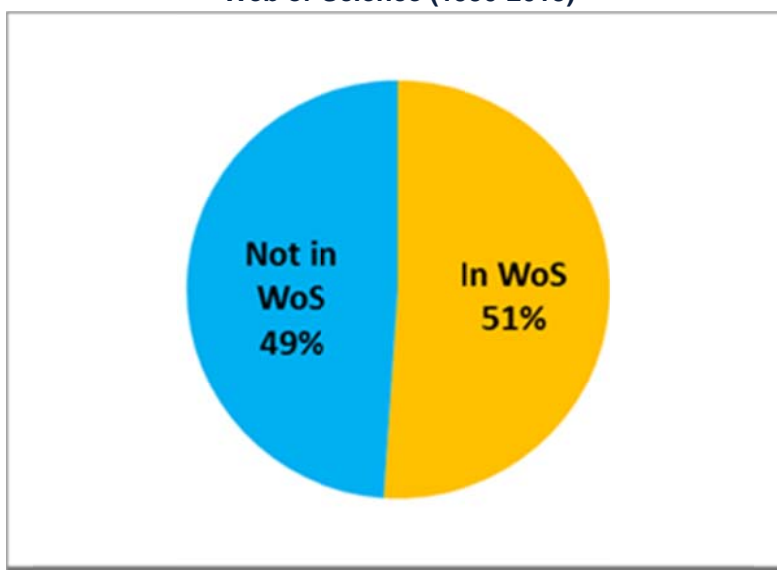

This is extremely relevant, although the following issues should be taken into consideration:

- The different natures of GS and WoS as databases: GS covers academic documents (scientific, technical, educational…) published by all kinds of different sources and in all sorts of communication channels (books, theses, reports…), whereas the coverage in Web of Science Core Collection is oriented towards a more limited range of academic publications, i.e. journal articles and conference communications. This would confirm our hypothesis that GS measures a different kind of impact than the one measured by scientific databases: the academic impact.
- If we want to identify the most influential documents in the academic-scientific sphere, we must use GS.
- GS also identifies the most relevant scientific documents with a fair amount of reliability.

Furthermore, no significant differences are appreciated between 1950 and 2003 (Figure 18). However, the last decade suffers the consequences of the phenomenon we encountered in question 2: the overrepresentation of books in the last years caused by Google Scholar's policy of taking the latest edition of books as their primary version.

Since Web of Science's coverage of books is still very limited, it is not surprising that the reduction in the percentage of documents indexed in WoS in the last years closely matches the reduction in the number journal articles during the same years (Figure 9).

**Figure 18. Percentage of highly cited documents in Google Scholar that are also indexed in the Web of Science, broken down by decades (1950-2013)**

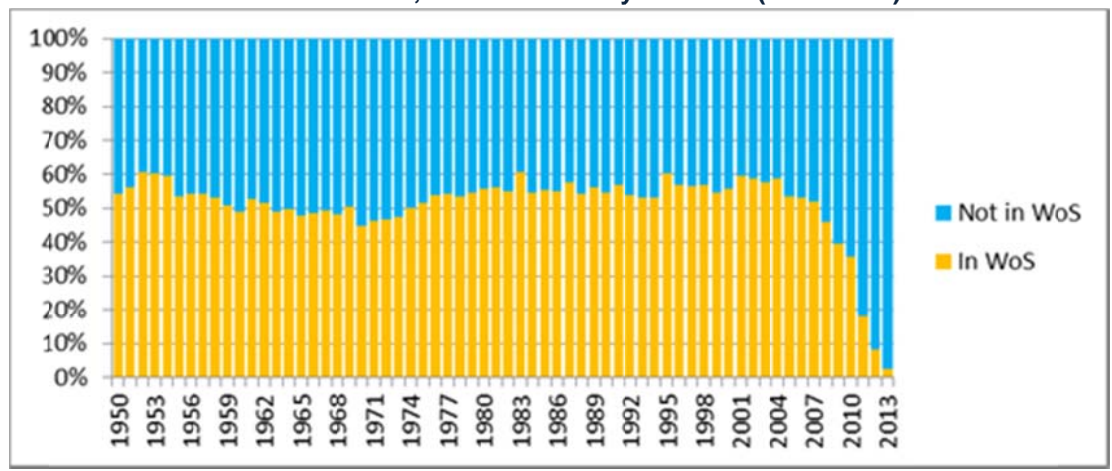

## *Discussions & Limitations*

1.  Is the GS-WoS connection correctly implemented?

    A more in-depth study should be carried out to determine potential flaws in the matching of documents and the frequency with which they occur:

    - False positives: a document in GS matched to a document in WoS even if they're not really the same documents. For example, a book in GS might be matched to a review of that book indexed in WoS. This is the case of the book "The discovery of grounded theory: Strategies for qualitative research", which was previously presented in Table 1.
    - False negatives: documents indexed both in GS and WoS for which a connection hasn't been established.

    As a first approximation, we have selected the 398 most cited WoS documents between 1950 and 2013 that, according to their WoS ID (accession number), weren't present in our GS sample. We have searched the titles of these documents on Google Scholar and found that 382 (96%) were in fact indexed in Google Scholar, and 300 of them were also connected to a different WoS record.

    Therefore, these mistakes arise from incorrect connections between Google Scholar and Web of Science records, caused by the existence of various records with the same name in WoS. For example, a case where a document in Google Scholar has been connected to the Correction of an article in WoS, and not to the article itself is shown in Figure 19.

**Figure 19. Incorrect connections between Google Scholar and Web of Science records**

2. Is it possible that some highly cited articles according to the Web of Science are not indexed on Google Scholar?

   As noted earlier in question 1, this may have happened in a very few cases, but not among the very highly cited (30,000 most cited documents in our sample).

3. The overrepresentation of books in the last decade

   Again, this is one of the flaws in our sample, since it has caused that many journal articles published in those last years of the sample (2003-2013) and that have received many citations, are being left out in favor of books that were first published many years ago.

## Question 6.

## Is there a correlation between the number of citations that these highly cited documents have received in GS and the number of citations they have received in WoS?

We have calculated Pearson's correlation coefficient for the number of citations that documents have received according to Google Scholar and the Web of Science, by year. The average correlation is 0.8 (calculated only for documents that are in both sources, which are 32,680). Figure 20 shows the Pearson correlation coefficient for each of the years in our sample.

**Figure 20. Pearson correlation coefficient between Google Scholar and Web of Science citations (1950-2013)**

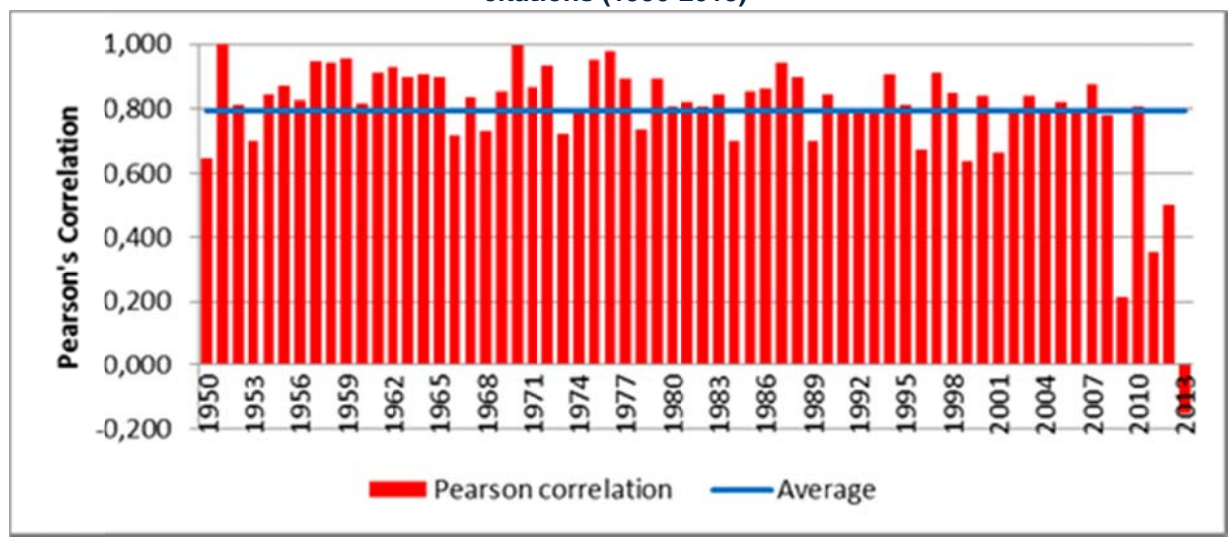

This finding is consistent with the results found in many previous studies (Sanderson 2008; Kousha, & Thelwall 2008; Meho & Rogers 2008; Franceschet, 2010; Delgado López-Cózar & Cabezas 2013; Delgado López-Cózar & Repiso 2013), who also found a high correlation among the journal indicators published by Google Scholar/Google Scholar Metrics and the Web of Science/Journal Citation Reports. However, none of these studies had analysed a sample as large as this one (32,680 documents).

It is common among the studies that compare Google Scholar and the Web of Science to quantify the number of citations they have been able to find for the documents they index. In our sample, 91.6% of the documents have received more citations in GS than in WoS. Only 3,079 documents (9.4%) have more citations according to WoS than in GS. Furthermore, the average number of citations per document in GS is 1,790, and 1,080 in WoS, which means that on average, GS has 70% more citations per document than WoS.

## *Discussions & Limitations*

1. As in question 5, the quality of the matching between GS and WoS plays an important part.
2. The instability of Google Scholar's indicators is also an important factor and should be further analysed.

   As an example, Lowry's classic article had 253,671 citations at the end of May, 2014, when we collected the data (see Table 1), but on August the 5[th] the count had gone down to 191,669 (Figure 21). WoS data seems to be much more stable, but it also went down from 304,893 citations in May, to 304,667 in August (See also Question 1, Figure 5).

**Figure 21. The most cited scientific article in history, according to Google Scholar (top), and WoS (bottom). Screen capture from 7th of August, 2014**

# Question 7.
# How many versions of these highly cited documents has GS detected?

One of the most interesting features of Google Scholar as an academic search engine is its ability to identify and connect all the different instances of the same document that have been deposited across the Web. We should bear in mind that a document can be stored in various locations: the journal publisher's webpage (Cell), databases (Pubmed), aggregators (Ingenta), library catalogues (Dialnet), subject or institutional repositories, and authors' personal or institutional web pages. Moreover, documents might go through various versions and revisions, and they can be cited in different forms. Google acknowledges this reality and tries to find a solution.

Excerpt from Verstak, AA and Acharya, A (2013). _Identifying multiple versions of documents_. U.S. Patent No. 8,589,784. Washington, DC: U.S. Patent and Trademark Office:

> **"[…] it is typical that a particular document or portion thereof, appears in a number of different versions or forms in various online repositories. This generally results in multiple versions of a document being included in the search results for any given query. Because the inclusion of different versions of the same document does not provide additional useful information, this increase in the number of the search results does not benefit users. Also, search results including different versions of the same document may crowd out diverse contents that should be included. These problems have seriously affected the quality of a search result provided by a search engine.**
>
> **Another problem arises in systems in which there are multiple versions of documents present. Documents in a document collection will have a number of citations to it by other documents. This is particularly the case for academic documents, legal documents, and the like. The number of citations (citation count) to a document is often reflective of the importance, significance, or quality of the document. Where there are different versions of a document present in a repository, each with its own citation count, a user does not have an accurate assessment of the actual significance, importance or quality of the document based on the individual citation counts.**
>
> **For these reasons, it would be desirable to identify documents that are different versions of the same document in a document collection. It would also be desirable to manage these documents in an efficient manner such that the search engine can furnish the most appropriate and reliable search result."**

83% of the documents in our sample have more than one version, whereas 40% have 6 or more versions, 19% have 10 or more versions, and 200 documents have more than 100 versions (0.1%). The distribution of documents according their number of versions can be observed in Table 7:

**Table 7. Distribution of documents according to their number of versions**

| Nº of versions | Nº of doc. | Accumulated | Acc. % |
|---|---|---|---|
| 1 | 10771 | 10771 | 16,83 |
| 2 | 6075 | 16846 | 26,32 |
| 3 | 6903 | 23749 | 37,11 |
| 4 | 6814 | 30563 | 47,75 |
| 5 | 5539 | 36102 | 56,41 |
| 6 | 4781 | 40883 | 63,88 |
| 7 | 3746 | 44629 | 69,73 |
| 8 | 2940 | 47569 | 74,33 |
| 9 | 2429 | 49998 | 78,12 |
| 10 | 1929 | 51927 | 81,14 |
| 11-15 | 5243 | 57170 | 89,33 |
| 16-25 | 3585 | 60755 | 94,93 |
| 26-50 | 2202 | 62957 | 98,37 |
| 51-100 | 762 | 63719 | 99,56 |
| 101-200 | 202 | 63921 | 99,88 |
| 201-300 | 40 | 63961 | 99,94 |
| 301-400 | 16 | 63977 | 99,96 |
| 401-500 | 9 | 63986 | 99,98 |
| More than 501 | 14 | 64000 | 100,00 |

## *Discussions & Limitations:*

1. Does GS correctly identify all versions of a same document? Does it make mistakes, like linking a document with a different document (i.e., a review of that document, or a citation found in the list of references of another document), or failing to connect two records that refer to the same document? How frequently does it make these mistakes?

   In order to successfully answer these questions, we would need to analyse a sample of documents and study all their versions individually. While we carry out this study, we present, by way of an example, an illustrative example in Appendix B.

## Question 8.

## Is there a correlation between the number of versions GS has detected for these documents, and the number citations they have received?

Using Pearson's correlation coefficient, we have been able to determine that there is no correlation whatsoever between the number of citations of a document in Google Scholar and its number of versions (r = 0.2**). Calculating it by year of publication yields similar results (Figure 22).

**Figure 22. Pearson's correlation between the nº of citations and nº of versions in Google Scholar documents (64,000 most cited documents in Google Scholar; 1950-2013)**

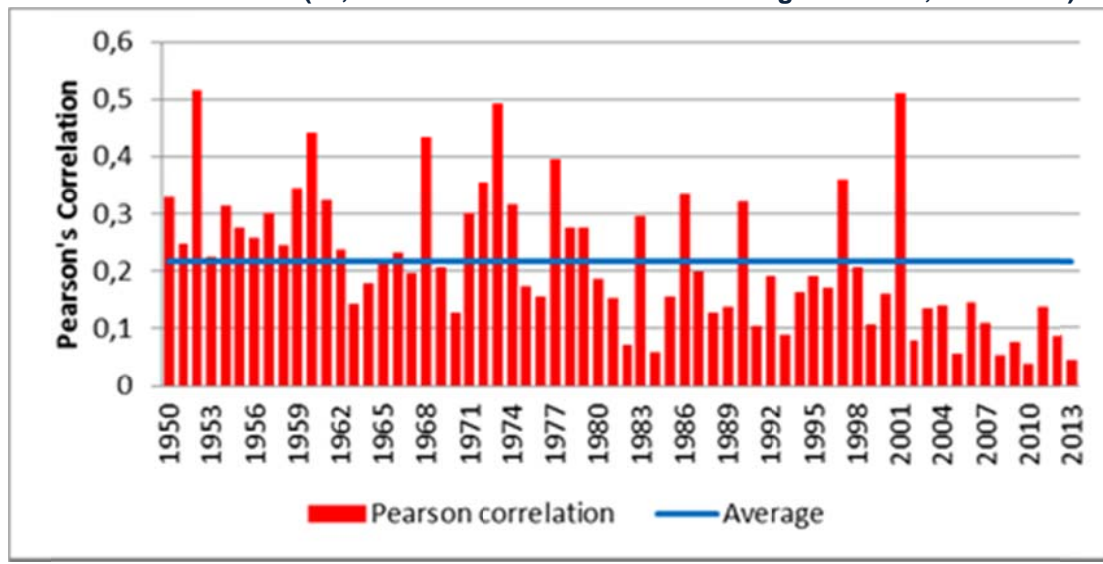

# Question 9.

# Is there a correlation between the number of versions Google Scholar has detected for these documents, and their position in the result pages?

Using Pearson's correlation coefficient, we also have determined that there is no correlation whatsoever between the number of versions of a document in Google Scholar and the position it occupies in the search engine results page (Figure 23). The average correlation for the results we collected from 64 queries is r = -0.2**.

**Figure 23. Pearson's correlation between the number of versions of the documents in Google Scholar and their rank in the SERP**

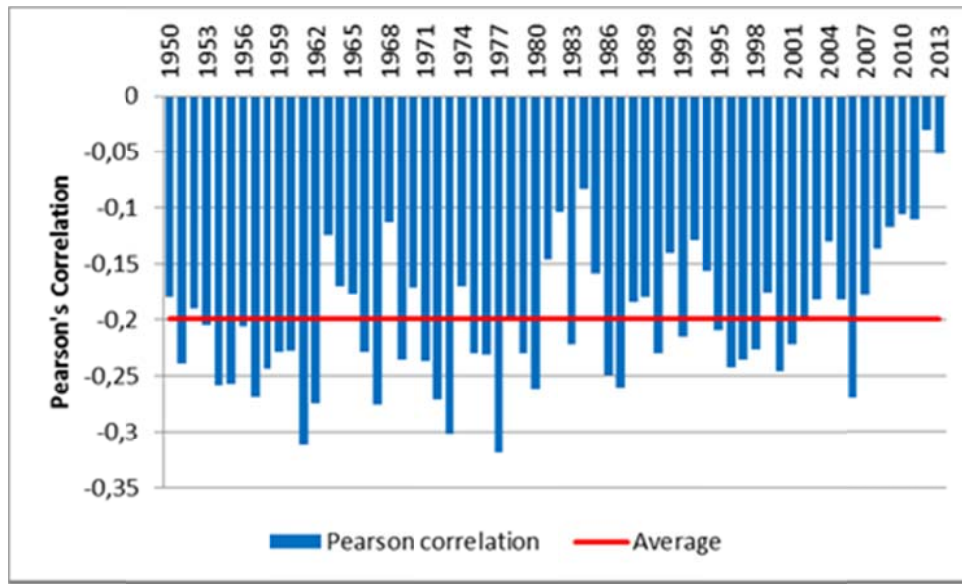

## Question 10.

## Is there some relation between the positions these documents occupy in the search engine result pages, and the number of citations they have received?

After calculating the Pearson correlation for each of the years in our queries, we obtained an average r = 0.9** (Figure 24). These results confirm that the most important factor in the calculation of the position a document will occupy in Google Scholar's SERP is its citation count, confirming the statement of Google Scholar in this regard.

**Figure 24. Pearson correlation between the number of citations of documents in Google Scholar and the position they occupy in the Search Engine Result Page**

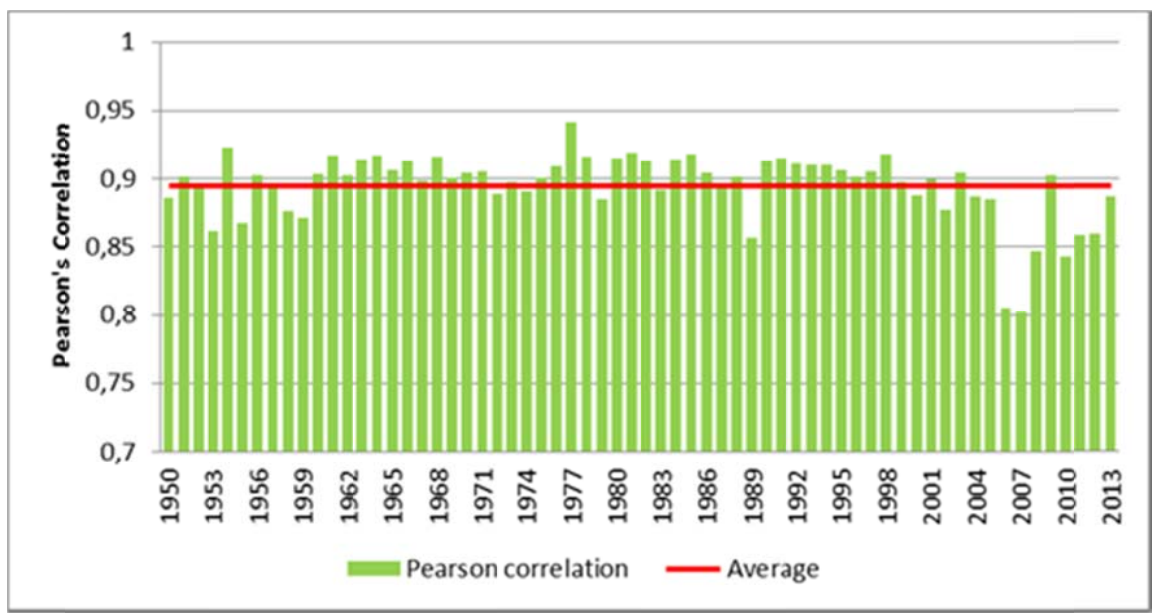

Moreover, according to the scatterplot in Figure 25, the correlation is almost perfect until we reach the last 100 results of the queries, but then the correlation becomes much more tenuous. If we calculate the Pearson correlation for the first 900 and the last 100 results of each query separately, the average correlation for all years is 0.97** and 0.61** respectively. Clearly, the problem is restricted to the tail of the distribution.

**Figure 25. Relationship between the number of citations of documents in Google Scholar and the position they occupy in the Search Engine Result Page**

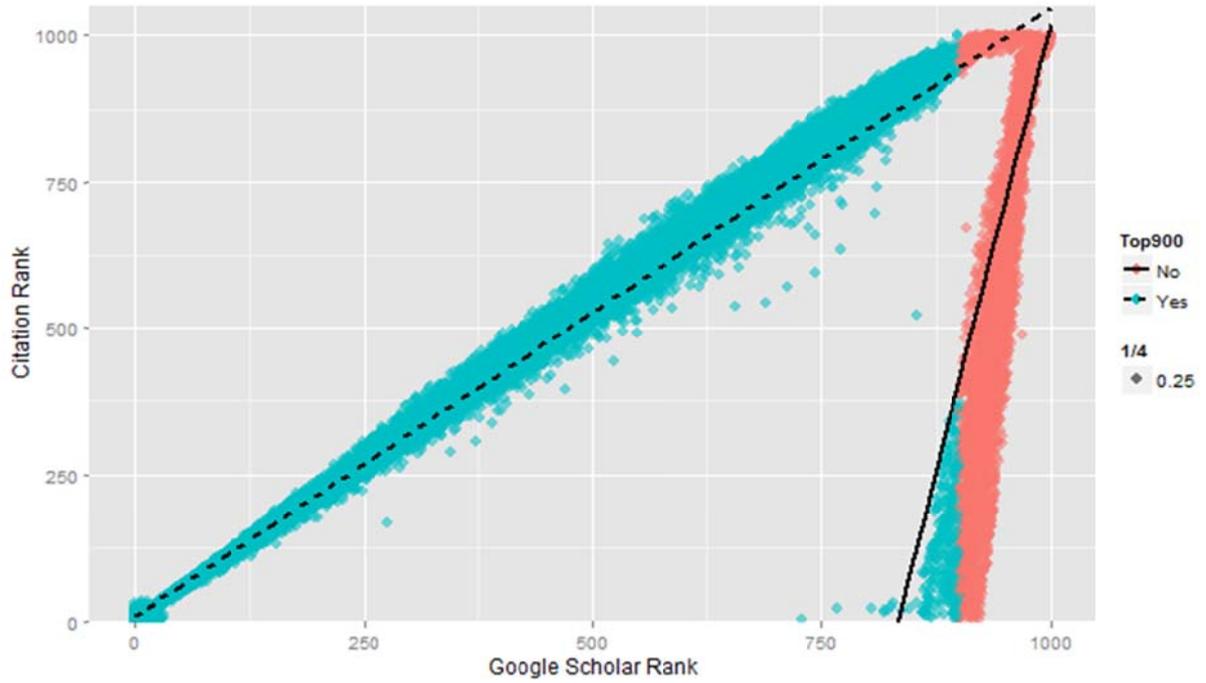

# 5. CONCLUSIONS

As we've seen, the analysis of GS provides a very different vision to the question of which are the most influential academic, scientific and technical documents for the scientific, professional and educational community. This fact can be explained by Google Scholar's own nature:

- Google Scholar's crawlers sweep the entire academic web: the most well-known scholarly publishers (such as Elsevier, Springer, Sage, Willey, Taylor & Francis, IEEE, ACS, ACM, Macmillan, Wiley, Oxford University Press); their digital hosts/facilitators (such as HighWire Press, MetaPress, Ingenta); societies and other scholarly organizations (such as the American Physical Society, American Chemical Society, ACM), government agencies (National Institute of Health, National Oceanic and Atmospheric Administration, U.S. Geological Survey), databases (Pubmed, ERIC), disciplinary repositories (such as arXiv.org, Astrophysics Data System, RePEc, SSRN, CiteBase), institutional repositories from universities or research centers, library catalogs (Dialnet), as well as personal web pages from researchers, professors, research groups, departments, faculties… hosted inside the servers of the university or research center they belong to.
- While traditional citation-based databases deal with the strictly scientific world (mainly journal articles, conference communications, and some books), Google Scholar's aim is to index all kinds of scientific documents (scientific and professional journals, conferences, books, working papers, reports…), as well as educational documents (master's and doctoral theses, teaching materials…), and technical and professional documents (reports, patents, american case laws, annuals…) circulating in the Web.
- It covers documents written in all languages and from all countries.

In conclusion, thanks to the wide and varied sources from which GS feeds, we are able to measure not only scientific impact, but also educational and professional impact in the broadest sense of the term (Kousha and Thelwall, 2008).

At the same time, as regards strict scientific impact, the analysis of GS data provides very similar results to the results obtained from traditional citation-based databases, with the advantage of being able to retrieve a larger and more varied number of citations, since they come from a wider range of document types, different geographical environments, and languages different to English.

The profile of the average highly cited document is: a book or journal article written in English and available online in PDF format.

The rest of the findings of this study can be summarised as follows:

- 40% of the highly cited documents in GS are freely accessible, mostly from educational institutions (mainly universities), and other non-profit organizations. The availability of these documents is essential for GS as a search engine.
- Almost half of these highly cited documents are not indexed in Web of Science, which for many years has has been considered the most prestigious scientific information database.
- There is a high correlation (r = 0.8) between the number of citations of these documents in GS and their citations in WoS.
- GS has detected more than one version for the 83.17% of the documents in our sample.
- There is no correlation between the number of versions GS has detected, and the number citations they have received.
- There is no correlation between the number of versions GS has detected for these documents, and their position in the result pages (SERPs).

- There is a high correlation (r = 0.9) between the positions these documents occupy in the result pages and the number of citations they have received, at least in queries that only use the filtering option to select the documents published in a given year.

# FUNDING ACKNOLEDGMENTS:

Alberto Martín-Martín enjoys a four-year doctoral fellowship (FPU2013/05863) granted by the Spanish Ministry of Education, Culture and Sports. Juan Manuel Ayllón enjoys a four-year doctoral fellowship (BES-2012-054980) granted by the Spanish Ministry of Economy and Competitiveness. This study has also been funded under project HAR2011-30383-C02-02 from Dirección General de Investigación y Gestión del Plan Nacional de I+D+I (Ministry of Economy and Competitiveness), of which Emilio Delgado López-Cózar is the principal investigator, and project APOSTD/2013/002 from the Regional Ministry of Education, Culture and Sport (Generalitat Valenciana, Spain), awarded to Enrique Orduña-Malea.

# APPENDIX A

| Document type | Bibliographic reference | 1st ed. Pub. Year | GS Citations |
|---|---|---|---|
| J | LOWRY, O.H. et al., (1951). Protein measurement with the Folin phenol reagent.The Journal of biological chemistry, 193(1), 265-275. | 1951 | 253671 |
| J | LAEMMLI, U.K. (1970). Cleavage of structural proteins during the assembly of the head of bacteriophage T4. Nature, 227(5259), 680-685. DOI: 10.1038/227680a0 | 1970 | 221680 |
| J | BRADFORD, M.M. (1976). A rapid and sensitive method for the quantitation of microgram quantities of protein using the principle of protein dye binding.Analytical Biochemistry, 72, 248-254. DOI: 10.1006/abio.1976.9999 | 1976 | 185749 |
| B | SAMBROOK, J., FRITSCH, E. F., & MANIATIS, T. (1982). Molecular cloning: a laboratory manual. New York, Cold Spring Harbor Laboratory Press. | 1982 | 171004 |
| B | AMERICAN PSYCHIATRIC ASSOCIATION. (1952). Diagnostic and statistical manual: mental disorders. Washington, American Psychiatric Assn., Mental Hospital Service. | 1952 | 129473 |
| B | PRESS, W. H. (1986). Numerical recipes: the art of scientific computing. Cambridge [Cambridgeshire], Cambridge University Press. | 1986 | 108956 |
| B | YIN, R. K. (1984). Case study research: design and methods. Beverly Hills, Calif, Sage Publications. | 1984 | 82538 |
| B | ABRAMOWITZ, M., & STEGUN, I. A. (1964). Handbook of mathematical functions: with formulas, graphs, and mathematical tables. Washington, Government printing office. | 1964 | 80482 |
| B | KUHN, T. S. (1962). The structure of scientific revolutions. Chicago, University of Chicago Press. | 1962 | 70662 |
| B | ZAR, J. H. (1974). Biostatistical analysis. Englewood Cliffs, Prentice Hall international. | 1974 | 68267 |
| J | SHANNON, C.E. (1948). A mathematical theory of communication. The Bell System Technical Journal, 27, 379-423. | 1948 | 66851 |
| J | CHOMCZYNSKI, , & SACCHI, N. (1987). Single-step method of RNA isolation by acid guanidinium thiocyanate-phenol-chloroform extraction. Analytical Biochemistry, 162, 156-159. DOI: 10.1006/abio.1987.9999 | 1987 | 63871 |
| J | SANGER F, NICKLEN S, & COULSON AR. (1977). DNA sequencing with chain-terminating inhibitors. Proceedings of the National Academy of Sciences of the United States of America. 74, 5463-7. DOI: 10.1073/pnas.74.12.5463 | 1977 | 63767 |
| B | COHEN, J. (1969). Statistical power analysis for the behavioral sciences. New York, Academic Press. | 1969 | 63766 |
| B | GLASER, B. G., & STRAUSS, A. L. (1967). The discovery of grounded theory: strategies for qualitative research. New York, Aldine de Gruyter. | 1967 | 61158 |
| B | NUNNALLY, J. C. (1967). Psychometric Theory. New York , McGraw-Hill. | 1967 | 60725 |
| B | GOLDBERG, D. E. (1989). Genetic algorithms in search, optimization, and machine learning. Reading, Mass, Addison-Wesley Pub. Co. | 1989 | 59764 |

| Document type | Bibliographic reference | 1st ed. Pub. Year | GS Citations |
|---|---|---|---|
| B | ROGERS, E. M. (1962). Diffusion of Innovations. Pxiii. 367. Free Press of Glencoe, New York; Macmillan, New York: London. | 1962 | 55738 |
| J | BECKE, A.D. (1993). Density Functional Thermochemistry III The Role of Exact Exchange. J. Chem. Phys., 98, 5648-5652. DOI: 10.1063/1.464913 | 1993 | 54642 |
| J | LEE, C., YANG, W. & PARR, R.G., 1988. Development of the Colle-Salvetti correlation-energy formula into a functional of the electron density. Physical Review B, 37(2), 785-789. DOI: 10.1103/PhysRevB.37.785 | 1988 | 52316 |
| J | MURASHIGE, T. & SKOOG, F. (1962). A revised medium for rapid growth and bio assays with tobacco tissue cultures. Physiologia Plantarum, 15, 473–497. DOI: 10.1111/j.1399-3054.1962.tb08052.x | 1962 | 52011 |
| B | ANDERSON, B. R. O. (1983). Imagined communities: reflections on the origin and spread of nationalism. London , Verso. | 1983 | 51177 |
| J | FOLSTEIN, M.F., FOLSTEIN, S.E. & MCHUGH, R., (1975). "Mini-mental state." Journal of Psychiatric Research, 12(3), 189-198. DOI: 10.1016/0022-3956(75)90026-6 | 1975 | 51150 |
| J | TOWBIN, H., STAEHELIN, T. & GORDON, J. (1979). Electrophoretic transfer of proteins from polyacrylamide gels to nitrocellulose sheets: procedure and some applications. Proceedings of the National Academy of Sciences of the United States of America, 76(9), 4350-4354. DOI: 10.1073/pnas.76.9.4350 | 1979 | 50608 |
| B | PAXINOS, G., & WATSON, C. (1982). The rat brain in stereotaxic coordinates. Sydney [etc.], Academic Press. | 1982 | 50471 |
| J | ALTSCHUL, S.F. et al. (1990). Basic local alignment search tool. Journal of molecular biology, 215(3), 403-410. DOI: 10.1006/jmbi.1990.9999 | 1990 | 50437 |
| J | ALTSCHUL, S.F. et al. (1997). Gapped BLAST and PSI-BLAST: A new generation of protein database search programs. Nucleic Acids Research, 25(17), 3389-3402. DOI: 10.1093/nar/25.17.3389 | 1997 | 50052 |
| J | ZADEH, L.A. (1965). Fuzzy sets. Information and Control, 8(3), 338-353. DOI: 10.1016/S0019-9958(65)90241-X | 1965 | 49496 |
| B | GAREY, M. R., & JOHNSON, D. S. (1979). Computers and intractability: a guide to the theory of NP-completeness. San Francisco, W.H. Freeman. | 1979 | 48816 |
| B | RAWLS, J. (1971). A theory of justice. Cambridge, MA, Belknap Press of Harvard University Press. | 1971 | 48792 |
| J | THOMPSON, J.D., HIGGINS, D.G. & GIBSON, T.J. (1994). CLUSTAL W: improving the sensitivity of progressive multiple sequence alignment through sequence weighting, position-specific gap penalties and weight matrix choice. Nucleic acids research, 22(22), 4673-4680. DOI: 10.1093/nar/22.22.4673 | 1994 | 47907 |
| B | SIEGEL, S. (1956). Nonparametric statistics for the behavioral sciences. New York, McGraw-Hill. | 1956 | 47805 |
| B | VYGOTSKY, L. S. (1978). Mind in society: the development of higher psychological processes. Cambridge, Mass, Harvard University Press. | 1978 | 47664 |
| B | BORN, M., & WOLF, E. (1959). Principles of optics: | 1959 | 47486 |

| Document type | Bibliographic reference | 1st ed. Pub. Year | GS Citations |
|---|---|---|---|
| | electromagnetic theory of propagation, interference and diffraction of light. London, Pergamon Press. | | |
| B | GOLUB, G. H., & VAN LOAN, C. F. (1983). Matrix computations. Baltimore, Md, The Johns Hopkins University Press. | 1983 | 47083 |
| J | FOLCH, J. et al. (1957). A simple method for the isolation and purification of total lipids from animal tissues. J Biol Chem, 226(1), 497-509. DOI: 10.1007/s10858-011-9570-9 | 1957 | 45728 |
| J | SHELDRICK, G.M. (2007). A short history of SHELX. Acta Crystallographica Section A: Foundations of Crystallography, 64(1), 112-122. DOI: 10.1107/S0108767307043930 | 2007 | 45208 |
| B | MILES, M. B., & HUBERMAN, A. M. (1984). Qualitative data analysis: a sourcebook of new methods. London, Sage Publications. | 1984 | 45137 |
| J | BARON, R.M. & KENNY, D.A. (1986). The moderator-mediator variable distinction in social psychological research: conceptual, strategic, and statistical considerations. Journal of personality and social psychology, 51(6), 1173-1182. DOI:10.1037/0022-3514.51.6.1173 | 1986 | 44043 |
| B | GREENE, W. H. (1990). Econometric analysis. New York, Macmillan. | 1990 | 43955 |
| B | TABACHNICK, B. G., & FIDELL, L. S. (1983). Using multivariate statistics. New York, Harper & Row. | 1983 | 43474 |
| J | KAPLAN, E.L. & MEIER, (1958). Nonparametric Estimation from Incomplete Observations. Journal of the American Statistical Association, 53(282), 457-481. DOI:10.2307/2281868 | 1958 | 43293 |
| B | GRADŠTEJN, I. S. et al. (1965). Table of integrals, series, and products. New York, Academic Press. | 1965 | 42948 |
| B | BANDURA, A. (1986). Social foundations of thought and action: a social cognitive theory. Englewood Cliffs, N.J., Prentice-Hall. | 1986 | 42791 |
| J | JENSEN, M.C. & MECKLING, W.H. (1976). Theory of the firm: Managerial behavior, agency costs and ownership structure. Journal of Financial Economics, 3(4), 305-360. DOI:10.1016/0304-405X(76)90026-X | 1976 | 42702 |
| B | HAIR, J. F. et al. (1998). Multivariate data analysis. Upper Saddle River, NJ: Pearson Prentice Hall. | 1998 | 41984 |
| B | FREIRE, , FREIRE, , & FREIRE, (1970). Pedagogy of the oppressed. New York (N.Y.), Seabury. | 1970 | 41463 |
| B | FELLER, W. (1950). An introduction to probability theory and its applications. New York, Wiley. | 1950 | 41135 |
| B | FOUCAULT, M. (1977). Discipline and punish: the birth of the prison. New York, Pantheon Books. | 1977 | 41076 |
| J | PERDEW, J., BURKE, K. & ERNZERHOF, M. (1996) . Generalized Gradient Approximation Made Simple. Physical review letters, 77(18), 3865-3868. DOI: 10.1103/PhysRevLett.78.1396 | 1996 | 40868 |
| B | HOLLAND, J. H. (1975). Adaptation in natural and artificial systems an introductory analysis with applications to biology, control, and artificial intelligence. Ann Arbor, University of Michigan Press | 1975 | 40031 |

| Document type | Bibliographic reference | 1st ed. Pub. Year | GS Citations |
|---|---|---|---|
| B | GIANNETTI, F., & LUISE, M. (2007). Spread Spectrum Signals for Digital Communications. In : Handbook of Computer Networks: Key Concepts, Data Transmission, and Digital and Optical Networks, Volume 1, 675-691. | 2007 | 39891 |
| J | SHELDRICK, G.M. et al. (1993). The application of direct methods and Patterson interpretation to high-resolution native protein data. Acta crystallographica. Section D, Biological crystallography, 49(Pt 1), 18-23. DOI: 10.1107/S0907444992007364 | 1993 | 39807 |
| B | LINCOLN, Y. S., & GUBA, E. G. (1985). Naturalistic inquiry. Beverly Hills, Calif, Sage. | 1985 | 37883 |
| J | LIVAK, K.J. & SCHMITTGEN, T.D. (2001). Analysis of relative gene expression data using real-time quantitative PCR and the 2(-Delta Delta C(T)) Method. Methods (San Diego, Calif.), 25(4), 402-408. DOI: 10.1006/meth.2001.1262 | 2001 | 37688 |
| B | LAVE, J., & WENGER, E. (1991). Situated learning legitimate peripheral participation. Cambridge, England, Cambridge University. | 1991 | 37459 |
| J | DEMPSTER, A., LAIRD, N.M. & RUBIN, D.B. (1977). Maximum Likelihood from Incomplete Data Via Em Algorithm. Journal of the Royal Statistical Society Series BMethodological, 39(1), 1-38. | 1977 | 37353 |
| B | SZE, S. N. (1969). Physics of semiconductor devices. New York , J. Wiley and Sons. | 1969 | 37134 |
| B | STRAUSS, A., & CORBIN, J. (1990). Basics of qualitative research: grounded theory procedures and techniques. Newbury Park, Sage. | 1990 | 36986 |
| J | COX, D.R. (1972). Regression models and life tables. Journal of the Royal Statistical Society. Series B:, 34(2), 187-220. | 1972 | 36953 |
| B | SENGE, M. (1990). The fifth discipline: the art and practice of the learning organization. New York, Doubleday/Currency. | 1990 | 36478 |
| J | SAITOU, N. & NEI, M. (1987). The neighbor-joining method: a new method for reconstructing phylogenetic trees. Molecular biology and evolution, 4(4), 406-425. | 1987 | 36207 |
| B | SCHÖN, D. A. (1983). The reflective practitioner how professionals think in action. New York, Basic Books. | 1983 | 35852 |
| B | JACKSON, J. D. (1962). Classical electrodynamics. New York, Wiley. | 1962 | 35849 |
| J | BLIGH, E.G. & DYER, W.J. (1959). A rapid method for total lipid extraction and purification. Canadian Journal of Biochemistry and Physiology 37, no. 8: 911-917. | 1959 | 35095 |
| B | CORMEN, T. H., LEISERSON, C. E., & RIVEST, R. L. (1990). Introduction to algorithms. Cambridge, Mass, MIT Press. | 1990 | 35050 |
| B | COVER, T. M., & THOMAS, J. A. (1991). Elements of information theory. New York, J. Wiley. | 1991 | 34674 |
| B | HAYKIN, S. S. (1994). Neural networks a comprehensive foundation. New York, Macmillan. | 1994 | 34522 |
| J | KOTLER, (2011). Reinventing Marketing to Manage the Environmental Imperative. Journal of Marketing, 75(4), 132-135. | 2011 | 34479 |
| B | WINER, B. J. (1962). Statistical principles in experimental design. | 1962 | 34118 |

| Document type | Bibliographic reference | 1st ed. Pub. Year | GS Citations |
|---|---|---|---|
| | New York, McGraw-Hill. | | |
| J | BARNEY, J. (1991). Firm Resources and Sustained Competitive Advantage.Journal of Management, 17(1), 99-120. DOI:10.1177/014920639101700108 | 1991 | 33976 |
| B | VAPNIK, V. N. (1995). The nature of statistical learning theory. New York, Springer-Verlag. | 1995 | 33506 |
| B | HOFSTEDE, G. (1980). Culture's consequences: international differences in work-related values. Beverly Hill, Sage Publications. | 1980 | 33340 |
| B | HOSMER, D. W., & LEMESHOW, S. (1989). Applied logistic regression. New York, John Wiley & Sons. | 1989 | 33306 |
| B | CRESWELL, J. W. (1994). Research design: qualitative and quantitative approaches. Thousand Oaks, Calif, Sage. | 1994 | 33111 |
| J | BANDURA, A. (1977). Self-efficacy: toward a unifying theory of behavioral change.Psychological review, 84(2), 191-215. DOI: 10.1037/0033-295X.84.2.191 | 1977 | 33038 |
| B | GEERTZ, C. (1973). The interpretation of cultures selected essays. New York, Basic Books. | 1973 | 33003 |
| B | BERGER, L., & LUCKMANN, T. (1966). The social construction of reality a treatise in the sociology of knowledge. Garden City, N.Y., Doubleday. | 1966 | 32710 |
| J | KOHN, W. & SHAMh, L.J. (1965). Self-consistent equations including exchange and correlation effects. Physical Review, 140(4A). DOI:10.1103/PhysRev.140.A1133 | 1965 | 32699 |
| B | CHEMICAL RUBBER COMPANY (CLEVELAND, OHIO). (1913). CRC Handbook of chemistry and physics: a ready-reference book of chemical and physical data. Cleveland, Chemical Rubber Co. | 1913 | 32542 |
| B | BANDURA, A. (1997). Self-efficacy: the exercise of control. New York, W. H. Freeman. | 1997 | 32393 |
| J | IIJIMA, S. (1991). Helical microtubules of graphitic carbon. Nature, 354(6348), 56-58. DOI:10.1038/354056a0 | 1991 | 32338 |
| B | GOFFMAN, E. (1959). The Presentation of Self in Everyday Life. New York, Doubleday Anchor Books. | 1959 | 32251 |
| B | BOX, G. E. , & JENKINS, G. M. (1970). Time series analysis forecasting and control. San Francisco, Holden-Day. | 1970 | 32139 |
| B | GAMMA, E. et al. (1994). Design patterns: elements of reusable object-oriented software. Reading, Mass, Addison-Wesley. | 1994 | 32067 |
| J | SOUTHERN, E. (1975). Detection of specific sequences among DNA fragments separated by gel electrophoresis. Journal of Molecular Biology. 98, 503-517. | 1975 | 31950 |
| B | PORTER, M. E. (1980). Competitive strategy: techniques for analyzing industries and competitors. New York, Free Press. | 1980 | 31532 |
| B | WILLIAMSON, O. E. (1985). The economic institutions of capitalism: firms, markets, relational contracting. New York, The Free Press. | 1985 | 31394 |
| J | KIRKPATRICK, S., GELATT, C.D. & VECCHI, M. (1983). Optimization by Simulated Annealing. Science, 220(4598), 671-680. DOI: 10.1126/science.220.4598.671 | 1983 | 31026 |
| B | NORTH, D. C. (1990). Institutions, institutional change, and economic performance. New York, Cambridge University | 1990 | 31019 |

| Document type | Bibliographic reference | 1st ed. Pub. Year | GS Citations |
|---|---|---|---|
| | Press. | | |
| B | BOURDIEU, P. (1979). Distinction: a social critique of the judgement of taste. London, Routledge & Kegan Paul. | 1979 | 30870 |
| B | PORTER, M. E. (1985). Competitive advantage: creating and sustaining superior performance. New York, The Free Press. | 1985 | 30532 |
| J | MOSMANN, T. (1983). Rapid colorimetric assay for cellular growth and survival: application to proliferation and cytotoxicity assays. Journal of immunological methods, 65(1-2), 55-63. DOI:10.1016/0022-1759(83)90303-4 | 1983 | 30514 |
| B | PATTON, M. Q. (1980). Qualitative evaluation methods. Beverly Hills, Calif, Sage. | 1980 | 30258 |
| J | THOMPSON, J.D. et al. (1997). The CLUSTAL X windows interface: Flexible strategies for multiple sequence alignment aided by quality analysis tools.Nucleic Acids Research, 25(24), 4876-4882. DOI: 10.1093/nar/25.24.4876 | 1997 | 30123 |
| B | MASLOW, A. H. (1954). Motivation and personality. New York, Harper & Row. | 1954 | 30095 |
| J | DUBOIS, M. et al.(1956). Colorimetric method for determination of sugars and related substances. Analytical Chemistry, 28(3), 350-356. DOI: 10.1021/ac60111a017 | 1956 | 30045 |
| B | PORTER, M. E. (1985). Competitive advantage: Creating and sustaining superior performance. New York , Free Press | 1985 | 29924 |
| B | LAZARUS, R. S., & FOLKMAN, S. (1984). Stress, appraisal, and coping. New York, Springer Publishing Company. | 1984 | 29844 |
| J | SHANNON, R.D. (1976). Revised effective ionic radii and systematic studies of interatomic distances in halides and chalcogenides. Acta Crystallographica Section A, 32(5), 751-767. DOI: 10.1107/S0567739476001551 | 1976 | 29796 |
| J | BECKE, A.D. (1988). Density-functional exchange-energy approximation with correct asymptotic behavior. Physical Review A, 38(6), 3098-3100. DOI: 10.1103/PhysRevA.38.3098 | 1988 | 29764 |
| B | COHEN, J., & COHEN, (1975). Applied multiple regression/correlation analysis for the behavioral sciences. Hillsdale, N.J., Lawrence Erlbaum Associates. | 1975 | 29609 |
| B | KITTEL, C. (1953). Introduction to solid state physics. New York, John Wiley & Sons, Inc. | 1953 | 29486 |
| B | CARSLAW, H. S., & JAEGER, J. C. (1947). Conduction of heat in solids. Oxford, Clarendon Press. | 1947 | 29426 |
| B | KNUTH, D. E. (1968). The art of computer programming. Reading Mass, Addison-Wesley. | 1968 | 29396 |
| B | MANDELBROT, B. B. (1977). The fractal geometry of nature. New York, W.H. Freeman. | 1977 | 29270 |
| B | LAKOFF, G., & JOHNSON, M. (1980). Metaphors we live by. Chicago, University of Chicago Press. | 1980 | 29211 |
| J | TVERSKY, A. & KAHNEMAN, D. (1974). Judgment under Uncertainty: Heuristics and Biases. Science (New York, N.Y.), 185(4157), 1124-1131. DOI:10.1126/science.185.4157.1124 | 1974 | 29152 |
| J | BLAND, J.M. & ALTMAN, D.G. (1986). Statistical methods for assessing agreement between two methods of clinical measurement. Lancet, 1(8476), 307-310. | 1986 | 28934 |

| Document type | Bibliographic reference | 1st ed. Pub. Year | GS Citations |
|---|---|---|---|
| | Press. | | |
| B | BANDURA, A. (1971). Social learning theory. Morristown, N.J., General Learning Press. | 1971 | 27016 |
| B | FOUCAULT, M. (1985). The use of pleasure : volume 2 of the history of sexuality. Harmondsworth, Middlesex, England, Viking. | 1985 | 26955 |
| J | RADLOFF, L.S. (1977). The CES-D Scale: A Self Report Depression Scale for Research in the General. Applied Psychological Measurement, 1, 385-401. DOI: 10.1177/014662167700100306 | 1977 | 26787 |
| B | CRESWELL, J. W. (1997). Qualitative inquiry and research design: choosing among five traditions. London, SAGE. | 1997 | 26706 |
| B | ALLEN, M. , & TILDESLEY, D. J. (1987). Computer simulation of liquids. Oxford, Clarendon press | 1987 | 26703 |
| B | CRANK, J. (1956). The mathematics of diffusion. Clarendon Press, Oxford. | 1956 | 26633 |
| B | SCHUMPETER, J. A., & SWEDBERG, R. (1942). Capitalism, socialism and democracy. London, Routledge. | 1942 | 26603 |
| J | FEYERABEND, (1955). Wittgenstein's Philosophical Investigations. The Philosophical Review. 64, 449-483. | 1955 | 26576 |
| O | FRISCH, M. et al. (2004). Gaussian 03, revision c. 02; Gaussian. Inc., Wallingford, CT, 4. | 2004 | 26531 |
| B | BARD, A. J., & FAULKNER, L. R. (1980). Electrochemical methods: fundamentals and applications. New York, N.Y., J. Wiley and Sons. | 1980 | 26494 |
| B | JAMES, W. (1890). The principles of psychology. New York, H. Holt and Company. | 1890 | 26472 |
| B | PATTON, M. Q. (1980). Qualitative research & evaluation methods. Thousand Oaks, Calif, Sage Publications. | 1980 | 26382 |
| B | MARX, K. (1886). Capital: a critical analysis of capitalist production. London, William Glaisher. | 1886 | 26242 |
| B | NELSON, R. R., & WINTER, S. G. (1982). An evolutionary theory of economic change. Cambridge, Mass, Belknap Press of Harvard University Press. | 1982 | 26145 |
| J | AJZEN, I. (1991). The theory of planned behavior. Organizational Behavior and Human Decision Processes, 50(2), 179-211. DOI:10.1016/0749-5978(91)90020-T | 1991 | 26144 |
| B | DRAPER, N. R., & SMITH, H. (1966). Applied regression analysis. New York, John Wiley and Sons, Inc. | 1966 | 26109 |
| B | DARWIN, C. (1859). The origin of species by means of natural selection, or, The preservation of favoured races in the struggle for life. London, John Murray, Albemarle Street. | 1859 | 25970 |
| B | AUSTIN, J. L. (1962). How to do things with words. Cambridge, Harvard University Press. | 1962 | 25949 |
| B | EFRON, B., & TIBSHIRANI, R. J. (1993). An introduction to the bootstra London, Angleterre, Chapman and Hall. | 1993 | 25940 |
| J | OTWINOWSKI, Z. & MINOR, W. (1997). Processing of X-ray diffraction data collected in oscillation mode. Methods in Enzymology, 276, 307-326. DOI:10.1016/S0076-6879(97)76066-X | 1997 | 25800 |
| B | AMERICAN PUBLIC HEALTH ASSOCIATION. (1900). Standard methods for the examination of water and wastewater. | 1900 | 25714 |

| Document type | Bibliographic reference | 1st ed. Pub. Year | GS Citations |
|---|---|---|---|
| | Washington, APHA-AWWA-WPCF. | | |
| B | BARDIN, L. (1977). Analise de conteudo. Lisboa, Edições 70. | 1977 | 25644 |
| B | BOURDIEU, P. (1977). Outline of a theory of practice. Cambridge, U.K., Cambridge University Press. | 1977 | 25613 |
| B | PAULING, L. (1939). The nature of the chemical bond and the structure of molecules and crystals; an introduction to modern structural chemistry. Ithaca, N.Y., Cornell University Press. | 1939 | 25506 |
| J | DIMAGGIO, J. & POWELL, W.W. (1983). The Iron Cage Revisited: Institutional Isomorphism and Collective Rationality in Organizational Fields. American Sociological Review, 48(2), 147. DOI: 10.2307/2095101 | 1983 | 25488 |
| J | EISENHARDT, K.M. (1989). Building Theories from Case Study Research. Academy of Management Review, 14(4), 532-550. DOI:10.2307/258557 | 1989 | 25411 |
| B | FESTINGER, L. (1957). A theory of cognitive dissonance. Stanford, Calif, Stanford university press. | 1957 | 25299 |
| J | FELSENSTEIN, J. (1985). Confidence limits on phylogenies: an approach using the bootstra Evolution, 783–791. DOI: 10.2307/2408678 | 1985 | 25221 |
| J | BECK, A.T. et al. (1961) . An inventory for measuring depression. Archives of general psychiatry, 4, 561-571. | 1961 | 25085 |
| B | VYGOTSKY, L. S. (1962). Thought and language. Cambridge, The Massachusetts Institute of Technology. | 1962 | 24996 |
| J | COLEMAN, J.S. (1988). Social Capital in the Creation of Human Capital. American Journal of Sociology, 94(s1), S95. DOI:10.1086/228943 | 1988 | 24994 |
| J | LANDIS, J.R. & KOCH, G.G. (1977). The measurement of observer agreement for categorical data. Biometrics, 33(1), 159-174. DOI:10.2307/2529310 | 1977 | 24981 |
| B | KOLB, D. A. (1984). Experiential learning: experience as the source of learning and development. Englewood Cliffs, N.J., Prentice-Hall. | 1984 | 24860 |
| B | MCCULLAGH, , & NELDER, J. (1983). Generalized Linear Models. London, Chapman and Hall. | 1983 | 24694 |
| B | MILLER, J. H. (1972). Experiments in molecular genetics. New York, Cold Spring Harbor Laboratory. | 1972 | 24682 |
| B | PATANKAR, S. V. (1980). Numerical heat transfer and fluid flow. Washington, D.C., New York, Hemisphere. Taylor and Francis. | 1980 | 24663 |
| J | BLACK, F. & SCHOLES, M. (1973). The Pricing of Options and Corporate Liabilities. Journal of Political Economy, 81(3), 637. DOI:10.1086/260062 | 1973 | 24582 |
| J | GRANOVETTER, M. (1985). Economic-action and social-structure - the problem of embeddedness. American Journal of Sociology, 91(3), 481-510. DOI:10.1086/228311 | 1985 | 24324 |
| B | BREIMAN, L. et al.(1984). Classification and regression trees. Pacific Grove, Calif, Wadsworth & Brooks-Cole advanced books & software. | 1984 | 24299 |
| J | LOWE, D.G. (2004). Distinctive image features from scale-invariant keypoints. International Journal of Computer | 2004 | 24234 |

| Document type | Bibliographic reference | 1st ed. Pub. Year | GS Citations |
|---|---|---|---|
| | Vision, 60(2), 91-110. DOI: 10.1023/B:VISI.0000029664.99615.94 | | |
| B | STRAUSS, A. L., & CORBIN, J. (1998). Basics of qualitative research: techniques and procedures for developing grounded theory. Thousand Oaks , SAGE Publications. | 1998 | 24209 |
| B | PAPOULIS, A. (1965). Probability, Random variables, and stochastic processes. New York: McGraw-Hill. | 1965 | 24099 |
| J | AKAIKE, H. (1974). A new look at the statistical model identification. IEEE Transactions on Automatic Control, 19(6). DOI:10.1109/TAC.1974.1100705 | 1974 | 24061 |
| B | QUINLAN, J. R. (1993). C4.5: Programs for machine learning. San Mateo, CA: Morgan Kaufmann. | 1993 | 24050 |
| B | LAKOWICZ, J. R. (1983). Principles of fluorescence spectroscopy. New York: Plenum Press. | 1983 | 23977 |
| B | SWOFFORD, D. (1995). PAUP 4.0 phylogenetic analysis using parsimony. [Sunderland, Mass.], Sinauer Associates. | 1995 | 23957 |
| B | HORN, R. A., & JOHNSON, C. R. (1985). Matrix analysis - Roger A. Horn, Charles R. Johnson. Cambridge, Cambridge university press. | 1985 | 23908 |
| B | BECKER, G. S. (1964). Human capital: a theoretical and empirical analysis, with special reference to education. New York, National Bureau of Economic Research. | 1964 | 23879 |
| B | MEAD, G. H., & MORRIS, C. W. (1934). Mind, self, and society: from the standpoint of a social behaviorist. Chicago, The University of Chicago Press. | 1934 | 23824 |
| J | HARDIN, G.(1968). The Tragedy of the Commons. Science, 162(3859), 1243-1248. DOI: | 1968 | 23737 |
| B | ARGYRIS, C., & SCHÖN, D. A. (1978). Organisational learning: a theory of action perspective. Reading, Mass. [etc.], Addison-Wesley Publishing company. | 1978 | 23329 |
| B | CHOMSKY, N. (1965). Aspects fo the theory of syntax. Cambridge, Mass, MIT Press. | 1965 | 23178 |
| J | COASE, R.H. (1960). The Problem of Social Cost. The Journal of Law and Economics, 3(1), 1. DOI:10.1086/466560 | 1960 | 23141 |
| B | BURNHAM, K. , ANDERSON, D. R., & BURNHAM, K. (2002). Model selection and multimodel inference: a practical information-theoretic approach. New York, Springer | 2002 | 23046 |
| B | BECK, U., & RITTER, M. (1992). Risk society: towards a new modernity. London, Sage Publications. | 1992 | 22990 |
| B | GIDDENS, A. (1991). Modernity and self-identity: self and society in the late modern age. Cambridge, U.K., Polity Press in association with Basil Blackwell. | 1991 | 22977 |
| J | ENGLE, R.F. & GRANGER, C.W.J. (1987). Co-integration and Error Correction: Representation, Estimation, and Testing. Econometrica, 55(2), 251-76. DOI: 10.2307/1913236 | 1987 | 22851 |
| B | ZHU, F., WU, R., HU, Y., & JIANG, Z. (1995). Zhu Futang shi yong er ke xue. Chinamaxx Digital Library. Beijing Shi, Ren min wei sheng chu ban she. | 1995 | 22810 |
| J | WATTS, D. & STROGATZ, S. (1998). Collective dynamics of "small-world"networks. nature, 393(6684), 440-442. DOI: 10.1038/30918 | 1998 | 22681 |

| Document type | Bibliographic reference | 1st ed. Pub. Year | GS Citations |
|---|---|---|---|
| B | FLORY, J. (1953). Principles of polymer chemistry. Ithaca, N.Y., Cornell Univ. Pr. | 1953 | 22680 |
| J | TAMURA, K. et al. (2007). MEGA4: Molecular Evolutionary Genetics Analysis (MEGA) software version 4.0. Molecular Biology and Evolution, 24(8), 1596-1599. DOI: 10.1093/molbev/msm092 | 2007 | 22680 |
| B | FALCONER, D. S. (1960). Introduction to quantitative genetics. Edinburgh, Oliver and Boyd. | 1960 | 22654 |
| B | GRICE, H. (1970). Logic and conversation. Cambridge, Mass, Harvard Univ. | 1970 | 22608 |
| B | RUSSELL, S. J., NORVIG, , & CANNY, J. F. (1995). Artificial intelligence: a modern approach. Englewood Cliffs, Prentice-Hall International. | 1995 | 22577 |
| J | CRONBACH, L.J. (1951). Coefficient alpha and the internal structure of tests.Psychometrika, 16(3), 297-334. DOI: | 1951 | 22531 |
| B | AJZEN, I., & FISHBEIN, M. (1980). Understanding attitudes and predicting social behavior. Englewood Cliffs, N.J., Prentice-Hall. | 1980 | 22419 |
| B | BHABHA, H. K. (1994). The Location of culture. London, Routledge. | 1994 | 22414 |
| B | COHEN, W.M. & LEVINTHAL, D.A. (1990). Absorptive Capacity: A New Perspective on Learning and Innovation W. H. Starbuck & S. Whalen, eds. Administrative Science Quarterly, 35(1), 128-152. | 1990 | 22301 |
| B | CASTELLS, M. (1996). The rise of the network society. Oxford, Blackwell Publishers. | 1996 | 22207 |
| B | DAUBECHIES, I. (1992). Ten lectures on wavelets. Philadelphia, Pa, Soc. for Industrial and Applied Mathematics. | 1992 | 22165 |
| B | AXELROD, R. (1984). The evolution of cooperation. New York, Basic Books. | 1984 | 22109 |
| B | AIKEN, L. S., WEST, S. G., & RENO, R. R. (1991). Multiple regression: testing and interpreting interactions. Newbury Park, CA, Sage Publications. | 1991 | 22036 |
| B | NAKAMOTO, K. (1970). Infrared and Raman spectra of inorganic and coordination compounds. New York, Wiley. | 1970 | 22022 |
| J | BENJAMINI, Y. & HOCHBERG, Y. (1995). Controlling the False Discovery Rate: A Practical and Powerful Approach to Multiple Testing. Journal of the Royal Statistical Society. Series B (Methodological), 57(1), 289 - 300. DOI: | 1995 | 21933 |
| J | PRAHALAD, C.K. & HAMEL, G. (1990). The core competencies of the corporation. Harvard Business Review, 68(3), 79-91. | 1990 | 21868 |
| B | WITTEN, I. H., & FRANK, E. (1999). Data mining: practical machine learning tools and techniques with Java implementations. San Francisco, Calif, Morgan Kaufmann. | 1999 | 21828 |
| B | COLEMAN, J. S. (1990). Foundations of social theory. Cambridge, Mass, Harvard University Press. | 1990 | 21791 |
| J | ROSS, R. (1999). Atherosclerosis--an inflammatory disease. The New England journal of medicine, 340(2), 115-126. DOI:10.1016/S0002-8703(99)70266-8 | 1999 | 21741 |
| B | KEYNES, J. M. (1936). The General theory of employment interest and money. London, Macmillan and Co. | 1936 | 21690 |

| Document type | Bibliographic reference | 1st ed. Pub. Year | GS Citations |
|---|---|---|---|
| B | BIRD, R. B., STEWART, W. E., & LIGHTFOOT, E. N. (1960). Transport phenomena. New York, J.Wiley. | 1960 | 21628 |
| B | ISRAELACHVILI, J. N. (1985). Intermolecular and surface forces: with applications to colloidal and biological systems. London, Academic Press. | 1985 | 21522 |
| B | BISHOP, C. M. (1995). Neural networks for pattern recognition. Oxford, UK, Oxford University Press. | 1995 | 21458 |
| B | COTTON, F. A. :. W., G. (1962). Advanced Inorganic Chemistry. London, Wiley. | 1962 | 21450 |
| J | KRESSE, G., & FURTHMÜLLER, J. (1996). Efficient iterative schemes for ab initio total-energy calculations using a plane-wave basis set. Physical Review B, 54(16), 11169-11186. DOI: 10.1103/PhysRevB.54.11169 | 1996 | 21248 |
| J | MARQUARDT, D.W. (1963). An Algorithm for Least-Squares Estimation of Nonlinear Parameters. Journal of the Society for Industrial and Applied Mathematics, 11(2), 431-441. DOI: | 1963 | 21246 |
| B | HOFSTEDE, G. (1991). Cultures and organizations: sofware of the mind. London, McGraw-Hill Book Company. | 1991 | 21232 |
| J | MCKHANN, G. et al. (1984). Clinical diagnosis of Alzheimer's disease: report of the NINCDS-ADRDA Work Group under the auspices of Department of Health and Human Services Task Force on Alzheimer's Disease. Neurology, 34(7), 939-944. | 1984 | 21172 |
| B | BRYK, A. S., & RAUDENBUSH, S. W. (1992). Hierarchical linear models: Applications and data analysis methods. London, Sage Publications. | 1992 | 21134 |
| B | DOWNS, A. (1957). An economic theory of democracy. New York, Harper. | 1957 | 21017 |
| J | WARE, J.E. & SHERBOURNE, C.D. (1992). The MOS 36-item short-form health survey (SF-36). I. Conceptual framework and item selection. Medical care, 30(6), 473-483. DOI: 10.1097/00005650-199206000-00002 | 1992 | 20975 |
| B | HAN, J., & KAMBER, M. (2000). Data mining: concepts and techniques. London, Harcourt Publishers, a subsidiary of Harcourt International Ltd. | 2000 | 20974 |
| B | MONTGOMERY, D. C. (1976). Design and analysis of experiments. New York, John Wiley & Sons. | 1976 | 20905 |
| J | REYNOLDS, E. S. (1963). The use of lead citrate at high pH as an electron-opaque stain in electron microscopy. The Journal of cell biology, 17(1), 208-212. DOI: 10.1083/jcb.17.1.208 | 1963 | 20890 |
| B | BRONFENBRENNER, U. (1979). The ecology of human development: experiments by nature and design. Cambridge - Mass. & London, Harvard University Press. | 1979 | 20865 |
| J | SCHWARZ, G. (1978). Estimating the Dimension of a Model. The Annals of Statistics, 6(2), 461-464. DOI: 10.1214/aos/1176344136 | 1978 | 20828 |
| B | GOODMAN, L. S., & GILMAN, A. (1941). The pharmacological basis of therapeutics: a textbook of pharmacology, toxicology and therapeutics for physicians and medical students. New York, Macmillan. | 1941 | 20827 |
| B | NIELSEN, M. A., & CHUANG, I. (2000). Quantum computation | 2000 | 20803 |

| Document type | Bibliographic reference | 1st ed. Pub. Year | GS Citations |
|---|---|---|---|
| | & Vol. III. London, Charles Griffin. | | |
| B | FLEISS, J. L. (1973). Statistical methods for rates and proportions. New York, John Wiley & Sons. | 1973 | 20196 |
| B | ROSENBERG, M. (1965). Society and the adolescent self-image. Princeton, No.J., Princeton University Press. | 1965 | 20191 |
| J | FORNELL, C. & LARCKER, D.F. (1981). Evaluating Structural Equation Models with Unobservable Variables and Measurement Error. Journal of Marketing Research (JMR). Feb1981, 18(1), 39-50. 12 1 Diagram. DOI:10.2307/3151312 | 1981 | 20042 |
| B | FISHBEIN, M., & AJZEN, E. (1975). Belief, attitude, intention and behavior: an introduction to theory and research. Reading, Mass. ; Don Mills, Ont, Addison-Wesley. | 1975 | 20030 |
| B | HEBB, D. O. (1949). The organization of behavior a neuropsychological approach. New York, NY, John Wiley & Sons. | 1949 | 19954 |
| B | GARDNER, H. (1983). Frames of mind: the theory of multiple intelligences. New York, Basic books. | 1983 | 19928 |
| J | MONKHORST, H. J., & PACK, J. D. (1976). Special points for Brillouin-zone integrations. Physical Review B. 13, 5188-5192. DOI: 10.1103/PhysRevB.13.5188 | 1976 | 19904 |
| J | CULLITY, B.D. (1957). Elements of X-Ray Diffraction. American Journal of Physics, 25(6), 394. DOI: | 1957 | 19875 |
| J | GRYNKIEWICZ, G., POENIE, M. & TSIEN, R.Y. (1985). A new generation of Ca2+ indicators with greatly impoved fluorescence properties. Journal of Biological Chemistry, 260(6), 3440-3450. | 1985 | 19871 |
| J | BARABISI, A.L., & ALBERT, R. (1999). Emergence of Scaling in Random Networks.Science. 286, 509. DOI: 10.1126/science.286.5439.509 | 1999 | 19806 |
| B | KEETON, W. , & PROSSER, W. L. (1941). Prosser and Keeton on the law of torts. St. Paul, Minn, West publishing. | 1941 | 19725 |
| B | MITRA, G. (1988). Mathematical models for decision support: Advanced study institute : Papers. | 1988 | 19713 |
| B | WOOLDRIDGE, J. M. (2002). Econometric analysis of cross section and panel data. Cambridge, Mass, MIT Press. | 2002 | 19698 |
| J | HAMILTON, M. (1960). A rating scale for depression. Journal of neurology, neurosurgery, and psychiatry, 23, 56-62. DOI: 10.1136/jnn23.1.56 | 1960 | 19690 |
| B | PEARL, J. (1988). Probabilistic reasoning in intelligent systems: networks of plausible inference. San Mateo, Calif, Morgan Kaufmann Publishers. | 1988 | 19578 |
| B | GIBSON, J. J. (1979). The ecological approach to visual perception. Boston, Houghton Mifflin company. | 1979 | 19555 |
| B | DEWEY, J. (1916). Democracy and Education : An introduction to the philosophy of education. New York, The Macmillan Company. | 1916 | 19420 |
| J | DAVIS, F.D. (1989). Perceived usefulness, perceived ease of use, and user acceptance of information technology. MIS Quarterly, 13(3), 319-340. DOI: 10.2307/249008 | 1989 | 19355 |
| B | POPPER, K. R. (1959). The logic of scientific discovery. London, Hutchinson. | 1959 | 19325 |
| B | ESPING-ANDERSEN, G. (1990). The three worlds of welfare | 1990 | 19311 |

| Document type | Bibliographic reference | 1st ed. Pub. Year | GS Citations |
|---|---|---|---|
| | capitalism. Cambridge, Polity press. | | |
| B | HASTIE, T., HASTIE, T., TIBSHIRANI, R., & FRIEDMAN, J. H. (2001). The elements of statistical learning: data mining, inference, and prediction. New York, Springer. | 2001 | 19242 |
| B | PENROSE, E. T. (1959). The theory of growth of the firm. Oxford, Basil Blackwell. | 1959 | 19180 |
| B | RAPPAPORT, T. S. (1996). Wireless communications: principles and practice. Upper Saddle River (New Jersey), Prentice Hall PTR. | 1996 | 19175 |
| B | BLOOM, B. S. (1956). Taxonomy of educational objectives : the classification of educational goals Handbook 1 Handbook 1. New York, McKay. | 1956 | 19139 |
| B | JOLLIFFE, I. T. (1986). Principal component analyses. New York, Springer-Verlag. | 1986 | 19120 |
| B | WEINER, I. B., & CRAIGHEAD, W. E. (1984). The Corsini encyclopedia of psychology. Hoboken, NJ, Wiley. | 1984 | 19119 |
| J | COHEN, J. (1960). A coefficient of agreement of nominal scales. Educational and Psychological Measurement, 20(1), 37-46. DOI: 10.1177/001316446002000104 | 1960 | 19116 |
| B | LEZAK, M. D. (1976). Neuropsychological assessment. New York, Oxford University Press. | 1976 | 19081 |
| B | MILLER, G. A. (1956). The magical number seven, plus or minus two: some limits on our capacity for processing information. Indiana, Bobbs-Merrill. DOI: 10.1037//0033-295X.101.2.343 | 1956 | 19070 |
| B | HARVEY, D. (1989). The condition of postmodernity: an enquiry into the origins of cultural change. Oxford ; Cambridge, Mass, Blackwell. | 1989 | 19053 |
| J | ROGERS, S. (1996). Adaptive filter theory. Control Engineering Practice, 4(11), 1629-1630. | 1996 | 19012 |
| J | CANNY, J. (1986). A computational approach to edge detection. IEEE transactions on pattern analysis and machine intelligence, 8(6), 679-698. | 1986 | 18958 |
| J | RABINER, L.R. (1989). A tutorial on hidden Markov models and selected applications in speech recognition. Proceedings of the IEEE, 77(2), 257-286. DOI: 10.1109/5.18626 | 1989 | 18920 |
| B | BOLLEN, K.A. (1998). Structural Equation Models. Encyclopedia of Biostatistics. 7. | 1998 | 18905 |
| J | WHITE, H. (1980). A heteroskedasticity-consistent covariance matrix estimator and a direct test for heteroskedasticity. Econometrica, 48(4), 817-838. DOI:10.2307/1912934 | 1980 | 18878 |
| B | MARCH, J. G., & SIMON, H. A. (1958). Organizations. New York, Wiley. | 1958 | 18835 |
| J | O'FARRELL, H. (1975). High resolution two-dimensional electrophoresis of proteins. The Journal of biological chemistry, 250(10), 4007-4021. | 1975 | 18835 |
| B | POLANYI, M. (1966). The tacit dimension. Garden City, N. Y, Doubleday. | 1966 | 18818 |
| J | HANAHAN, D., & WEINBERG, R. A. (2000). The Hallmarks of Cancer. Cell. 100, 57. DOI:10.1016/S0092-8674(00)81683-9 | 2000 | 18740 |

| Document type | Bibliographic reference | 1st ed. Pub. Year | GS Citations |
|---|---|---|---|
| J | NELDER, J.A. & MEAD, R. (1965). A Simplex Method for Function Minimization. The Computer Journal, 7(4), 308-313. DOI: | 1965 | 18732 |
| B | SEN, A. (1999). Development as freedom. Oxford, Oxford University Press. | 1999 | 18728 |
| B | GOFFMAN, E. (1963). Stigma; notes on the management of spoiled identity. Englewood Cliffs, N.J., Prentice-Hall. | 1963 | 18727 |
| J | AKERLOF, G. A. (1970). The Market for "Lemons": Quality Uncertainty and the Market Mechanism. The Quarterly Journal of Economics, 84(3), 488-500. DOI:10.2307/1879431 | 1970 | 18684 |
| J | DUNCAN, D. (1955). Multiple range and multiple F tests. Biometrics, 11(1), 1-42. DOI:10.2307/3001478 | 1955 | 18682 |
| B | LYOTARD, J.F., BENNINGTON, G., MASSUMI, B., & JAMESON, F. (1984). The postmodern condition: a report on knowledge. Manchester, Manchester University Press. | 1984 | 18672 |
| B | FOUCAULT, M. (1972). The archaeology of knowledge: and the discourse on language. New York, Pantheon Books. | 1972 | 18668 |
| J | MALLAT, S.G. (1989). Theory for multiresolution signal decomposition: the wavelet representation. IEEE Transactions on Pattern Analysis and Machine Intelligence, 11(7), 674-693. DOI: 10.1109/34.192463 | 1989 | 18662 |
| B | FOUCAULT, M., & GORDON, C. (1980). Power/knowledge: selected interviews and other writings, 1972-1977. Brighton, Sussex, Harvester Press. | 1980 | 18643 |
| B | KLINE, R. B. (1998). Principles and practice of structural equation modeling. New York, Guilford Press. | 1998 | 18587 |
| B | OSTROM, E. (1990). Governing the commons: the evolution of institutions for collective action. Cambridge, UK, Cambridge University. | 1990 | 18575 |
| B | LJUNG, L.. (1998).System identification. In:  Signal analysis and prediction. Boston, Birkhauser. | 1998 | 18509 |
| J | NOVOSELOV, K.S. et al. (2004). Electric field effect in atomically thin carbon films.Science (New York, N.Y.), 306(5696), 666-669. DOI: 10.1126/science.1102896 | 2004 | 18489 |
| B | SPIELBERGER, C. D., GORSUCH, R. L., & LUSHENE, R. E. (1970). STAI manual: for the State-Trait Anxiety Inventory ("self-evaluation questionnaire"). Palo Alto, Consulting Psychologists Press. | 1970 | 18481 |
| B | CYERT, R. M., & MARCH, J. G. (1963). A behavioral theory of the firm. New Jersey, Prentice Hall. | 1963 | 18465 |
| B | SIMON, H. A., & BARNARD, C. I. (1947). Administrative behavior: a study of decision-making processes in administrative organization. New York, Macmillan Co. | 1947 | 18430 |
| B | SCHLICHTING, H., & KESTIN, J. (1955). Boundary layer theory. New York, McGraw-Hill. | 1955 | 18383 |
| J | OLDFIELD, R. (1971). The assessment and analysis of handedness: The Edinburgh inventory. Neuropsychologia. 9, 97-113. DOI:10.1016/0028-3932(71)90067-4 | 1971 | 18325 |
| B | ARTIKUNTO, S. (2006). Prosedur Penelitian Suatu Pendekatan Praktik. Jakarta: Rineka Cipta. | 2006 | 18302 |
| J | TEECE, D. J., PISANO, G., & SHUEN, A. (1997). Dynamic | 1997 | 18291 |

| Document type | Bibliographic reference | 1st ed. Pub. Year | GS Citations |
|---|---|---|---|
|  | Capabilities and Strategic Management. Strategic Management Journal. 18, 509. DOI:10.1002/(SICI)1097-0266(199708)18:7<509::AID-SMJ882>3.0.CO;2-Z |  |  |
| B | WASSERMAN, S., & FAUST, K. (1994). Social network analysis: methods and applications. New York, Cambridge University Press. | 1994 | 18267 |
| B | BOURDIEU, (1986). The forms of capital. In Handbook of Theory and Research for the Sociology of Education. p 241-258. | 1986 | 18248 |
| B | ALLPORT, G. W. (1954). The nature of prejudice Semesterhylde. Reading, Mass, Addison-Wesley. | 1954 | 18215 |
| J | HECKMAN, J. (1979). Sample selection bias as a specification error. Econometrica. 47, 153-161. DOI:10.2307/1912352 | 1979 | 18160 |
| B | ERIKSON, E. H. (1968). Identity: youth and crisis. New York, Norton. | 1968 | 18140 |
| B | KHALIL, H. K. (1992). Nonlinear systems. New York, Macmillan Pub. Co. | 1992 | 18072 |
| B | THOMPSON, J. D. (1967). Organizations in action. New York, McGraw Hill. | 1967 | 17926 |
| J | KYTE, J. & DOOLITTLE, R.F. (1982). A simple method for displaying the hydropathic character of a protein. Journal of molecular biology, 157(1), 105-132. DOI: 10.1016/0022-2836(82)90515-0 | 1982 | 17888 |
| C | PERKINS, C.E. & ROYER, E.M. (1999). Ad-hoc on-demand distance vector routing. In Proceedings - WMCSA'99: 2nd IEEE Workshop on Mobile Computing Systems and Applications. p 90-100. DOI: | 1999 | 17879 |
| B | DAWKINS, R. (1976). The selfish gene. New York, Oxford University Press. | 1976 | 17849 |
| J | ROTTER JB. (1966). Generalized expectancies for internal versus external control of reinforcement. Psychological Monographs. 80, 1-28. | 1966 | 17834 |
| B | ROCKAFELLAR, R. T. (1970). Convex analysis. Princeton, N.J., Princeton University Press. | 1970 | 17831 |
| B | ABRAGAM, A. (1961). The principles of nuclear magnetism. Oxford, Clarendon Press. | 1961 | 17826 |
| J | KALNAY, E. et al. (1996). The NCEP/NCAR 40-year reanalysis project. Bulletin of the American Meteorological Society, 77(3), 437-471. DOI:10.1175/1520-0477(1996)077<0437:TNYRP>2.0.CO;2 | 1996 | 17798 |
| J | MARKOWITZ, H. (1952). Portfolio Selection. Journal of Finance, 7, 77-91. DOI: 10.2307/2975974 | 1952 | 17768 |
| B | FREIRE, (1997). Pedagogía de la autonomía: saberes necesarios para la práctica educativa. México, Siglo XXI. | 1997 | 17723 |
| C | HALL, T. (1999). BioEdit: a user-friendly biological sequence alignment editor and analysis program for Windows 95/98/NT. Nucleic Acids Symposium Series, 41, 95-98. | 1999 | 17716 |
| B | SCHÖN, D. A. (1987). Educating the reflective practitioner: toward a new design for teaching and learning in the professions. San Francisco, Jossey-Bass. | 1987 | 17705 |
| B | DUDA, R. O., & HART, E. (1973). Pattern classification and scene analysis. New York, N.Y., J. Wiley and Sons. | 1973 | 17702 |

| Document type | Bibliographic reference | 1st ed. Pub. Year | GS Citations |
|---|---|---|---|
| J | CHARNES, A., COOPER, W., & RHODES, E. (1978). Measuring the efficiency of decision making units. European Journal of Operational Research. 2, 429-444. DOI: 10.1016/0377-2217(78)90138-8 | 1978 | 17667 |
| J | BRÜNGER AT, et al. (1998). Crystallography & NMR system: A new software suite for macromolecular structure determination. Acta Crystallographica. Section D, Biological Crystallography. 54, 905-21. DOI:10.1107/S0907444998003254 | 1998 | 17614 |
| B | PFEFFER, J., & SALANCIK, G. R. (1978). The external control of organizations: a resource dependence perspective. New York, Harper & Row. | 1978 | 17608 |
| J | FARRUGIA, L. J. (1997). ORTEP-3 for Windows - a version of ORTEP-III with a Graphical User Interface (GUI). JOURNAL OF APPLIED CRYSTALLOGRAPHY. 30, 565. DOI: | 1997 | 17602 |
| J | O'REGAN, B. & GRÄTZEL, M. (1991). A low-cost, high-efficiency solar cell based on dye-sensitized colloidal $TiO_2$ films. Nature, 353(6346), 737-740. DOI: 10.1038/353737a0 | 1991 | 17565 |
| J | ROMER, M. (1990). Endogenous Technological Change. Journal of Political Economy, 98(S5), S71. DOI: 10.1086/261725 | 1990 | 17554 |
| J | WERNERFELT, B. (1984). A resource-based view of the firm. Strategic Management Journal, 5(2), 171-180. DOI: 10.1002/smj.4250050207 | 1984 | 17545 |
| J | SHELDRICK, G.M. (1990). Phase annealing in SHELX-90: direct methods for larger structures. Acta Crystallographica Section A Foundations of Crystallography, 46(6), 467-473. DOI: 10.1107/S0108767390000277 | 1990 | 17444 |
| B | SEARLE, J. R. (1969). Speech acts: an essay in the philosophy of language. London, Cambridge Univ. Press. | 1969 | 17327 |
| B | MARSCHNER, H., & MARSCHNER, (1986). Marschner's mineral nutrition of higher plants. London, Academic Press. | 1986 | 17305 |
| B | HALLIDAY, M. A. K. (1985). An introduction to functional grammar. London, E. Arnold. | 1985 | 17303 |
| B | DAMASIO, A. R. (1994). Descartes' error: emotion, reason, and the human brain. New York, Putnam. | 1994 | 17298 |
| B | WEBER, M., PARSONS, T., & GIDDENS, A. (1930). The Protestant ethic and the spirit of capitalism. London, George Allen & Unwin. | 1930 | 17288 |
| J | PARKIN, D. M. et al. (2005). Global Cancer Statistics, 2002. CA: A Cancer Journal for Clinicians. 55, 74-108. | 2005 | 17287 |
| J | BERMAN, H.M. et al (2000) . The Protein Data Bank. Nucleic acids research, 28(1), 235-242. DOI: 10.1093/nar/28.1.235 | 2000 | 17283 |
| B | POLING, B. E., PRAUSNITZ, J. M., & O'CONNELL, J. (1958). The properties of gases and liquids. Boston, McGraw-Hill. | 1958 | 17255 |
| B | COCHRAN, W. G. (1953). Sampling techniques. New York, John Wiley & sons. | 1953 | 17243 |
| J | SAIKI, R. et al. (1988). Primer-directed enzymatic amplification of DNA with a thermostable DNA polymerase. Science. 239, 487-491. DOI:10.1126/science.2448875 | 1988 | 17239 |
| B | LANDAU, L. D. et al .(1960). Electrodynamics of continuous media. Oxford, Pergamon Press. | 1960 | 17234 |
| B | 马克思, & 恩格斯. (2007). 马克思恩格斯全集: 书信 (1848 年 3 | 2007 | 17218 |

| Document type | Bibliographic reference | 1st ed. Pub. Year | GS Citations |
|---|---|---|---|
| | 月--1851 年 12 月). 第四十八卷. 人民出版社. | | |
| B | HALLIWELL, B., & GUTTERIDGE, J. M. C. (1985). Free radicals in biology and medicine. Oxford, Clarendon Press. | 1985 | 17206 |
| B | KATŌ, T. (1966). Perturbation theory for linear operators. Berlin, Springer-Verlag. | 1966 | 17161 |
| B | CHOMSKY, N. (1992). A minimalist program for linguistic theory. Cambridge, MA, Distributed by MIT Working Papers in Linguistics, Dept. of Linguistics and Philosophy, Massachusetts Institute of Technology. | 1992 | 17151 |
| J | CHANG C.C., & LIN C.J. (2011). LIBSVM: A Library for support vector machines. ACM Transactions on Intelligent Systems and Technology. 2. DOI:10.1145/1961189.1961199 | 2011 | 17145 |
| J | HAMILL, O. et al. (1981). Improved patch-clamp techniques for high-resolution current recording from cells and cell-free membrane patches. Pflügers Archiv : European Journal of Physiology. 391, 85-100. DOI: 10.1007/BF00656997 | 1981 | 17143 |
| B | BEVINGTON, R., & ROBINSON, D. K. (1969). Data reduction and error analysis for the physical sciences. New York, McGraw-Hill. | 1969 | 17137 |
| B | SUTTON, R. S., & BARTO, A. G. (1998). Reinforcement learning: an introduction. Cambridge, Mass, MIT Press. | 1998 | 17131 |
| B | GOLEMAN, D. (1995). Emotional intelligence: [why it can matter more than IQ]. New York, Bantam Books. | 1995 | 17123 |
| J | ROMER, M. (1986). Increasing Returns and Long-Run Growth. Journal of Political Economy, 94(5), 1002. DOI:10.1086/261420 | 1986 | 17082 |
| B | HIRSJÄRVI, S., REMES, , & SAJAVAARA, (1997). Tutki ja kirjoita. Helsinki, Kirjayhtymä. | 1997 | 16993 |
| B | PARSONS, T. (1951). The social system. Glencoe, The Free Press. | 1951 | 16977 |
| J | LASKOWSKI, R. A. et al. (1993). PROCHECK: a program to check the stereochemical quality of protein structures.Journal of Applied Crystallography. 26, 283-291. DOI:10.1107/S0021889892009944 | 1993 | 16975 |
| B | CAO, Z., & WENG, L. (1999). Zhonghua fu chan ke xue = Chinese obstetrics and gynecology. Beijing Shi, Ren min wei sheng chu ban she. | 1999 | 16947 |
| B | DENZIN, N. K., & LINCOLN, Y. S. (1994). Handbook of qualitative research. Thousand Oaks, Calif, Sage. | 1994 | 16946 |
| J | KALMAN, R.E. (1960). A New Approach to Linear Filtering and Prediction Problems. Transactions of the ASME-Journal of Basic Engineering, 82(Series D), 35-45. | 1960 | 16881 |
| B | GOODMAN, J. W. (1968). Introduction to fourier optics. San Francisco, Mcgraw-Hill. | 1968 | 16852 |
| B | RUMELHART, D. E., & MCCLELLAND, J. L. (1986). Parallel distributed processing: explorations in the microstructure of cognition. Cambridge, Mass, MIT Press. | 1986 | 16827 |
| J | BERNERS-LEE, T., HENDLER, J. & LASSILA, O. (2001). The Semantic Web. Scientific American, 284(5), 34-43. | 2001 | 16826 |
| B | BOHREN, C. F., & HUFFMAN, D. R. (1983). Absorption and scattering of light by small particles. New York, Wiley. | 1983 | 16796 |
| J | THE DIABETES CONTROL AND COMPLICATIONS TRIAL RESEARCH | 1993 | 16794 |

| Document type | Bibliographic reference | 1st ed. Pub. Year | GS Citations |
|---|---|---|---|
| | GROUP, (1993). The effect of intensive treatment of diabetes on the development and progression of long-term complications in insulin-dependent diabetes mellitus. The New England Journal of Medicine, 329, 977-86. | | |
| B | GRAMSCI, A., & BOOTHMAN, D. (1995). Further selections from the prison notebooks. Minneapolis, University of Minnesota Press. | 1995 | 16774 |
| B | BOGDAN, R. C., & BIKLEN, S. K. (1982). Qualitative research for education: an introduction to theory and methods. Boston, Mass, Allyn and Bacon. | 1982 | 16743 |
| B | CERTEAU, M. D. et al. (1998). The practice of everyday life, volume 2. Minneapolis, University of Minnesota Press. | 1998 | 16734 |
| B | HIRSCHFELDER, J. O., CURTISS, C. F., & BIRD, R. B. (1954). Molecular theory of gases and liquids. New York, J. Wiley & sons. | 1954 | 16732 |
| J | ZIGMOND, A.S. & SNAITH, R. (1983). The hospital anxiety and depression scale.Acta psychiatrica Scandinavica, 67(6), 361-370. DOI: 10.1111/j.1600-0447.1983.tb09716.x | 1983 | 16667 |
| B | RUMELHART, D. E., HINTON, G. E., & WILLIAMS, R. J. (1985). Learning internal representations by error propagation. La Jolla, Calif, Institute for Cognitive Science, University of California, San Diego. | 1985 | 16629 |
| B | WEBER, M. (1922). Wirtschaft und Gesellschaft: Grundriß der verstehenden Soziologie. Tübingen, Mohr. | 1922 | 16607 |
| B | TALAIRACH, J., & TOURNOUX, (1988). Co-planar stereotaxic atlas of the human brain: 3-dimensional proportional system : an approach to cerebral imaging. Stuttgart, G. Thieme. | 1988 | 16584 |
| B | CAMPBELL, D. T., & STANLEY, J. C. (1963). Experimental and quasi-experimental design for research. Chicago, Rand McNally. | 1963 | 16557 |
| J | BECK, A. T., WARD, C., & MENDELSON, M. (1961). Beck depression inventory (BDI). Arch Gen Psychiatry, 4(6), 561-571. | 1961 | 16508 |
| J | LANDER, E.S. et al. (2001). Initial sequencing and analysis of the human genome. Nature, 409(6822), 860-921. DOI: 10.1038/35057062 | 2001 | 16503 |
| B | MICHALEWICZ, Z. (1992). Genetic algorithms + data structures = evolution programs. Berlin, Springer. | 1992 | 16501 |
| B | PERRY, R. H., & GREEN, D. W. (1934). Perry's chemical engineers' handbook. New York, McGraw-Hill. | 1934 | 16496 |
| B | LAKOFF, G. (1986). Women, fire, and dangerous things: what categories reveal about the mind. Chicago, University of Chicago Press. | 1986 | 16469 |
| B | JACKSON, M. L. (1956). Soil Chemical analysis: advances course; a manual of methods useful for instruction and research in soil chemistry, physical chemistry of soils, soil fertility, and soil genesis. Madison, Wis. | 1956 | 16452 |
| B | PETERS, T., & WATERMAN, R. H. (1982). In search of excellence: lessons from America's best-run companies. Cambridge, Mass. and London, Harper & Row. | 1982 | 16439 |
| J | SMITH, et al. (1985). Measurement of protein using | 1985 | 16405 |

| Document type | Bibliographic reference | 1st ed. Pub. Year | GS Citations |
|---|---|---|---|
| | Association 1987 revised criteria for the classification of rheumatoid arthritis. Arthritis and rheumatism, 31(3), 315-324. DOI: 10.1002/art.1780310302 | | |
| J | KÖHLER, G. & MILSTEIN, C. (2005). Continuous cultures of fused cells secreting antibody of predefined specificity. 1975. Journal of immunology (Baltimore, Md. : 1950), 174(5), 2453-2455. DOI: 10.1038/256495a0 | 2005 | 16117 |
| B | NETER, J., & WASSERMAN, W. (1974). Applied linear statistical models: regression, analysis of variance, and experimental designs. Homewood, Ill, Richard D. Irwin. | 1974 | 16099 |
| B | MERLEAU-PONTY, M., & SMITH, C. (1962). Phenomenology of perception. New York, Humanities Press. | 1962 | 16091 |
| J | MONCADA, S., PALMER, R.M. & HIGGS, E.A. (1991). Nitric oxide: physiology, pathophysiology, and pharmacology. Pharmacological reviews, 43(2), 109-142. | 1991 | 16088 |
| B | DUNFORD, N. et al. (1957). Linear operators. Part I, Part I. New York, Interscience. | 1957 | 16083 |
| B | GILBARG, D., & TRUDINGER, N. S. (1977). Elliptic partial differential equations of second order. Berlin, Springer-Verlag. | 1977 | 16072 |
| J | ANDERSON, J. C., & GERBING, D. W. (1988). Structural equation modeling in practice: A review and recommended two-step approach. Psychological Bulletin. 103, 411-423. DOI: 10.1037/0033-2909.103.3.411 | 1988 | 16069 |
| J | ENGLE, R. (1982). Autoregressive conditional heteroscedasticity with estimates of the variance of United Kingdom inflation. Econometrica. 50, 987-1007. DOI: 10.2307/1912773 | 1982 | 16029 |
| B | POLANYI, M. (1958). Personal knowledge: Towards a post-critical philosophy. Chicago, Chicago University Press. | 1958 | 15949 |
| B | JENSEN, M.C. (1986). Agency Costs of Free Cash Flow , Corporate Finance , and Takeovers Agency Costs of Free Cash Flow , Corporate Finance , and Takeovers. American Economic Review, 76(2), 323-329. DOI: | 1986 | 15902 |
| B | FRANCIS, G., & FIELD, A. (2000). Discovering statistics using SPSS. London, SAGE. | 2000 | 15900 |
| B | KVALE, S., & BRINKMANN, S. (1996). InterViews: learning the craft of qualitative research interviewing. Los Angeles, Sage Publications. | 1996 | 15891 |
| O | WECHSLER, D. (1981). WAIS-R Wechsler adult intelligence scale. New York, N.Y., Psychological Cor | 1981 | 15882 |
| J | DUNNING, T. H. (1989). Gaussian basis sets for use in correlated molecular calculations. I. The atoms boron through neon and hydrogen. The Journal of Chemical Physics. 90, 1007. DOI: 10.1063/1.456153 | 1989 | 15875 |
| J | PITTENGER M.F. et al. (1999). Multilineage potential of adult human mesenchymal stem cells. Science (New York, N.Y.). 284, 143-7. DOI: 10.1126/science.284.5411.143 | 1990 | 15836 |
| J | BADDELEY, A. (1992). Working memory. Science, 255(5044), 556-559. DOI: 10.1126/science.1736359 | 1992 | 15828 |
| B | SNEATH, H. A., & SOKAL, R. R. (1973). Numerical taxonomy (by) Peter H.A. Sneath (and) Robert R. Sokal: the principles and | 1973 | 15804 |

| Document type | Bibliographic reference | 1st ed. Pub. Year | GS Citations |
|---|---|---|---|
| | practice of numerical classification. San Francisco, W.H. Freeman. | | |
| B | CHOMSKY, N. (1981). Lectures on government and binding the Pisa lectures. Dordrecht, Foris publications. | 1981 | 15763 |
| B | DEWEY, J. (1938). Experience and education. New York, Collier Books. | 1938 | 15754 |
| B | CSIKSZENTMIHALYI, M. (1990). Flow: the psychology of optimal experience. New York, Harper and Row. | 1990 | 15734 |
| B | FIRST, M. B. et al. (1997). Structured clinical interview for DSM-IV axis I disorders SCID-I: clinician version, scoresheet. Washington, DC, American Psychiatric Press. | 1997 | 15731 |
| B | POSTONE, M., HABERMAS, J. & MCCARTHY, T. (1984). The Theory of Communicative Action: Vol. 2: Lifeworld and System: A Critique of Functionalist Reason.Contemporary Sociology, 19(2), 170. | 1984 | 15727 |
| J | GEMAN, S. & GEMAN, D. (1984). Stochastic Relaxation, Gibbs Distributions, and the Bayesian Restoration of Images. IEEE Transactions on Pattern Analysis and Machine Intelligence, PAMI-6(6). | 1984 | 15703 |
| J | MATTHEWS, D.R. et al. (1985). Homeostasis model assessment: insulin resistance and beta-cell function from fasting plasma glucose and insulin concentrations in man. Diabetologia, 28(7), 412-419. DOI: 10.1007/BF00280883 | 1985 | 15690 |
| J | SOLOW, R.M. (1956). A Contribution to the Theory of Economic Growth. The Quarterly Journal of Economics, 70(1), 65-94. | 1956 | 15685 |
| B | MCLUHAN, M. (1964). Understanding Media: the extensions of man. New York, McGraw-Hill. | 1964 | 15609 |
| B | WITTGENSTEIN, L. (1922). Tractatus logico-philosophicus. London, Routledge & Kegan Paul. | 1922 | 15602 |
| B | AGRESTI, A. (1990). Categorical data analysis. New York, John Wiley & Sons. | 1990 | 15593 |
| B | FERRY, J. D. (1961). Viscoelastic properties of polymers. New York, John Wiley & sons inc. | 1961 | 15582 |
| B | CONOVER, W. J. (1971). Practical nonparametric statistics. New York , John Wiley and Sons. | 1971 | 15571 |
| B | BUTLER, J. (1993). Bodies that matter: on the discursive limits of "sex". New York, Routledge. | 1993 | 15549 |
| J | KRESSE, G., & JOUBERT, D. (1999). Electronic structure: Wide-band, narrow-band, and strongly correlated systems - From ultrasoft pseudopotentials to the projector augmented-wave method. Physical Review. B, Condensed Matter. 59, 1758. DOI: 10.1103/PhysRevB.59.1758 | 1999 | 15532 |
| B | CHOMSKY, N. (1957). Syntactic structures. London, Mouton. | 1957 | 15489 |
| B | COOK, T. D., & CAMPBELL, D. T. (1979). Quasi-experimentation: design & analysis issues for field settings. Chicago, Rand McNally College. | 1979 | 15482 |
| B | BILLINGSLEY, (1968). Convergence of probability measures. New York, John Wiley and sons. | 1968 | 15415 |
| J | BOLLERSLEV, T. (1986). Generalized autoregressive conditional heteroskedasticity.Journal of Econometrics, 31(3), 307-327. | 1986 | 15359 |

| Document type | Bibliographic reference | 1st ed. Pub. Year | GS Citations |
|---|---|---|---|
| | DOI: 10.1016/0304-4076(86)90063-1 | | |
| B | LITTLE, R. J. A., & RUBIN, D. B. (1987). Statistical Analysis with Missing Data. New York [NY], John Wiley and Sons. | 1987 | 15342 |
| J | BLÖCHL, E. (1994). Projector augmented-wave method. Physical Review B. 50, 17953-17979. DOI:10.1103/PhysRevB.50.17953 | 1994 | 15332 |
| J | WATSON, D., CLARK, L. A., & TELLEGEN, A. (1988). Development and validation of brief measures of positive and negative affect: The PANAS scales. Journal of Personality and Social Psychology. 54, 1063-1070. DOI: 10.1037/0022-3514.54.6.1063 | 1988 | 15320 |
| J | PERDEW, J. , & WANG, Y. (1992). Accurate and simple analytic representation of the electron-gas correlation energy. Physical Review B. 45, 13244-13249. DOI: 10.1103/PhysRevB.45.13244 | 1992 | 15302 |
| B | COHEN, L., & MANION, L. (1979). Research methods in education. London, Croom Helm. | 1979 | 15298 |
| B | BATESON, G. (1972). Steps to an ecology of mind. New York, Ballantine Books. | 1972 | 15288 |
| B | BIMBOIM, H.C. & DOLY, J. (1979). A rapid alkaline extraction procedure for screening recombinant plasmid DNA. Nucleic Acids Research, 7(6), 1513-1523. | 1979 | 15253 |
| B | FISHER, R. A. (1930). The genetical theory of natural selection. Oxford, The Clarendon Press. | 1930 | 15246 |
| B | TOCQUEVILLE, A. D. (1945). Democracy in America. New York, A. A. Knopf. | 1945 | 15187 |
| B | BLAU, M. (1964). Exchange and Power in Social Life. New York, John Wiley & Sons,Inc. | 1964 | 15152 |
| B | PARASURAMAN, A., ZEITHAML, V., & BERRY, L. L. (1986). Servqual: a multiple-item scale for measuring customer perceptions of service quality. Cambridge, Mass, MSI. | 1986 | 15137 |
| B | ADAMSON, A. W., & GAST, A. (1960). Physical chemistry of surfaces. New York , John Wiley & Sons. | 1960 | 15118 |
| B | CHANDLER, A. D. (1962). Strategy and structure: Chapters in the history of the industrial enterprise. Cambridge, Mass, The M. I. T. Press. | 1962 | 15039 |
| J | OTSU, N. (1979). A Threshold Selection Method from Gray-Level Histograms IEEE Transactions on Systems, Man and Cybernetics, 9(1), (January 1979), p 62-66. DOI:10.1109/tsmc.1979. | 1979 | 14992 |
| B | WEBER, M., ROTH, G., & FISCHOFF, E. (1968). Economy and society: an outline of interpretive sociology. New York, Bedminster Press. | 1968 | 14986 |
| J | MILLER, S.A., DYKES, D.D. & POLESKY, H.F. (1988). A simple salting out procedure for extracting DNA from human nucleated cells. Nucleic Acids Research, 16(3), 1215. DOI: 10.1093/nar/16.3.1215 | 1988 | 14917 |
| B | SILVERMAN, B. W. (1986). Density estimation for statistics and data analysis. London, Chapman and Hall. | 1986 | 14905 |
| J | JORGENSEN, W. L. et al. (1983). Comparison of simple potential functions for simulating liquid water. The Journal of | 1983 | 14881 |

| Document type | Bibliographic reference | 1st ed. Pub. Year | GS Citations |
|---|---|---|---|
| | Chemical Physics. 79, 926. DOI: 10.1063/1.445869 | | |
| J | KRESSE, G. & FURTHMÜLLER, J. (1996). Efficiency of ab-initio total energy calculations for metals and semiconductors using a plane-wave basis set.Computational Materials Science, 6(1), 15-50. DOI: 10.1016/0927-0256(96)00008-0 | 1996 | 14861 |
| J | HODGKIN, A.L. & HUXLEY, A.F. (1952). A quantitative description of membrane current and its applications to conduction and excitation in nerve. Journal of Physiology, 117(1-2), 500-544. | 1952 | 14857 |
| J | BURTON, K. (1956). A study of the conditions and mechanism of the diphenylamine reaction for the colorimetric estimation of deoxyribonucleic acid. Biochemical journal, 62(2), 315. | 1956 | 14854 |
| J | BELLMAN, R. (1956). Dynamic programming and lagrange multipliers.Proceedings of the National Academy of Sciences. 42, 767-769. DOI: 10.1073/pnas.42.10.767 | 1956 | 14840 |
| B | HOPCROFT, J. E., & ULLMAN, J. D. (1979). Introduction to automata theory, languages, and computation. Reading, Mass, Addison-wesley. | 1979 | 14830 |
| B | COURANT, R., & HILBERT, D. (1937). Methods of mathematical physics. New York, Interscience Publ. | 1937 | 14794 |
| B | MALLAT, S. G. (1998). A wavelet tour of signal processing. San Diego , Academic Press. | 1998 | 14786 |
| B | WEBER, M., HENDERSON, A. M., & PARSONS, T. (1947). The theory of social and economic organization. New York, Oxford University Press. | 1947 | 14768 |
| B | ZHENG, X. (2002). Hua xue yao pin he zhi liao yong sheng wu zhi pin yan jiu zhi dao yuan ze: shi xing. Beijing Shi, Zhongguo yi yao ke ji chu ban she. | 2002 | 14736 |
| B | FREEMAN, R. E. (1984). Strategic management: A stakeholder approach. Boston, Mass, Pitman. | 1983 | 14736 |
| B | HIRSCHMAN, A. O. (1970). Exit, voice, and loyalty; responses to decline in firms, organizations, and states. Cambridge, Mass, Harvard University Press. | 1970 | 14696 |
| B | BURT, R. S. (1992). Structural holes the social structure of competition. Cambridge, Mass, Harvard University. | 1992 | 14686 |
| B | DELEUZE, G., & GUATTARI, F. (1987). A thousand plateaus: capitalism and schizophrenia. Minneapolis, MN, University of Minnesota Press. | 1987 | 14661 |
| B | MERRIAM, S. B. (1998). Qualitative research and case study applications in education: revised and expanded from case study research in education. San Francisco, Jossey-Bass. | 1998 | 14660 |
| B | ANDERSON, T. W. (1958). An introduction to multivariate statistical analysis. New York, John Wiley & Sons. | 1958 | 14653 |
| B | MARSHALL, A. (1890). Principles of economics: an introductory volume. London, Macmillan and Co. | 1890 | 14611 |
| B | BEAUCHAMP, T. L. (1979). Principles of biomedical ethics. New York, Oxford University Press. | 1979 | 14593 |
| J | KRAULIS, J. (1991). MOLSCRIPT: a program to produce both detailed and schematic plots of protein structures. Journal of Applied Crystallography. 24, 946-950. | 1991 | 14576 |
| B | BECK, A.T. & WEISHAAR, M. (1989). Cognitive therapy. In: | 1989 | 14569 |

| Document type | Bibliographic reference | 1st ed. Pub. Year | GS Citations |
|---|---|---|---|
| | Comprehensive handbook of cognitive therapy. New York, Plenum Press. | | |
| B | GIL, A. C. (1987). Como elaborar projetos de pesquisa. Sao Paulo, Atlas. | 1987 | 14536 |
| B | JOHNSON, R. A., & WICHERN, D. (1982). Applied multivariate statistical analysis. Englewood Cliffs, Prentice-Hall. | 1982 | 14525 |
| J | DICKEY, D. A., & FULLER, W. A. (1979). Distribution of the Estimators for Autoregressive Time Series with a Unit Root. Journal of the American Statistical Association. 74, 427-431. DOI: 10.2307/2286348 | 1979 | 14492 |
| B | DAVENPORT, T. H., & PRUSAK, L. (1997). Working knowledge: How organizations manage what they know. Boston, Harvard Business Business Press. | 1997 | 14469 |
| B | HARTLEY, R., & ZISSERMAN, A. (2000). Multiple view geometry in computer vision. Cambridge, UK, Cambridge University Press. | 2000 | 14426 |
| J | YANISCH-PERRON, C., VIEIRA, J., & MESSING, J. (1985). Improved M13 phage cloning vectors and host strains: nucleotide sequences of the M13mpl8 and pUC19 vectors. Gene. 33, 103-119. DOI: 10.1016/0378-1119(85)90120-9 | 1985 | 14390 |
| J | KIMURA, M. (1980). A simple method for estimating evolutionary rates of base substitutions through comparative studies of nucleotide sequences. Journal of molecular evolution, 16(2), 111-120. DOI: 10.1007/BF01731581 | 1980 | 14375 |
| B | GOFFMAN, E. (1974). Frame analysis: an essay on the organization of experience. Cambridge,Ma, Harvard University Press. | 1974 | 14372 |
| J | AGRAWAL, R., IMIELIŃSKI, T., & SWAMI, A. (1993). Mining association rules between sets of items in large databases. ACM SIGMOD Record. 22, 207-216. | 1993 | 14333 |
| B | ROTHMAN, K. J. (1986). Modern epidemiology. Boston, Little, Brown. | 1986 | 14328 |
| B | JOHNSON, N. L., KOTZ, S., & BALAKRISHNAN, N. (1970). Continuous univariate distributions Volumes 1-2. Volumes 1-2. New York, Wiley. | 1970 | 14311 |
| J | JOHANSEN, S. (1988). Statistical analysis of cointegration vectors. Journal of Economic Dynamics and Control, 12(2-3), 231-254. DOI: 10.1016/0165-1889(88)90041-3 | 1988 | 14303 |
| J | (1998). Intensive blood-glucose control with sulphonylureas or insulin compared with conventional treatment and risk of complications in patients with type 2 diabetes (UKPDS 33). The Lancet. 352, 837-853. | 1998 | 14295 |
| B | FOUCAULT, M. (1970). The order of things; an archaeology of the human sciences. New York, Vintage Books. | 1970 | 14286 |
| B | BELLAMY, L. J. (1954). The infra-red spectra of complex molecules. London, Methuen. | 1954 | 14283 |
| B | NEWELL, A., & SIMON, H. A. (1972). Human problem solving. Englewood Cliffs, N.J., Prentice-Hall. | 1972 | 14248 |
| B | MADDALA, G. S. (1983). Limited-dependent and qualitative variables in econometrics. Cambridge, Cambridge | 1983 | 14238 |

**70**

| Document type | Bibliographic reference | 1st ed. Pub. Year | GS Citations |
|---|---|---|---|
| | University Press. | | |
| B | INCROPERA, F. , LAVINE, A.S. & DEWITT, D. (1981). Fundamentals of heat transfer. New York, Wiley. | 1981 | 14236 |
| B | OPPENHEIM, A.V., BUCK, J.R.& SCHAFER, R.W. (1989). Discrete-time signal processing. Englewood Cliffs, N.J., Prentice-Hall. | 1989 | 14231 |
| C | RIVEST, R.L., SHAMIR, A. & ADLEMAN, L. (1978). A method for obtaining digital signatures and public-key cryptosystems. Communications of the ACM, 21(2), 120-126. DOI: 10.1145/359340.359342 | 1978 | 14225 |
| J | ELLMAN, G.L. et al. (1961). A new and rapid colorimetric determination of acetylcholinesterase activity. Biochemical Pharmacology, 7(2), 88-95. DOI: 10.1016/0006-2952(61)90145-9 | 1961 | 14219 |
| B | DARWIN, C. (1871). The descent of man, and selection in relation to sex. London, John Murray, Albemarle Street. | 1871 | 14216 |
| B | PARR, R. G., & YANG, W. (1994). Density-functional theory of atoms and molecules. Oxford , Oxford University Press. | 1994 | 14205 |
| O? | ATTORNEY, A. (1999). or Firm—. | 1999 | 14187 |
| J | WILLIAMS, J.G. et al. (1990). DNA polymorphisms amplified by arbitrary primers are useful as genetic markers. Nucleic acids research, 18(22), 6531-6535. DOI: 10.1093/nar/18.22.6531 | 1990 | 14160 |
| B | KATZ, D., & KAHN, R. L. (1966). The social psychology of organizations. New York, Wiley. | 1966 | 14154 |
| B | HAMILTON, J. D. (1994). Time series analysis. Princeton, N.J., Princeton University Press. | 1994 | 14154 |
| B | MACINTYRE, A. C. (1981). After virtue: a study in moral theory. London, Gerald Duckworth. | 1981 | 14152 |
| B | BATHE, K.J. (1996). Finite element procedures. Englewood Cliffs, N.J., Prentice-Hall. | 1996 | 14150 |
| J | JONES, T. A., ZOU, J.-Y., COWAN, S. W., & KJELDGAARD, M. (1991). Improved methods for building protein models in electron density maps and the location of errors in these models. Acta Crystallographica Section A. 47, 110-119. DOI:10.1107/S0108767390010224 | 1991 | 14126 |
| J | OHKAWA, H., OHISHI, N., & YAGI, K. (1979). Assay for lipid peroxides in animal tissues by thiobarbituric acid reaction. Analytical Biochemistry. 95, 351-358. DOI:10.1016/0003-2697(79)90738-3 | 1979 | 14122 |
| J | PARASURAMAN, A., ZEITHAML, V.A. & BERRY, L.L. (1985). A Conceptual Model of Service Quality and Its Implications for Future Research. Journal of Marketing, 49(4), 41-50. DOI: 10.2307/1251430 | 1985 | 14094 |
| J | EISEN , M.B. et al. (1998) Cluster analysis and display of genome-wide expression patterns. Proceedings of the National Academy of Sciences of the United States of America. 95, 14863-8. DOI:10.1073/pnas.95.25.14863 | 1998 | 14091 |
| J | NONAKA, I. (1994). A Dynamic Theory of Organizational Knowledge Creation.Organization Science, 5, 14-37. DOI: 10.1287/orsc.5.1.14 | 1994 | 14073 |
| B | FOUCAULT, M., VARELA, J., & ÁLVAREZ-URÍA, F. (1978). Microfísica del poder. Madrid, Ediciones de la | 1978 | 14067 |

| Document type | Bibliographic reference | 1st ed. Pub. Year | GS Citations |
|---|---|---|---|
| | Piqueta. | | |
| J | HANSEN, M., ANDERKO, K., & SALZBERG, H. W. (1958). Constitution of Binary Alloys. Journal of The Electrochemical Society. 105. | 1958 | 14066 |
| J | HOPFIELD, J. J. (1982). Neural Networks and Physical Systems with Emergent Collective Computational Abilities. Proceedings of the National Academy of Sciences.79, 2554-2558. DOI: 10.1073/pnas.79.8.2554 | 1982 | 14065 |
| J | KRESGE, C. T. et al. (1992). Ordered mesoporous molecular sieves synthesized by a liquid-crystal template mechanism. Nature. 359, 710-712. | 1992 | 14047 |
| B | KRIPPENDORFF, K. (1980). Content analysis: an introduction to its methodology. Beverly Hills, Calif, Sage. | 1980 | 14027 |
| B | STAKE, R. E. (1995). The art of case study research. Thousand Oaks, Sage Publications. | 1995 | 13980 |
| B | ADAMS, R. A. (1975). Sobolev spaces. New York, N.Y. , Academic press. | 1975 | 13975 |
| J | ARELLANO, M., & BOND, S. R. (1991). Some tests of specification for panel data: Monte Carlo evidence and an application to employment equations. The Review of Economic Studies (New York). 58, 277-297. DOI:10.2307/2297968 | 1991 | 13966 |
| J | DEVEREUX, J., HAEBERLI, , & SMITHIES, O. (1984). A comprehensive set of sequence analysis programs for the VAX. Nucleic Acids Research. 12, 387-395. DOI:10.1093/nar/12.1Part1.387 | 1984 | 13939 |
| B | GUCKENHEIMER, J., & HOLMES, (1983). Nonlinear oscillations, dynamical systems, and bifurcations of vector fields. New York, Springer-Verlag. | 1983 | 13916 |
| B | MUTHÉN, L. K., & MUTHÉN, B. O. (2001). Mplus: statistical analysis with latent variables : User's Guide. Los Angeles, Muthén & Muthén. | 2001 | 13913 |
| J | VOSKO, S. H., WILK, L., & NUSAIR, M. (1980). Accurate spin-dependent electron liquid correlation energies for local spin density calculations: a critical analysis. Canadian Journal of Physics. 58, 1200-1211. DOI: | 1980 | 13899 |
| B | COHEN, J. (1969). Statistical power analysis for the behavioral sciences. New York, Academic Press. | 1969 | 13894 |
| B | DEMING, W. E. (1989). Calidad, productividad y competitividad: la salida de la crisis. Madrid, Ediciones Díaz de Santos. | 1989 | 13890 |
| B | WEICK, K. E. (1995). Sensemaking in organizations. Thousand Oaks, Calif, Sage. | 1995 | 13890 |
| J | HUBER, K. & HERZBERG, G. (1979). Constants of Diatomic Molecules. InMolecular Spectra and Molecular Structure. | 1979 | 13886 |
| B | GUJARATI, D. N. (1978). Basic econometrics. New York, McGraw-Hill. | 1978 | 13885 |
| J | FARRUGIA, L. J. (1999). WinGX suite for small-molecule single-crystal crystallography. Journal of Applied Crystallography. 32, 837-838. | 1999 | 13877 |
| J | COHEN, J. (1992). A power primer. Psychological bulletin, 112(1), 155-159. DOI: 10.1037/0033-2909.112.1.155 | 1992 | 13875 |
| B | DE GENNES, G. (1974). The physics of liquid crystals. Oxford, | 1974 | 13870 |

| Document type | Bibliographic reference | 1st ed. Pub. Year | GS Citations |
|---|---|---|---|
| | Clarendon Press. | | |
| B | HILLE, B. (1984). Ionic channels of excitable membranes. Sunderland, Mass, Sinauer associates. | 1984 | 13847 |
| B | WEICK, K. E. (1969). The social psychology of organizing. Reading, Mass, Addison-Wesley Pub. | 1969 | 13842 |
| B | BLUMER, H. (1969). Symbolic interactionism; perspective and method. Englewood Cliffs, N.J., Prentice-Hall. | 1969 | 13839 |
| J | RONQUIST, F. (2003). MrBayes 3: Bayesian phylogenetic inference under mixed models. Bioinformatics. 19, 1572-1574. DOI: 10.1093/bioinformatics/btg180 | 2003 | 13812 |
| B | MARSHALL, C., & ROSSMAN, G. B. (1989). Designing qualitative research. Newbury Park, Calif, Sage. | 1989 | 13786 |
| J | CORTES, C. & VAPNIK, V. (1995). Support-vector networks. Machine Learning, 20(3), 273-297. DOI: 10.1023/A:1022627411411 | 1995 | 13769 |
| J | BARTEL D (2004). MicroRNAs: genomics, biogenesis, mechanism, and function. Cell.116, 281-97. | 2004 | 13766 |
| B | ARROW, K. J. (1951). Social choice and individual values. New York, John Wiley & sons. | 1951 | 13764 |
| B | BOYD, S. Et al. (1994). Linear matrix inequalities in system and control theory. Philadelphia, Society for Industrial and Applied Mathematics. | 1994 | 13763 |
| J | MILLER, G. L. (1959). Use of dinitrosalicylic acid reagent for determination of reducing sugar. Analitical Chemistry. 31, 426-428. DOI: 10.1021/ac60147a030 | 1959 | 13743 |
| B | R Development Core Team, R. (2011). R: A Language and Environment for Statistical Computing R. D. C. Team, ed. R Foundation for Statistical Computing, 1(2.11.1), 409. | 2011 | 13724 |
| J | ALBERT, R., & BARABASI, A.L. (2002). Statistical mechanics of complex networks.Reviews of Modern Physics. 74, 47-98. DOI: 10.1103/RevModPhys.74.47 | 2002 | 13707 |
| J | MAXAM, A.M. , & GILBERT, W. (1980). Sequencing end-labeled DNA with base-specific chemical cleavages. Methods in Enzymology. 65, 499-560. | 1980 | 13690 |
| J | KERR JF, WYLLIE AH, & CURRIE AR. (1972). Apoptosis: a basic biological phenomenon with wide-ranging implications in tissue kinetics. British Journal of Cancer.26, 239-57. DOI: 10.1038/bjc.1972.33 | 1972 | 13681 |
| B | ALTMAN, D. G. (1991). Practical statistics for medical research. London, Chapman and Hall. | 1991 | 13672 |
| B | LAW, A. M., & KELTON, W. D. (1982). Simulation Modeling and Analysis. New York, McGraw-Hill. | 1982 | 13656 |
| B | DOUGLAS, M. (1966). Purity and danger: an analysis of concepts of pollution and taboo. London, Routledge & Kegan Paul. | 1966 | 13575 |
| B | WATSON, G. N. (1922). A treatise on the theory of Bessel functions. Cambridge , The University press. | 1922 | 13570 |
| B | BERLE, A. A., & MEANS, G. C. (1932). Modern corporation and private property. New York, Commerce Clearing House, Loose leaf Service division of the Corporation Trust Company. | 1932 | 13564 |
| B | MORGAN, G. (1986). Images of organization. Beverly Hills, Ca, | 1986 | 13547 |

| Document type | Bibliographic reference | 1st ed. Pub. Year | GS Citations |
|---|---|---|---|
| | Sage Pulications. | | |
| B | WHITE, T.J. et al., (1990). Amplification and direct sequencing of fungal ribosomal RNA genes for phylogenetics. In PCR Protocols: A Guide to Methods and Applications. Academic Press, p 315-322. | 1990 | 13540 |
| J | LORENZ, E. N. (1963). Deterministic Nonperiodic Flow. Journal of the Atmospheric Sciences. 20, 130-141. DOI: 10.1175/1520-0469(1963)020<0130:DNF>2.0.CO;2 | 1963 | 13528 |
| B | KAPLAN, R.S., & NORTON, D. (2007). Balanced scorecard. In : Das Summa Summarum des Management die 25 wichtigsten Werke für Strategie, Führung und Veränderung. Wiesbaden, Gabler. | 2007 | 13522 |
| B | KERLINGER, F. N. (1964). Foundations of behavioral research: educational and psychological inquiry. New York, N.Y., Holt, Rinehart & Winston. | 1964 | 13489 |
| B | MACARTHUR, R. H., & WILSON, E. O. (1967). The theory of island biogeography. Princeton, N.J, Princeton University Press. | 1967 | 13489 |
| B | HOBSBAWM, E. J. E., & RANGER, T. O. (1983). The Invention of tradition. Cambridge, Cambridge University Press. | 1983 | 13485 |
| B | ARENDT, H. (1958). The Human Condition. Chicago, The University of Chicago Press. | 1958 | 13482 |
| J | JEMAL, A. et al. (2009). Cancer statistics, 2009. CA: a Cancer Journal for Clinicians, 59(4), 225. DOI:10.3322/caac.20006 | 2009 | 13454 |
| J | BERENDSEN, H. J. C. et al. (1984). Molecular dynamics with coupling to an external bath. The Journal of Chemical Physics. 81, 3684. DOI:10.1063/1.448118 | 1984 | 13447 |
| J | CHARLSON, M.E. et al. (1987). A new method of classifying prognostic comorbidity in longitudinal studies: development and validation. Journal of chronic diseases, 40(5), 373-383. DOI: 10.1016/0021-9681(87)90171-8 | 1987 | 13440 |
| J | BELL, D. (1976 ). The coming of the post-industrial society. In The Educational Forum (Vol. 40, No. 4, p 574-579). Taylor & Francis Grou | 1976 | 13430 |
| J | MORGAN, R.M. & HUNT, S.D. (1994). The Commitment-Trust Theory of Relationship Marketing. Journal of Marketing, 58, 20-38. DOI: 10.2307/1252308 | 1994 | 13408 |
| B | MENEZES, A. J., VAN OORSCHOT, C., & VANSTONE, S. A. (1996). Handbook of applied cryptography. Boca Raton, Fla, CRC. | 1996 | 13393 |
| B | MACINTYRE, R. J. (1985). Molecular evolutionary genetics. New York, Plenum. | 1985 | 13375 |
| B | NIE, N. H., BENT, D. H., & HULL, C. H. (1970). SPSS: Statistical Package for the Social Sciences. New York [etc.], McGraw-Hill Book Company. | 1970 | 13369 |
| B | KRUEGER, R. A. (1988). Focus groups a practical guide for applied research. Newbury Park, Calif, Sage Publications. | 1988 | 13369 |
| J | VANDERBILT, D. (1990). Soft self-consistent pseudopotentials in a generalized eigenvalue formalism. Physical Review B. 41(11), 7892-7895. DOI: 10.1103/PhysRevB.41.7892 | 1990 | 13361 |
| B | RECHTSCHAFFEN, A., & KALES, A. (1968). A manual of standardized terminology, techniques and scoring system | 1968 | 13338 |

| Document type | Bibliographic reference | 1st ed. Pub. Year | GS Citations |
|---|---|---|---|
| | for sleep stages of human subjects. Bethesda, Md, United States Department of Health and Human Services. Public Health Service. National Institutes of Healt | | |
| J | GEIM, A.K. & NOVOSELOV, K.S.,(2007). The rise of graphene. Nature materials, 6(3), 183-191. DOI: 10.1038/nmat1849 | 2007 | 13278 |
| B | DERRIDA, J. (1976). Of grammatology. Baltimore, Johns Hopkins University Press. | 1976 | 13262 |
| J | PERDEW, J. & ZUNGER, A. (1981). Self-interaction correction to density-functional approximations for many-electron systems. Physical Review B, 23(10), 5048-5079. DOI: 10.1103/PhysRevB.23.5048 | 1981 | 13261 |
| B | BARNARD, C. J. (1938). The functions of the executive. Cambridge, Mass, Harvard University Press. | 1938 | 13257 |
| J | QUINLAN, J. R. (1986). Induction of decision trees. Machine learning, 1(1), 81-106. | 1986 | 13252 |
| B | HULST, H.C.V.D. (1957). Light scattering by small particles. New York, J. Wiley. | 1957 | 13248 |
| J | SHARPE, W. F. (1964). Capital asset prices: a theory of market equilibrium under conditions of risk*. The Journal of Finance. 19, 425-442. DOI: 10.2307/2977928 | 1964 | 13242 |
| J | HSU SM, RAINE L, & FANGER H. (1981). Use of avidin-biotin-peroxidase complex (ABC) in immunoperoxidase techniques: a comparison between ABC and unlabeled antibody (PAP) procedures. The Journal of Histochemistry and Cytochemistry : Official Journal of the Histochemistry Society. 29, 577-80. | 1981 | 13214 |
| B | GELMAN, A. et al. (1995). Bayesian data analysis. London [etc.], Chapman & Hall. | 1995 | 13208 |
| J | DEWAR, M. J. S. et al.(1985). Development and use of quantum mechanical molecular models. 76. AM1: a new general purpose quantum mechanical molecular model. Journal of the American Chemical Society, 107, 3902-3909. DOI: 10.1021/ja00299a024 | 1985 | 13178 |
| J | BROWN, J. S., COLLINS, A., & DUGUID, (1989). Situated Cognition and the Culture of Learning. Educational Researcher. 18, 32-42. | 1989 | 13178 |
| B | TRIBE, L. H. (1978). American constitutional law. Mineola, NY, The Foundation Press. | 1978 | 13169 |
| J | TAMURA, K. et al. (2011). MEGA5: Molecular Evolutionary Genetics Analysis Using Maximum Likelihood, Evolutionary Distance, and Maximum Parsimony Methods. Molecular Biology and Evolution. 28, 2731-2739. DOI: 10.1093/molbev/msr121 | 2011 | 13145 |
| J | DIJKSTRA, E.W. (1959). A note on two problems in connexion with graphs.Numerische Mathematik, 1(1), 269-271. | 1959 | 13125 |
| J | MODIGLIANI, F., MILLER, M.H. & MODIGLIANI, F. (1958). The cost of capital, corporation finance and the theory of investmient. The American economic, 48(3), 261-297. | 1958 | 13123 |
| J | HUELSENBECK, J. , & RONQUIST, F. (2001). MRBAYES: Bayesian inference of phylogenetic trees. Bioinformatics. 17, 754-755. DOI: 10.1093/bioinformatics/17.8.754 | 2001 | 13120 |

| Document type | Bibliographic reference | 1st ed. Pub. Year | GS Citations |
|---|---|---|---|
| J | BINNIG, G. & QUATE, C.F. (1986). Atomic Force Microscope. Physical Review Letters, 56(9), 930-933. DOI: 10.1103/PhysRevLett.56.930 | 1986 | 13115 |
| J | BURGES, C.C.J.C. (1998). A Tutorial on Support Vector Machines for Pattern Recognition U. Fayyad, ed. Data Mining and Knowledge Discovery, 2(2), 121-167. | 1998 | 13088 |
| B | REED, M., & SIMON, B. (1975). Fourier analysis, self-adjointness. New York, Academic Press. | 1975 | 13086 |
| J | TAKAGI, T., & SUGENO, M. (1985). Fuzzy identification of systems and its applications to modeling and control. IEEE Transactions on Systems, Man, and Cybernetics. SMC-15, 116-132. | 1995 | 13081 |
| B | BECKER, G. S. (1981). A treatise on the family. Cambridge, Mass, Harvard University Press. | 1981 | 13073 |
| B | FUKUYAMA, F. (1995). Trust: the social virtues and the creation of prosperity. New York, The Free Press. | 1995 | 13069 |
| B | NOZICK, R. (1974). Anarchy, state and utopia. New York, Basic Books. | 1974 | 13060 |
| J | BARTLETT, G.R. (1959). Phosphorus assay in column chromatography. The Journal of Biological Chemistry. 234, 466-8. | 1959 | 13053 |
| B | RAO, C. R. (1965). Linear statistical inference and its applications. New York, J. Wiley & Sons. | 1965 | 13052 |
| J | YABLONOVITCH, E. (1987). Inhibited Spontaneous Emission in Solid-State Physics and Electronics. Physical Review Letters. 58(20), 2059-2062. DOI: 10.1103/PhysRevLett.58.2059 | 1987 | 13051 |
| B | SPERBER, D. et al. (2001). Relevance: communication and cognition = Guan lian xing : jiao ji yu ren zhi. Beijing, Bu lai ke wei er chu ban she. | 2001 | 13048 |
| B | WILSON, W. J. (1987). The truly disadvantaged: the inner city, the underclass, and public policy. Chicago, University of Chicago Press. | 1987 | 13040 |
| B | ARMAREGO, W. L. F., CHAI, C. & PERRIN, D. R. (1966). Purification of laboratory chemicals. Oxford, Butterworth-Heinemann. | 1966 | 13036 |
| J | BEDNORZ, J. G., & MULLER, K. A. (1986). Possible highT c superconductivity in the Ba?La?Cu?O system. Zeitschrift Fur Physik B Condensed Matter. 64(2), 189-193. | 1986 | 13025 |
| J | PERDEW, J. et al. (1992). Atoms, molecules, solids, and surfaces: Applications of the generalized gradient approximation for exchange and correlation. Physical Review B. 46 (11), 6671-6687. | 1992 | 13021 |
| J | DERSIMONIAN, R. & LAIRD, N. (1986). Meta-analysis in clinical trials. Controlled clinical trials, 7(3), 177-188. DOI: 10.1016/0197-2456(86)90046-2 | 1986 | 13016 |
| J | BROWNE, M. W., & CUDECK, R. (1993). Alternative Ways of Assessing Model Fit. SAGE FOCUS EDITIONS. 154, 136. | 1993 | 13007 |
| J | BOYS, S., & BERNARDI, F. (2002). The calculation of small molecular interactions by the differences of separate total energies. Some procedures with reduced errors. Molecular Physics. 100, 65-73. DOI: 10.1080/00268977000101561 | 2002 | 13005 |

| Document type | Bibliographic reference | 1st ed. Pub. Year | GS Citations |
|---|---|---|---|
| B | ZIEGLER, J. F., BIERSACK, J. , & LITTMARK, U. (1985). The stopping and range of ions in solids. New York, Pergamon Press. | 1985 | 12978 |
| J | PORTA, R.L. et al. (1998). Law and Finance. Journal of Political Economy, 106(6), 1113. | 1998 | 12974 |
| B | BALANIS, C. A. (1982). Antenna theory: analysis and design. New York, Harper & Row publisher. | 1982 | 12966 |
| B | BEAR, J. (1972). Dynamics of fluids in porous media. New York, Dover Publications. | 1972 | 12962 |
| J | BÖYUM A. (1968). Isolation of mononuclear cells and granulocytes from human blood. Isolation of monuclear cells by one centrifugation, and of granulocytes by combining centrifugation and sedimentation at 1 g. Scandinavian Journal of Clinical and Laboratory Investigation. Supplementum. 97, 77-89. | 1968 | 12946 |
| J | HAMMER, M., & CHAMPY, J. (1993). Reengineering the corporation: A manifesto for business revolution. Business Horizons. 36(5), 90-91. | 1993 | 12943 |
| J | REAVEN, G. M. (1988). Role of Insulin Resistance in Human Disease. Diabetes. 37(12), 1595-1607. | 1988 | 12922 |
| B | BEZDEK, J. C. (1981). Pattern recognition with fuzzy objective function algorithms. New York, N.Y., Plenum Press. | 1981 | 12917 |
| B | GUYTON, A. C., & HALL, J. E. (1998). Pocket companion to Textbook of medical physiology. Philadelphia, Pa, Saunders. | 1998 | 12898 |
| B | BECK, U. (1986). Risikogesellschaft: auf dem Weg in eine andere Moderne. Frankfurt am Main, Suhrkam | 1986 | 12895 |
| B | CRISTIANINI, N., & SHAWE-TAYLOR, J. (2000). An introduction to support vector machines: and other kernel-based learning methods. New York, Cambridge University Press. | 2000 | 12877 |
| B | DILLMAN, D. A. (2000). Mail and internet surveys the tailored design method. New York, Wiley. | 2000 | 12872 |
| B | PATTERSON, D. A., GOLDBERG, D., & HENNESSY, J. L. (1990). Computer architecture: a quantitative approach. San Mateo, Calif, Morgan Kaufman. | 1990 | 12865 |
| B | VROOM, V. H. (1964). Work and motivation. New York, Wiley. | 1964 | 12840 |
| C | AKAIKE, H. (1973). Information theory and an extension of the maximum likelihood principle. In B. N. PETRAN & F. CSAKI, eds. International Symposium on Information Theory. Akademiai Kiadi, p 267-281 | 1973 | 12828 |
| B | NOCEDAL, J., & WRIGHT, S. J. (1999). Conjugate gradient methods. In: Numerical optimization. New York, Springer. | 1999 | 12815 |
| J | ALCHIAN, A.A. & DEMSETZ, H. (1972). Production, information and economic organization. American Economic Review, 62(5), 777-795. | 1972 | 12813 |
| C | MACQUEEN, J. (1967). Some methods for classification and analysis of multivariate observations. Proceedings of the Fifth Berkeley Symposium on Mathematical Statistics and Probability. 1. Statistics : Held at the Statistical Laboratory University of California June 21-July 18, 1965 and December 27, 1965-January 7, 1966 with the Support of University of California, National Science Foundation, National Institutes of Health... <Et Al. > / Edited by Lucien M. Le Cam and Jerzy | 1967 | 12808 |

| Document type | Bibliographic reference | 1st ed. Pub. Year | GS Citations |
|---|---|---|---|
| | Neyman. | | |
| J | ASHBURNER M, et al. (2000). Gene ontology: tool for the unification of biology. The Gene Ontology Consortium. Nature Genetics. 25, 25-9. | 2000 | 12802 |
| B | TAFLOVE, A., & HAGNESS, S. C. (1995). Computational electrodynamics: the finite-difference time-domain method. Boston, Artech House. | 1995 | 12799 |
| J | HUMPHREY, W., DALKE, A. & SCHULTEN, K. (1996). VMD: Visual molecular dynamics. Journal of Molecular Graphics, 14(1), 33-38. DOI: | 1996 | 12788 |
| J | CAMPBELL, D.T., & FISKE, D.W. (1959). Convergent and discriminant validation by the multitrait-multimethod matrix. Psychological Bulletin. 56(2), 81-105. DOI:10.1037/h0046016 | 1959 | 12780 |
| J | PFAFFL, M.W. & PFAFFL, M.W. (2001). A new mathematical model for relative quantification in real-time RT-PCR. Nucleic acids research, 29(9), e45.DOI: 10.1093/nar/29.9.e45 | 2001 | 12778 |
| B | SHAFER, G. (1976). A mathematical theory of evidence. Princeton, Princeton University Press. | 1976 | 12765 |
| J | HOFFMANN, M. R. et al. (1995). Environmental Applications of Semiconductor Photocatalysis. Chemical Reviews. 95, 69. | 1995 | 12743 |
| J | BONDI, A. (1964). van der Waals Volumes and Radii. The Journal of Physical Chemistry. 68(3), 441-451. DOI: 10.1021/j100785a001 | 1964 | 12731 |
| B | BISHOP, C. M. (2006). Pattern recognition and machine learning. New York, Springer. | 2006 | 12724 |
| J | FUJISHIMA, A., & HONDA, K. (1972). Electrochemical Photolysis of Water at a Semiconductor Electrode. Nature. 238 (5358), 37-38. DOI: 10.1038/238037a0 | 1972 | 12719 |
| J | EMSLEY, , & COWTAN, K. (2004). Coot: model-building tools for molecular graphics.Acta Crystallographica. Section D, Biological Crystallography,60 (12), 2126. DOI: 10.1107/S0907444904019158 | 2004 | 12715 |
| B | BARTLETT, F. C. (1932). Remembering: a study in experimental and social psychology. London, Cambridge University Press. | 1932 | 12710 |
| B | KVALE, S., & TORHELL, S.-E. (1997). Den kvalitativa forskningsintervjun. Lund, Studentlitteratur. | 1997 | 12708 |
| B | OSGOOD, C. E., SUCI, G. J., & TANNENBAUM, H. (1957). The measurement of meaning. Urbana, University of Illinois Press. | 1957 | 12706 |
| B | KOHONEN, T. (1984). Self-organization and associative memory. Berlin, Springer-Verlag. | 1984 | 12705 |
| B | LEFEBVRE, H. (1991). The production of space. Oxford, Blackwell. | 1991 | 12688 |
| J | OLIVER, W., & PHARR, G. (1992). An improved technique for determining hardness and elastic modulus using load and displacement sensing indentation experiments.Journal of Materials Research,7(6), 1564-1583. DOI: 10.1557/JMR.1992.1564 | 1992 | 12687 |
| B | FUKUNAGA, K. (1972). Introduction to statistical pattern recognition. New York, Academic Press. | 1972 | 12671 |

| Document type | Bibliographic reference | 1st ed. Pub. Year | GS Citations |
|---|---|---|---|
| B | BRANDRUP, J., IMMERGUT, E. H., & ELIAS, H.-G. (1966). Polymer handbook. New York, Interscience Publishers a Division of John Wiley & Sons. | 1966 | 12661 |
| B | ACHENBACH, T. M., & EDELBROCK, C. S. (1983). Manual for the child behavior checklist: and revised child behavior profile. Burlington, VT, University of Vermont, Department of Psychiatry. | 1983 | 12653 |
| B | HUNTINGTON, S. (1997). The clash of civilizations and the remaking of world order. New York, NY, Simon & Schuster. | 1997 | 12633 |
| J | MALKIEL, B. G., & FAMA, E. F. (1970). EFFICIENT CAPITAL MARKETS: A REVIEW OF THEORY AND EMPIRICAL WORK*. The Journal of Finance, 25(2), 383-417. DOI: 10.2307/2325486 | 1970 | 12623 |
| J | ALAMOUTI, S.M. (1998). A simple transmit diversity technique for wireless communications. IEEE Journal on Selected Areas in Communications, 16(8). DOI: 10.1109/49.730453 | 1998 | 12604 |
| B | BATCHELOR, G. K. (1967). An introduction to fluid dynamics. Cambridge, University Press. | 1967 | 12591 |
| B | LITTELL, R. C. et al. (1996). SAS system for mixed models. Cary, N.C., SAS Institute, Inc. | 1996 | 12569 |
| B | BRESLOW, N. E., & DAY, N. E. (1980). Statistical methods in cancer research. Lyon, International Agency for Research on Cancer. | 1980 | 12562 |

# APPENDIX B

# A Case study: *The Mathematical Theory of Communication* in Google Scholar

This work, because of its bibliographic and bibliometric complexity, collects and illustrates the problems posed by this working paper on the treatment of highly cited documents. Therefore, it has been taken as a special case study, to develop it further.

*Complexity*

"A mathematical theory of communication" constitutes an article by Claude Shannon in 1948 in the Bell System Technical Journal and that was divided in two parts published separately.

Later, in 1949, this work is expanded and reedited in book form, published by the University of Illinois Press. On this occasion, is published co-authored by Claude Shannon and Warren Weaver, and the title varies imperceptibly: "The mathematical theory of communication".

*Problematic*

Despite being two articles published in 1948 and a book published in 1949, this work appears in the results of our analysis, which we remind that is limited to the period 1950-2013. So this raises a key question: Why this document appears in our sample?

Additionally, the fact that it is composed of two distinct works (article and book), both before 1950, generating different editions and different citations, raises a number of additional issues, which affect the functioning of the versions in Google Scholar as well as a number of additional issues raised in this working paper, for example:

Has GS identified all editions of the same document? Were successfully linked all editions of the same document? Were all citations received by each edition successfully merged? Was each citation successfully linked to each of the different editions?

*Bibliographic search*

In Figure B1 we show the query search for the work in Google Scholar by identifying the result with a higher number of versions.

**Figure B1. Principal version of *The mathematical theory of communication* in Google Scholar**

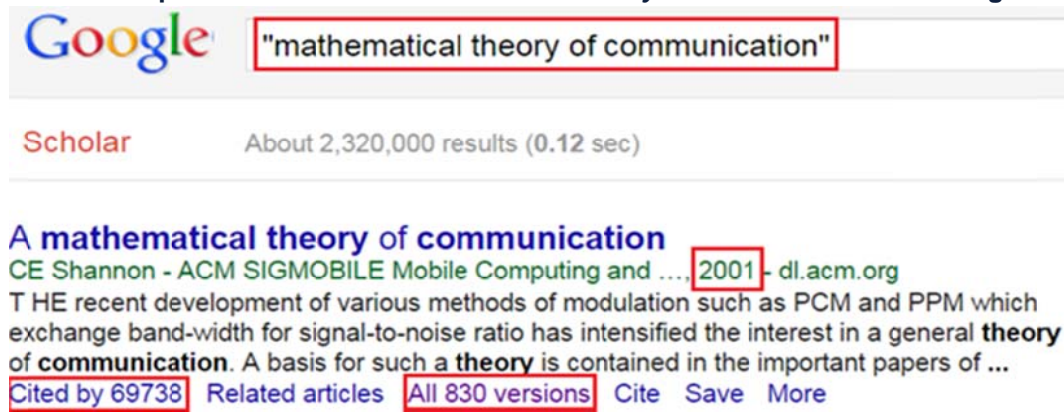

Even before trying to study the number of versions gathered, this point raises a fundamental issue: has Google Scholar merged all versions of this work?

To do this, we proceeded to refine this query (adding the search command author:Shannon), obtaining a total of 230 results, which have been analyzed to see which of them should not constitute a version (the raw data of this analysis is available in the complementary material).

Of the 230 results, 71.7% (165) are records that correspond to versions of the work, while the rest are not true versions, and they appear in the results of this query because they comment or review Shannon's work.

Of the 165 records, the first one includes a larger number of versions (shown in Figure B1). However, in the remaining 164 records, there are 3,714 potential citations (without eliminating possible duplicates).

*Number of versions for the main record*

On the one hand, we can observe the large amount of retrieved versions (830) and, on the other, that among these versions we can find both versions of the article, and versions of the book, although the latter are a minority.

Otherwise, a manual analysis gives us only 763, not the 830 displayed in Figure B1. That is, the figure shown is an approximation of the number of recovered versions. This effect has already been seen in the Hit Count Estimates to general queries (Orduna-Malea et al, 2014).

*Edition*

However, the biggest problem we found was the year of publication (2001). This is taken from the primary version, which corresponds with the last identified reprint of the book, although the dates for the rest of the versions themselves are properly identified (Figure B2).

**Figure B2. Versions grouped in Google Scholar about *The mathematical theory of communication***

 Therefore, as the primary version gives the publication year, this causes it to appear in the results of our sample, which should have been limited to the period 1950-2013.

Given the magnitude and global impact of the Shannon and Weaver work, the number of reprints and editions (in different languages) are very high, and we should also add some book reviews (as sometimes are taken as versions).

These other inquiries are answered in the remaining questions raised in the working paper, in a more detailed and comprehensive way, but especially detectable in this case study.

The decision of taking the publication date from the last available release is understandable from the point of view of the search engine service, although it limits its potential as an object of bibliometric analysis, and therefore should be considered. Nonetheless, it is likely that the number of highly cited works that are affected by this issue relatively low.

As a counterpoint to the model taken by Google Scholar to group different versions (Figure B2), an example from the library world (Figure B3) is offered, using a catalog where authority control is used (Worldcat).

As can be seen, the system recovers, after querying for the title "mathematical theory of communication" and author ("Shannon"), different versions of the book (in different languages), as well as the original article.

**Figure B3. *The mathematical theory of communication* in Worldcat**

*Analysis*

To further analyze this case, we have proceeded to download the 830 versions of the book "The Mathematical Theory of Communication", belonging to the primary version in Google Scholar, in order to understand and describe it.

The raw data for this analysis is available in the supplementary material, in a spreadsheet file. For each version, we have considered the same parameters that we have used for the 64,000 highly cited documents, fully described in the Introduction of the Working Paper.

Additionally, for each version, we manually checked if it was done correctly or not. And in those cases where the connection is unsuccessful, we have classified the different errors into categories, further detailing the reason for the error where needed.

Of the total 830 versions, Google Scholar has really returned only 763, of which 602 (78.9%) are working properly. In the remaining versions (161), the following problems occur:

- False positive: when a document has been identified as a version of another document, but actually it is not.
- Citation: false positive specific case, when the identified version is a citation rather than a document.
- Broken Link: If the link is not working properly.
- Unknown: when we have not been able to verify if the version was correct. This has occurred mainly in cases in which the files were available in PS (PostScript) file format.

In Table B1 we summarize all data about errors in the 830 different versions grouped.

**Table B1. Types of errors in the different versions of a document**

| Error Typology | Frequency |
| --- | --- |
| Citation | 23 |
| Broken link | 86 |
| False positive | 14 |
| Unknown | 38 |
| TOTAL | 161 |

# APPENDIX C

| Journal | Nº top cited articles |
|---|---|
| NATURE | 1518 |
| SCIENCE | 1437 |
| NEW ENGLAND JOURNAL OF MEDICINE | 848 |
| PHYSICAL REVIEW | 671 |
| PROCEEDINGS OF THE NATIONAL ACADEMY OF SCIENCES OF THE UNITED STATES OF AMERICA | 574 |
| CELL | 483 |
| JOURNAL OF BIOLOGICAL CHEMISTRY | 452 |
| PHYSICAL REVIEW LETTERS | 432 |
| LANCET | 363 |
| JOURNAL OF THE AMERICAN CHEMICAL SOCIETY | 328 |
| JAMA-JOURNAL OF THE AMERICAN MEDICAL ASSOCIATION | 251 |
| AMERICAN ECONOMIC REVIEW | 244 |
| ECONOMETRICA | 217 |
| PSYCHOLOGICAL REVIEW | 210 |
| REVIEWS OF MODERN PHYSICS | 206 |
| CHEMICAL REVIEWS | 203 |
| JOURNAL OF POLITICAL ECONOMY | 200 |
| JOURNAL OF PHYSIOLOGY-LONDON | 200 |
| PSYCHOLOGICAL BULLETIN | 194 |
| JOURNAL OF CHEMICAL PHYSICS | 187 |
| ASTROPHYSICAL JOURNAL | 183 |
| BIOCHEMICAL JOURNAL | 180 |
| PROCEEDINGS OF THE ROYAL SOCIETY OF LONDON SERIES A-MATHEMATICAL AND PHYSICAL SCIENCES | 180 |
| CIRCULATION | 174 |
| JOURNAL OF CLINICAL INVESTIGATION | 164 |
| JOURNAL OF PERSONALITY AND SOCIAL PSYCHOLOGY | 162 |
| PHYSICAL REVIEW D | 160 |
| JOURNAL OF EXPERIMENTAL MEDICINE | 153 |
| PHYSICAL REVIEW B | 149 |
| NUCLEIC ACIDS RESEARCH | 143 |
| JOURNAL OF MOLECULAR BIOLOGY | 141 |
| QUARTERLY JOURNAL OF ECONOMICS | 136 |
| PHYSIOLOGICAL REVIEWS | 131 |
| BIOCHIMICA ET BIOPHYSICA ACTA | 128 |
| COMMUNICATIONS OF THE ACM | 124 |

| Journal | Nº top cited articles |
|---|---|
| ANNUAL REVIEW OF BIOCHEMISTRY | 123 |
| ADMINISTRATIVE SCIENCE QUARTERLY | 116 |
| HARVARD BUSINESS REVIEW | 113 |
| ANALYTICAL BIOCHEMISTRY | 112 |
| JOURNAL OF FINANCE | 111 |
| JOURNAL OF THE AMERICAN STATISTICAL ASSOCIATION | 110 |
| AMERICAN SOCIOLOGICAL REVIEW | 109 |
| AMERICAN PSYCHOLOGIST | 106 |
| ANALYTICAL CHEMISTRY | 103 |
| JOURNAL OF CELL BIOLOGY | 102 |
| JOURNAL OF FLUID MECHANICS | 102 |
| NUCLEAR PHYSICS B | 100 |
| BRITISH MEDICAL JOURNAL | 97 |
| JOURNAL OF FINANCIAL ECONOMICS | 94 |
| JOURNAL OF MARKETING | 94 |
| IEEE TRANSACTIONS ON PATTERN ANALYSIS AND MACHINE INTELLIGENCE | 93 |
| ANNALS OF MATHEMATICAL STATISTICS | 93 |
| JOURNAL OF PHYSICAL CHEMISTRY | 90 |
| PHYSICS LETTERS B | 87 |
| HARVARD LAW REVIEW | 87 |
| ACADEMY OF MANAGEMENT REVIEW | 86 |
| CANCER RESEARCH | 84 |
| JOURNAL OF GEOPHYSICAL RESEARCH | 83 |
| AMERICAN JOURNAL OF SOCIOLOGY | 82 |
| BIOMETRIKA | 81 |
| ANNALS OF INTERNAL MEDICINE | 80 |
| ARCHIVES OF GENERAL PSYCHIATRY | 79 |
| JOURNAL OF BONE AND JOINT SURGERY-AMERICAN VOLUME | 79 |
| AMERICAN JOURNAL OF MEDICINE | 75 |
| STRATEGIC MANAGEMENT JOURNAL | 74 |
| ANNALS OF MATHEMATICS | 74 |
| NATURE GENETICS | 74 |
| BIOCHEMISTRY | 73 |
| MANAGEMENT SCIENCE | 73 |
| PHARMACOLOGICAL REVIEWS | 71 |
| IEEE TRANSACTIONS ON INFORMATION THEORY | 71 |
| JOURNAL OF NEUROPHYSIOLOGY | 69 |
| ACTA CRYSTALLOGRAPHICA | 66 |

| Journal | Nº top cited articles |
|---|---|
| REVIEW OF ECONOMIC STUDIES | 65 |
| ANGEWANDTE CHEMIE-INTERNATIONAL EDITION | 65 |
| JOURNAL OF APPLIED PHYSICS | 64 |
| PROCEEDINGS OF THE NATIONAL ACADEMY OF SCIENCES OF THE UNITED STATES OF AMERICA-BIOLOGICAL SCIENCES | 63 |
| CANCER | 63 |
| ECOLOGY | 63 |
| ANNALS OF PHYSICS | 63 |
| BIOMETRICS | 62 |
| JOURNAL OF APPLIED PHYSIOLOGY | 62 |
| PROCEEDINGS OF THE IEEE | 61 |
| AMERICAN JOURNAL OF PHYSIOLOGY | 61 |
| JOURNAL OF THE ACM | 60 |
| BLOOD | 60 |
| JOURNAL OF ECONOMIC LITERATURE | 60 |
| AMERICAN NATURALIST | 59 |
| BELL SYSTEM TECHNICAL JOURNAL | 58 |
| PHILOSOPHICAL MAGAZINE | 58 |
| METHODS IN ENZYMOLOGY | 57 |
| GEOCHIMICA ET COSMOCHIMICA ACTA | 57 |
| NATURE MEDICINE | 57 |
| PHYSICS REPORTS-REVIEW SECTION OF PHYSICS LETTERS | 57 |
| PROCEEDINGS OF THE SOCIETY FOR EXPERIMENTAL BIOLOGY AND MEDICINE | 57 |
| OPERATIONS RESEARCH | 56 |
| JOURNAL OF COMPARATIVE NEUROLOGY | 56 |
| APPLIED PHYSICS LETTERS | 55 |
| JOURNAL OF COMPUTATIONAL PHYSICS | 55 |
| JOURNAL OF IMMUNOLOGY | 55 |
| JOURNAL OF BACTERIOLOGY | 54 |
| ACADEMY OF MANAGEMENT JOURNAL | 54 |
| ARCHIVES OF BIOCHEMISTRY AND BIOPHYSICS | 53 |
| SOVIET PHYSICS JETP-USSR | 53 |
| JOURNAL OF APPLIED PSYCHOLOGY | 52 |
| GENETICS | 52 |
| ARTIFICIAL INTELLIGENCE | 52 |
| AMERICAN POLITICAL SCIENCE REVIEW | 52 |
| BIOCHEMICAL AND BIOPHYSICAL RESEARCH COMMUNICATIONS | 52 |
| JOURNAL OF THE NATIONAL CANCER INSTITUTE | 51 |
| JOURNAL OF PHARMACOLOGY AND EXPERIMENTAL THERAPEUTICS | 51 |

| Journal | Nº top cited articles |
|---|---|
| JOURNAL OF MARKETING RESEARCH | 50 |
| JOURNAL OF EXPERIMENTAL PSYCHOLOGY | 50 |
| GASTROENTEROLOGY | 49 |
| JOURNAL OF ABNORMAL AND SOCIAL PSYCHOLOGY | 48 |
| ADVANCES IN PHYSICS | 48 |
| JOURNAL OF GENERAL PHYSIOLOGY | 48 |
| COMMUNICATIONS ON PURE AND APPLIED MATHEMATICS | 47 |
| CHILD DEVELOPMENT | 47 |
| GEOLOGICAL SOCIETY OF AMERICA BULLETIN | 47 |
| JOURNAL OF THE OPTICAL SOCIETY OF AMERICA | 47 |
| ASTROPHYSICAL JOURNAL SUPPLEMENT SERIES | 46 |
| REVIEW OF ECONOMICS AND STATISTICS | 46 |
| CIRCULATION RESEARCH | 46 |
| BRAIN | 46 |
| JOURNAL OF BIOPHYSICAL AND BIOCHEMICAL CYTOLOGY | 45 |
| ANNUAL REVIEW OF IMMUNOLOGY | 45 |
| MONTHLY NOTICES OF THE ROYAL ASTRONOMICAL SOCIETY | 45 |
| NATURE REVIEWS CANCER | 45 |
| PHYSICAL REVIEW A | 44 |
| ECONOMIC JOURNAL | 44 |
| JOURNAL OF CLINICAL ONCOLOGY | 44 |
| JOURNAL OF THE ROYAL STATISTICAL SOCIETY SERIES B-STATISTICAL METHODOLOGY | 43 |
| JOURNAL OF CONSUMER RESEARCH | 43 |
| AMERICAN JOURNAL OF PSYCHIATRY | 43 |
| YALE LAW JOURNAL | 43 |
| IEEE TRANSACTIONS ON AUTOMATIC CONTROL | 42 |
| JOURNAL OF HISTOCHEMISTRY & CYTOCHEMISTRY | 42 |
| ACTA PHYSIOLOGICA SCANDINAVICA | 42 |
| NATURE REVIEWS MOLECULAR CELL BIOLOGY | 42 |
| ACTA METALLURGICA | 42 |
| BIOINFORMATICS | 41 |
| JOURNAL OF CLINICAL ENDOCRINOLOGY & METABOLISM | 41 |
| PEDIATRICS | 40 |
| GENES & DEVELOPMENT | 39 |
| ANNUAL REVIEW OF PLANT PHYSIOLOGY AND PLANT MOLECULAR BIOLOGY | 39 |
| JOURNAL OF CONSULTING AND CLINICAL PSYCHOLOGY | 38 |
| NATURE BIOTECHNOLOGY | 38 |
| ACCOUNTS OF CHEMICAL RESEARCH | 38 |

| Journal | Nº top cited articles |
|---|---|
| BRAIN RESEARCH | 38 |
| EVOLUTION | 37 |
| JOURNAL OF THEORETICAL BIOLOGY | 37 |
| AMERICAN JOURNAL OF HUMAN GENETICS | 37 |
| JOURNAL OF PHYSICS AND CHEMISTRY OF SOLIDS | 37 |
| ANNUAL REVIEW OF ECOLOGY AND SYSTEMATICS | 37 |
| NEUROLOGY | 36 |
| NATURE MATERIALS | 36 |
| JOURNAL OF COMPUTATIONAL CHEMISTRY | 36 |
| COGNITIVE PSYCHOLOGY | 36 |
| ECOLOGICAL MONOGRAPHS | 36 |
| PROGRESS OF THEORETICAL PHYSICS | 35 |
| EXPERIMENTAL CELL RESEARCH | 35 |
| JOURNAL OF ABNORMAL PSYCHOLOGY | 35 |
| PSYCHOMETRIKA | 35 |
| JOURNAL OF THE MECHANICS AND PHYSICS OF SOLIDS | 35 |
| JOURNAL OF NEUROSCIENCE | 35 |
| APPLIED OPTICS | 35 |
| ANNALS OF THE NEW YORK ACADEMY OF SCIENCES | 34 |
| JOURNAL OF ECONOMETRICS | 34 |
| ASTRONOMICAL JOURNAL | 34 |
| ENDOCRINOLOGY | 34 |
| MIS QUARTERLY | 33 |
| JOURNAL OF APPLIED CRYSTALLOGRAPHY | 33 |
| SCIENTIFIC AMERICAN | 33 |
| JOURNAL OF BUSINESS | 33 |
| ADVANCED MATERIALS | 33 |
| ANNUAL REVIEW OF PSYCHOLOGY | 33 |
| PROCEEDINGS OF THE ROYAL SOCIETY SERIES B-BIOLOGICAL SCIENCES | 33 |
| JOURNAL OF THE ATMOSPHERIC SCIENCES | 32 |
| DIABETES | 32 |
| IEEE TRANSACTIONS ON COMMUNICATIONS | 32 |
| COMMUNICATIONS IN MATHEMATICAL PHYSICS | 32 |
| JOURNAL OF THE AMERICAN COLLEGE OF CARDIOLOGY | 32 |
| TRANSACTIONS OF THE FARADAY SOCIETY | 32 |
| BELL JOURNAL OF ECONOMICS | 31 |
| IEEE TRANSACTIONS ON COMPUTERS | 31 |
| JOURNAL OF LABORATORY AND CLINICAL MEDICINE | 31 |

| Journal | Nº top cited articles |
|---|---|
| JOURNAL OF THE PHYSICAL SOCIETY OF JAPAN | 31 |
| REVIEW OF EDUCATIONAL RESEARCH | 31 |
| JOURNAL OF THE AMERICAN CERAMIC SOCIETY | 31 |
| INTERNATIONAL JOURNAL OF COMPUTER VISION | 30 |
| JOURNAL OF LAW & ECONOMICS | 30 |
| GENE | 30 |
| EMBO JOURNAL | 30 |
| QUARTERLY JOURNAL OF THE ROYAL METEOROLOGICAL SOCIETY | 30 |
| LIMNOLOGY AND OCEANOGRAPHY | 30 |
| ANNALS OF STATISTICS | 29 |
| ORGANIZATION SCIENCE | 29 |
| ANNALS OF NEUROLOGY | 29 |
| JOURNAL OF ECONOMIC THEORY | 29 |
| JOURNAL OF LIPID RESEARCH | 29 |
| PHILOSOPHICAL TRANSACTIONS OF THE ROYAL SOCIETY OF LONDON SERIES A-MATHEMATICAL AND PHYSICAL SCIENCES | 29 |
| AICHE JOURNAL | 29 |
| REPORTS ON PROGRESS IN PHYSICS | 29 |
| ARTHRITIS AND RHEUMATISM | 28 |
| LANGUAGE | 28 |
| MONTHL WEATHER REVIEW | 28 |
| NUCLEAR PHYSICS | 28 |
| COMPUTER | 28 |
| NATURE REVIEWS IMMUNOLOGY | 28 |
| ELECTROENCEPHALOGRAPHY AND CLINICAL NEUROPHYSIOLOGY | 28 |
| NEURON | 28 |
| NATURE REVIEWS GENETICS | 28 |
| ANNALS OF SURGERY | 28 |
| PHYSICA | 28 |
| BACTERIOLOGICAL REVIEWS | 28 |
| ARCHIVES OF INTERNAL MEDICINE | 28 |
| AMERICAN ANTHROPOLOGIST | 28 |
| ASTRONOMY & ASTROPHYSICS | 27 |
| JOURNAL OF ORGANIC CHEMISTRY | 27 |
| PROCEEDINGS OF THE INSTITUTE OF RADIO ENGINEERS | 27 |
| CHEMICAL PHYSICS LETTERS | 27 |
| JOURNAL OF GEOLOGY | 27 |
| IEEE TRANSACTIONS ON SYSTEMS MAN AND CYBERNETICS | 26 |
| DIABETES CARE | 26 |

| Journal | Nº top cited articles |
|---|---|
| HARVARD EDUCATIONAL REVIEW | 26 |
| MICROBIOLOGICAL REVIEWS | 26 |
| BRITISH JOURNAL OF PSYCHIATRY | 26 |
| ANNUAL REVIEW OF NEUROSCIENCE | 26 |
| JOURNAL OF ECONOMIC PERSPECTIVES | 26 |
| TRANSACTIONS OF THE AMERICAN MATHEMATICAL SOCIETY | 26 |
| JOURNAL OF VERBAL LEARNING AND VERBAL BEHAVIOR | 25 |
| ARCHIVES OF NEUROLOGY | 25 |
| JOURNAL OF NEUROCHEMISTRY | 25 |
| AMERICAN JOURNAL OF EPIDEMIOLOGY | 25 |
| ENDOCRINE REVIEWS | 25 |
| BIOLOGICAL REVIEWS OF THE CAMBRIDGE PHILOSOPHICAL SOCIETY | 25 |
| BULLETIN OF THE AMERICAN MATHEMATICAL SOCIETY | 25 |
| JOURNAL OF BONE AND JOINT SURGERY-BRITISH VOLUME | 25 |
| JOURNAL OF THE ROYAL STATISTICAL SOCIETY SERIES B-METHODOLOGICAL | 24 |
| JOURNAL OF THE ELECTROCHEMICAL SOCIETY | 24 |
| IEEE JOURNAL ON SELECTED AREAS IN COMMUNICATIONS | 24 |
| AMERICAN JOURNAL OF CLINICAL PATHOLOGY | 24 |
| VIROLOGY | 24 |
| JOURNAL OF PHYSICAL AND CHEMICAL REFERENCE DATA | 24 |
| WATER RESOURCES RESEARCH | 24 |
| IEEE TRANSACTIONS ON ACOUSTICS SPEECH AND SIGNAL PROCESSING | 24 |
| AMERICAN JOURNAL OF PATHOLOGY | 24 |
| ARCHIVE FOR RATIONAL MECHANICS AND ANALYSIS | 24 |
| AMERICAN JOURNAL OF SCIENCE | 24 |
| MEDICINE | 24 |
| JOURNAL OF MONETARY ECONOMICS | 23 |
| CA-A CANCER JOURNAL FOR CLINICIANS | 23 |
| CANADIAN JOURNAL OF PHYSICS | 23 |
| JOURNAL OF APPLIED MECHANICS-TRANSACTIONS OF THE ASME | 23 |
| PAIN | 23 |
| NATURE REVIEWS NEUROSCIENCE | 23 |
| ANNUAL REVIEW OF ASTRONOMY AND ASTROPHYSICS | 23 |
| JOURNAL OF GENERAL MICROBIOLOGY | 23 |
| COLD SPRING HARBOR SYMPOSIA ON QUANTITATIVE BIOLOGY | 23 |
| JOURNAL OF COMPARATIVE AND PHYSIOLOGICAL PSYCHOLOGY | 23 |
| MOLECULAR PHYSICS | 22 |
| HUMAN RELATIONS | 22 |

| Journal | Nº top cited articles |
|---|---|
| NATURE IMMUNOLOGY | 22 |
| AMERICAN JOURNAL OF CARDIOLOGY | 22 |
| CHEMICAL ENGINEERING SCIENCE | 22 |
| JOURNAL OF ACCOUNTING & ECONOMICS | 22 |
| AMERICAN JOURNAL OF CLINICAL NUTRITION | 22 |
| JOURNAL OF NEUROSURGERY | 22 |
| ZEITSCHRIFT FUR ELEKTROCHEMIE | 22 |
| AMERICAN JOURNAL OF MATHEMATICS | 22 |
| MOLECULAR BIOLOGY AND EVOLUTION | 21 |
| ACTA CRYSTALLOGRAPHICA SECTION D | 21 |
| PUBLIC OPINION QUARTERLY | 21 |
| SIAM JOURNAL ON COMPUTING | 21 |
| NUCLEAR PHYSICS A | 21 |
| TECHNOMETRICS | 21 |
| BEHAVIORAL AND BRAIN SCIENCES | 21 |
| LINGUISTIC INQUIRY | 21 |
| BULLETIN OF THE SEISMOLOGICAL SOCIETY OF AMERICA | 21 |
| CLINICAL ORTHOPAEDICS AND RELATED RESEARCH | 21 |
| CHEMICAL SOCIETY REVIEWS | 21 |
| INDUSTRIAL AND ENGINEERING CHEMISTRY | 21 |
| ACTA CRYSTALLOGRAPHICA SECTION A | 20 |
| APPLIED AND ENVIRONMENTAL MICROBIOLOGY | 20 |
| COGNITIVE SCIENCE | 20 |
| EARTH AND PLANETARY SCIENCE LETTERS | 20 |
| AUTOMATICA | 20 |
| COMPUTING SURVEYS | 20 |
| JOURNAL OF SOCIAL ISSUES | 20 |
| SLOAN MANAGEMENT REVIEW | 20 |
| NATURE NANOTECHNOLOGY | 20 |
| DOKLADY AKADEMII NAUK SSSR | 20 |
| SOIL SCIENCE | 20 |
| AAPG BULLETIN--AMERICAN ASSOCIATION OF PETROLEUM GEOLOGISTS | 20 |
| PSYCHOANALYTIC STUDY OF THE CHILD | 20 |
| COMPUTER JOURNAL | 19 |
| JOURNAL OF CHRONIC DISEASES | 19 |
| MATHEMATICS OF COMPUTATION | 19 |
| QUARTERLY REVIEW OF BIOLOGY | 19 |
| JOURNAL OF APPLIED MECHANICS | 19 |

| Journal | Nº top cited articles |
|---|---|
| AMERICAN JOURNAL OF OBSTETRICS AND GYNECOLOGY | 19 |
| REVIEW OF FINANCIAL STUDIES | 19 |
| ENVIRONMENTAL SCIENCE & TECHNOLOGY | 19 |
| COGNITION | 19 |
| COMPUTER PHYSICS COMMUNICATIONS | 19 |
| PLANT PHYSIOLOGY | 19 |
| NATURE REVIEWS DRUG DISCOVERY | 19 |
| AMERICAN HEART JOURNAL | 19 |
| TRANSACTIONS OF THE AMERICAN INSTITUTE OF MINING AND METALLURGICAL ENGINEERS | 19 |
| JOURNAL OF NEUROLOGY NEUROSURGERY AND PSYCHIATRY | 18 |
| MACHINE LEARNING | 18 |
| RADIOLOGY | 18 |
| SIAM REVIEW | 18 |
| STROKE | 18 |
| RESEARCH POLICY | 18 |
| IEEE-ACM TRANSACTIONS ON NETWORKING | 18 |
| JOURNAL OF CONSULTING PSYCHOLOGY | 18 |
| GENOME RESEARCH | 18 |
| AMERICAN JOURNAL OF RESPIRATORY AND CRITICAL CARE MEDICINE | 18 |
| NEUROIMAGE | 18 |
| INTERNATIONAL JOURNAL FOR NUMERICAL METHODS IN ENGINEERING | 18 |
| JOURNAL OF MAGNETIC RESONANCE | 18 |
| IBM JOURNAL OF RESEARCH AND DEVELOPMENT | 18 |
| PHILOSOPHICAL TRANSACTIONS OF THE ROYAL SOCIETY OF LONDON SERIES B-BIOLOGICAL SCIENCES | 18 |
| STAIN TECHNOLOGY | 18 |
| ANNUAL REVIEW OF PHYSIOLOGY | 18 |
| CLINICA CHIMICA ACTA | 18 |
| COLUMBIA LAW REVIEW | 18 |
| ACTA CHEMICA SCANDINAVICA | 18 |
| JOURNAL OF THE CHEMICAL SOCIETY | 18 |
| CALIFORNIA MANAGEMENT REVIEW | 17 |
| ADVANCES IN PROTEIN CHEMISTRY | 17 |
| FASEB JOURNAL | 17 |
| IEEE TRANSACTIONS ON SOFTWARE ENGINEERING | 17 |
| CLINICAL MICROBIOLOGY REVIEWS | 17 |
| BIOPHYSICAL JOURNAL | 17 |
| PROCEEDINGS OF THE AMERICAN MATHEMATICAL SOCIETY | 17 |

| Journal | Nº top cited articles |
|---|---|
| MACROMOLECULES | 17 |
| JOURNAL OF MATHEMATICAL PHYSICS | 17 |
| NUOVO CIMENTO | 17 |
| AMERICAN REVIEW OF RESPIRATORY DISEASE | 17 |
| CLINICAL SCIENCE | 17 |
| ACTA MATHEMATICA | 17 |
| MEDICAL CARE | 16 |
| CLINICAL CHEMISTRY | 16 |
| SCANDINAVIAN JOURNAL OF CLINICAL & LABORATORY INVESTIGATION | 16 |
| IEEE TRANSACTIONS ON IMAGE PROCESSING | 16 |
| QUARTERLY JOURNAL OF EXPERIMENTAL PSYCHOLOGY | 16 |
| JOURNAL OF THE ACOUSTICAL SOCIETY OF AMERICA | 16 |
| CHEST | 16 |
| JOURNAL OF EDUCATIONAL PSYCHOLOGY | 16 |
| AIAA JOURNAL | 16 |
| JOURNAL OF UROLOGY | 16 |
| BRITISH JOURNAL OF PHARMACOLOGY AND CHEMOTHERAPY | 16 |
| BEHAVIOUR RESEARCH AND THERAPY | 16 |
| TRENDS IN NEUROSCIENCES | 16 |
| ACM TRANSACTIONS ON COMPUTER SYSTEMS | 16 |
| NANO LETTERS | 16 |
| IEEE SIGNAL PROCESSING MAGAZINE | 16 |
| EUROPEAN HEART JOURNAL | 16 |
| CONTRIBUTIONS TO MINERALOGY AND PETROLOGY | 16 |
| MOLECULAR CELL | 16 |
| JOURNAL OF PHILOSOPHY | 16 |
| BOTANICAL REVIEW | 16 |
| INFORMATION AND CONTROL | 15 |
| JOURNAL OF IMMUNOLOGICAL METHODS | 15 |
| ANNUAL REVIEW OF SOCIOLOGY | 15 |
| EUROPEAN JOURNAL OF BIOCHEMISTRY | 15 |
| ADVANCED DRUG DELIVERY REVIEWS | 15 |
| JOURNAL OF PHYSICAL CHEMISTRY B | 15 |
| JOURNAL OF INTERNATIONAL BUSINESS STUDIES | 15 |
| ZEITSCHRIFT FUR PHYSIK | 15 |
| PSYCHOSOMATIC MEDICINE | 15 |
| INTERNATIONAL JOURNAL OF CANCER | 15 |
| INTERNATIONAL ORGANIZATION | 15 |

| Journal | Nº top cited articles |
|---|---|
| PROCEEDINGS OF THE ROYAL SOCIETY OF LONDON SERIES A-MATHEMATICAL PHYSICAL AND ENGINEERING SCIENCES | 15 |
| BRITISH JOURNAL OF SURGERY | 15 |
| PHILOSOPHICAL REVIEW | 15 |
| PSYCHOLOGICAL MEDICINE | 15 |
| INTERNATIONAL JOURNAL OF PSYCHOANALYSIS | 15 |
| SYSTEMATIC ZOOLOGY | 15 |
| WORD-JOURNAL OF THE INTERNATIONAL LINGUISTIC ASSOCIATION | 15 |
| JOURNAL OF MATHEMATICAL ANALYSIS AND APPLICATIONS | 15 |
| JOURNAL OF THE FISHERIES RESEARCH BOARD OF CANADA | 15 |
| INTERNATIONAL JOURNAL OF PSYCHO-ANALYSIS | 15 |
| FEBS LETTERS | 15 |
| JOURNAL OF THE AMERICAN PSYCHOANALYTIC ASSOCIATION | 15 |
| ARCHIVES OF OPHTHALMOLOGY | 15 |
| UNIVERSITY OF CHICAGO LAW REVIEW | 15 |
| CANCER CELL | 15 |
| JOURNAL OF POLYMER SCIENCE | 15 |
| AMERICAN JOURNAL OF ANATOMY | 15 |
| JOURNAL OF MANAGEMENT | 14 |
| JOURNAL OF THE SOCIETY FOR INDUSTRIAL AND APPLIED MATHEMATICS | 14 |
| BRITISH JOURNAL OF CANCER | 14 |
| ACM COMPUTING SURVEYS | 14 |
| IRE TRANSACTIONS ON INFORMATION THEORY | 14 |
| JOURNAL OF COLLOID AND INTERFACE SCIENCE | 14 |
| MOLECULAR AND CELLULAR BIOLOGY | 14 |
| JOURNAL OF HIGH ENERGY PHYSICS | 14 |
| PHYSICA D | 14 |
| PERSONNEL PSYCHOLOGY | 14 |
| JOURNAL OF ACCOUNTING RESEARCH | 14 |
| JOURNAL OF PETROLOGY | 14 |
| IEEE TRANSACTIONS ON NEURAL NETWORKS | 14 |
| JOURNAL OF INTERNATIONAL ECONOMICS | 14 |
| OPTICS LETTERS | 14 |
| GEOPHYSICAL JOURNAL OF THE ROYAL ASTRONOMICAL SOCIETY | 14 |
| JOURNAL OF EXPERIMENTAL PSYCHOLOGY-GENERAL | 14 |
| ZHURNAL EKSPERIMENTALNOI I TEORETICHESKOI FIZIKI | 14 |
| JOURNAL OF COMPUTER AND SYSTEM SCIENCES | 14 |
| ANNUAL REVIEW OF CELL AND DEVELOPMENTAL BIOLOGY | 14 |
| JOURNAL OF CLINICAL PATHOLOGY | 14 |

| Journal | Nº top cited articles |
|---|---|
| BULLETIN OF THE WORLD HEALTH ORGANIZATION | 14 |
| NEUROSCIENCE | 14 |
| SOCIAL PROBLEMS | 14 |
| SOLID STATE PHYSICS-ADVANCES IN RESEARCH AND APPLICATIONS | 14 |
| FEDERATION PROCEEDINGS | 14 |
| AMERICAN JOURNAL OF PSYCHOLOGY | 14 |
| PSYCHOLOGICAL MONOGRAPHS | 13 |
| JOURNAL OF RETAILING | 13 |
| IEEE COMMUNICATIONS MAGAZINE | 13 |
| JOURNAL OF DAIRY SCIENCE | 13 |
| NATURE-NEW BIOLOGY | 13 |
| JOURNAL OF PHYSICS C-SOLID STATE PHYSICS | 13 |
| FREE RADICAL BIOLOGY AND MEDICINE | 13 |
| TELLUS | 13 |
| MONOGRAPHS OF THE SOCIETY FOR RESEARCH IN CHILD DEVELOPMENT | 13 |
| ACM TRANSACTIONS ON PROGRAMMING LANGUAGES AND SYSTEMS | 13 |
| HEPATOLOGY | 13 |
| LABORATORY INVESTIGATION | 13 |
| IMMUNITY | 13 |
| ANGEWANDTE CHEMIE-INTERNATIONAL EDITION IN ENGLISH | 13 |
| CLINICAL INFECTIOUS DISEASES | 13 |
| NATURE NEUROSCIENCE | 13 |
| QUARTERLY OF APPLIED MATHEMATICS | 13 |
| SOLID STATE COMMUNICATIONS | 13 |
| NATURE CELL BIOLOGY | 13 |
| PSYCHIATRY | 13 |
| BRITISH JOURNAL OF PHARMACOLOGY | 13 |
| WORLD POLITICS | 13 |
| ANESTHESIOLOGY | 13 |
| JOURNAL OF MARINE RESEARCH | 13 |
| JOURNAL OF PATHOLOGY AND BACTERIOLOGY | 13 |
| ANTIBIOTICS AND CHEMOTHERAPY | 13 |
| SOIL SCIENCE SOCIETY OF AMERICA JOURNAL | 12 |
| ACTA CRYSTALLOGRAPHICA SECTION B | 12 |
| COORDINATION CHEMISTRY REVIEWS | 12 |
| IEEE TRANSACTIONS ON SIGNAL PROCESSING | 12 |
| ORGANIZATIONAL BEHAVIOR AND HUMAN PERFORMANCE | 12 |
| MAGNETIC RESONANCE IN MEDICINE | 12 |

| Journal | Nº top cited articles |
|---|---|
| PATTERN RECOGNITION | 12 |
| PROCEEDINGS OF THE INSTITUTE OF ELECTRICAL AND ELECTRONICS ENGINEERS | 12 |
| BIOSCIENCE | 12 |
| GEOPHYSICS | 12 |
| PACIFIC JOURNAL OF MATHEMATICS | 12 |
| BRITISH JOURNAL OF PSYCHOLOGY | 12 |
| JOURNAL OF CATALYSIS | 12 |
| SIAM JOURNAL ON NUMERICAL ANALYSIS | 12 |
| JOURNAL OF THE EXPERIMENTAL ANALYSIS OF BEHAVIOR | 12 |
| JOURNAL OF LEGAL STUDIES | 12 |
| JOURNAL OF ANIMAL ECOLOGY | 12 |
| PHYSICAL REVIEW A-GENERAL PHYSICS | 12 |
| AGRONOMY JOURNAL | 12 |
| SURFACE SCIENCE | 12 |
| PROCEEDINGS OF THE PHYSICAL SOCIETY OF LONDON SECTION A | 12 |
| SURGERY GYNECOLOGY & OBSTETRICS | 12 |
| BULLETIN OF THE AMERICAN METEOROLOGICAL SOCIETY | 11 |
| BIOCHEMICAL PHARMACOLOGY | 11 |
| CRITICAL CARE MEDICINE | 11 |
| ADVANCES IN EXPERIMENTAL SOCIAL PSYCHOLOGY | 11 |
| GERONTOLOGIST | 11 |
| JOURNAL OF HEALTH AND SOCIAL BEHAVIOR | 11 |
| NEURAL COMPUTATION | 11 |
| BIOLOGICAL CYBERNETICS | 11 |
| BRITISH JOURNAL OF NUTRITION | 11 |
| IEEE JOURNAL OF QUANTUM ELECTRONICS | 11 |
| REVIEWS OF GEOPHYSICS | 11 |
| ANNALS OF THE RHEUMATIC DISEASES | 11 |
| DEVELOPMENTAL PSYCHOLOGY | 11 |
| PROGRESS IN MATERIALS SCIENCE | 11 |
| ACCOUNTING REVIEW | 11 |
| TRENDS IN COGNITIVE SCIENCES | 11 |
| SOVIET JOURNAL OF NUCLEAR PHYSICS-USSR | 11 |
| ANNUAL REVIEW OF FLUID MECHANICS | 11 |
| INTERNATIONAL ECONOMIC REVIEW | 11 |
| INORGANIC CHEMISTRY | 11 |
| ANNUAL REVIEW OF MICROBIOLOGY | 11 |
| JOURNAL OF PEDIATRICS | 11 |

| Journal | Nº top cited articles |
|---|---|
| DISCUSSIONS OF THE FARADAY SOCIETY | 11 |
| LIFE SCIENCES | 11 |
| ECONOMICA | 11 |
| PHYSICS LETTERS A | 11 |
| JOURNAL OF ANATOMY | 11 |
| JOURNAL OF PUBLIC ECONOMICS | 11 |
| STANFORD LAW REVIEW | 11 |
| PROCEEDINGS OF THE PHYSICAL SOCIETY OF LONDON SECTION B | 11 |
| JOURNAL OF PERSONALITY | 11 |
| AMERICAN EDUCATIONAL RESEARCH JOURNAL | 11 |